\documentclass[sn-standardnature]{sn-jnl}

\usepackage{lineno}
\usepackage{siunitx}
\usepackage{mathtools}
\usepackage{bm}
\modulolinenumbers[1]
\usepackage{url}

\jyear{2022}

\unnumbered

\begin{document}

\title[ ]{Emulator-based Bayesian Inference on Non-Proportional Scintillation Models by Compton-Edge Probing}


\author*[1,2]{\fnm{David} \sur{Breitenmoser}}\email{david.breitenmoser@psi.ch}

\author[3]{\fnm{Francesco} \sur{Cerutti}}

\author[1]{\fnm{Gernot} \sur{Butterweck}}

\author[1]{\fnm{Malgorzata Magdalena} \sur{Kasprzak}}

\author[1]{\fnm{Sabine} \sur{Mayer}}

\affil[1]{\orgdiv{Department of Radiation Safety and Security}, \orgname{Paul~Scherrer~Institute (PSI)}, \orgaddress{\street{Forschungsstrasse~111}, \city{Villigen~PSI}, \postcode{5232}, \country{Switzerland}}}

\affil[2]{\orgdiv{Department of Physics}, \orgname{Swiss Federal Institute of Technology (ETH)}, \orgaddress{\street{Otto-Stern-Weg~5}, \city{Zurich}, \postcode{8093}, \country{Switzerland}}}

\affil[3]{\orgname{European Organization for Nuclear Research (CERN)}, \orgaddress{\street{Esplanade~des~Particules~1}, \\ \city{Geneva}, \postcode{1211}, \country{Switzerland}}}


\abstract{Scintillator detector response modeling has become an essential tool in various research fields such as particle and nuclear physics, astronomy or geophysics. Yet, due to the system complexity and the requirement for accurate electron response measurements, model inference and calibration remains a challenge. Here, we propose Compton edge probing to perform non-proportional scintillation model (NPSM) inference for inorganic scintillators.
We use laboratory-based gamma-ray radiation measurements with a NaI(Tl) scintillator to perform Bayesian inference on a NPSM. Further, we apply machine learning to emulate the detector response obtained by Monte Carlo simulations.
We show that the proposed methodology successfully constrains the NPSM and hereby quantifies the intrinsic resolution. Moreover, using the trained emulators, we can predict the spectral Compton edge dynamics as a function of the parameterized scintillation mechanisms.
The presented framework offers a novel way to infer NPSMs for any inorganic scintillator without the need for additional electron response measurements. 
}


\maketitle



\newpage
\section{Introduction}\label{sec:intro}

Inorganic scintillation detectors are a prevalent tool to measure ionizing radiation in various research fields such as nuclear and particle physics, astronomy or planetary science \citep{Cano-Ott1999,Bernabei2008,Adhikari2018AnDetectors,Paynter2021EvidenceBurst,Yang2022AOrigin,Lawrence1998,Trombka2000}. Other applications include radiation protection, medical diagnostics and homeland security \citep{Bashkirov2016NovelTomography,Curtis2020}. In almost all applications, the measured signal needs to be deconvolved to infer the properties of interest, e.g. the flux from a gamma-ray burst or the elemental composition on a comet. This deconvolution requires accurate detector response models and consequently detailed knowledge about the scintillation mechanisms themselves. 

Detector response models can either be derived empirically by radiation measurements or numerically using Monte Carlo simulations \citep{Knoll2010}. Regarding the numerical derivation, the most common approach to simulate the detector response is to use a proportional energy deposition model. In this model, the scintillation light yield is assumed to be proportional to the deposited energy \citep{Lawrence1998,Prettyman2011}. Consequently, the detector response characterization is reduced to a comparably simple energy deposition problem, which can be solved by any standard multi-purpose Monte Carlo code. 

However, thanks to the development of the Compton coincidence measurement technique \citep{Valentine1994}, recent studies could conclusively confirm the conjecture reported in earlier investigations \citep{Engelkemeir2004,Saito1981,Gardner2004} that not only organic but also inorganic scintillators exhibit a pronounced non-proportional relation between the deposited energy and the scintillation light yield \citep{Moses2008,Payne2009,Payne2011}. The origin of this scintillation non-proportionality seems to be linked to the intrinsic scintillation response to electrons and the different mechanisms associated with the creation and transport of excitation carriers in the scintillation crystal \citep{Moses2012,VasilEv2014MultiscaleCharacteristics}. Nonetheless, our understanding about these phenomena is still far from complete and, thanks to the advent of novel experimental techniques and the development of new scintillator materials, interest in scintillation physics has steadily grown over the past years \citep{Moses2008,Moses2012,VasilEv2014MultiscaleCharacteristics,Payne2009,Payne2011,Khodyuk2012TrendsNonproportionality,Payne2014,Payne2015,Beck2015}. 

Regarding the detector response modeling, the scintillation non-proportionality has two major implications. First, it leads to an intrinsic spectral broadening and thereby sets a lower limit on the spectral resolution achievable with the corresponding scintillator \citep{Zerby1961,Hill1966,Prescott1969ElectronNaITl,Valentine1998a,Cano-Ott1999}. Second, various studies stated the conjecture that specific spectral features such as the Compton edges are shifted and distorted as a result of the non-proportional scintillation response \citep{Saito1981,Shi2002,Gardner2004,Cano-Ott1999,Breitenmoser2022ExperimentalSpectrometry}. Furthermore, additional studies revealed a complex dependence of the scintillation non-proportionality on various scintillator properties including the activator concentration, the temperature and the crystal size, among others \citep{Murray1961b,Hill1966b,Zerby1961,Valentine1998a,Cano-Ott1999,Swiderski2006,Hull2009,Khodyuk2012TrendsNonproportionality,Payne2014}.

Based on these findings, we conclude that non-proportional scintillation models (NPSM) should be included in the detector response simulations to prevent systematic errors in the predicted spectral response. Non-proportional effects are known to increase with increasing crystal size \citep{Murray1961b,Zerby1961,Valentine1998a}. NPSMs are therefore particularly relevant for scintillators with large crystal volumes, e.g. in dark matter research, total absorption spectroscopy or remote sensing \citep{Cano-Ott1999,Bernabei2008,Adhikari2018AnDetectors,Paynter2021EvidenceBurst,Yang2022AOrigin,Lawrence1998,Trombka2000,Breitenmoser2022ExperimentalSpectrometry}. In addition, especially due to the sensitivity on the activator concentration and impurities \citep{Hull2009}, NPSMs need to be calibrated for each individual detector system. In case the scintillator properties change after detector deployment, e.g. due to radiation damage or temperature changes in space, this calibration should be repeated regularly. 

Currently, K-dip spectroscopy, the already mentioned Compton coincidence technique as well as electron beam measurements are the only available methods to calibrate NPSM \citep{Porter1966,Valentine1994,Wayne1998ResponseElectrons,Choong2008DesignNon-proportionality,Khodyuk2010}. Moreover, only a very limited number of laboratories are able to perform these measurements. Therefore, these methods are not readily available for extensive calibration campaigns of custom detectors, e.g. large satellite probes or scintillators for dark matter research. Additionally, they cannot be applied during detector deployment, which, as discussed above, might be important for certain applications such as deep space missions.

In this study, we propose Compton edge probing together with Bayesian inversion to infer and calibrate NPSMs. This approach is motivated by the already mentioned conjecture, that the Compton edge shifts as a result of the scintillation non-proportionality \citep{Saito1981,Shi2002,Gardner2004,Cano-Ott1999,Breitenmoser2022ExperimentalSpectrometry}. We obtain the spectral Compton edge data by gamma-ray spectrometry using a NaI(Tl) scintillator and calibrated radionuclide sources for photon irradiations under laboratory conditions. We apply Bayesian inversion with state-of-the-art Markov-Chain Monte Carlo algorithms \citep{Goodman2010EnsembleInvariance} to perform the NPSM inference with the gamma-ray spectral data. In contrast to traditional frequentist methods or simple data-driven optimization algorithms, a Bayesian approach offers a natural, consistent and transparent way of combining prior information with empirical data to infer scientific model properties using a solid decision theory framework \citep{Kennedy2001BayesianModels,Trotta2008BayesCosmology,Gelman2013BayesianAnalysis}. We simulate the detector response using a multi-purpose Monte Carlo radiation transport code in combination with parallel computing. To meet the required evaluation speed for the Bayesian inversion solver, we use machine learning trained polynomial chaos expansion (PCE) surrogate models to emulate the simulated detector response \citep{Torre2019}. This approach offers not only a novel way to calibrate NPSMs with minimal effort---especially during the detector deployment---but it also allows new insights into the non-proportional scintillation physics without the need for additional electron response measurements.


\section{Results}\label{sec:results}

\subsection{Compton edge probing}
\label{subsec:RESprobing}

\begin{figure}[t!]
\centerline{
\includegraphics[]{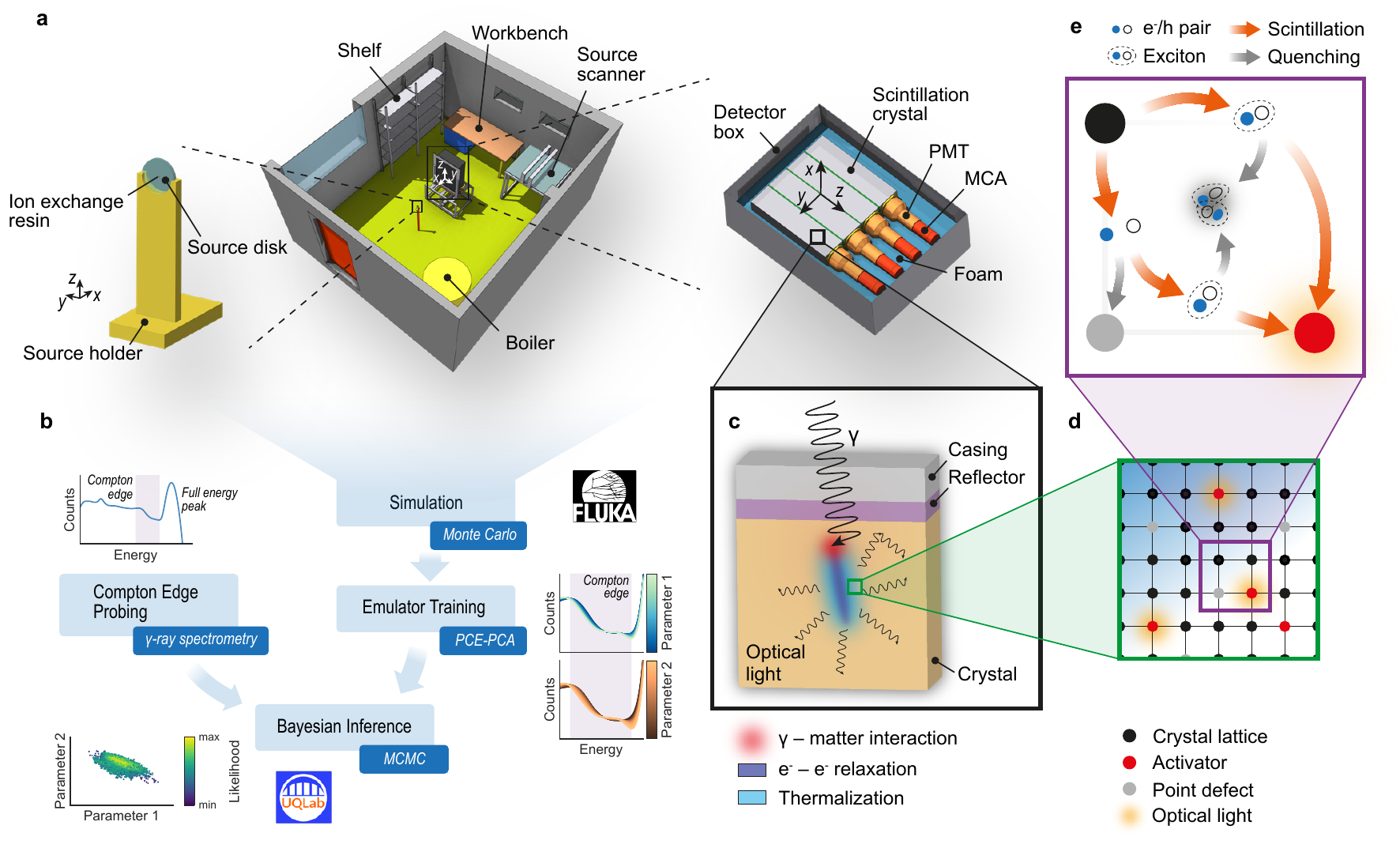}}
\caption{\textbf{ Compton edge probing to perform Bayesian inference on non-proportional scintillation models.} \textbf{a}~Monte Carlo mass model of the experimental setup to perform Compton edge probing with an inorganic gamma-ray scintillation spectrometer under laboratory conditions. The spectrometer consists of four individual 10.2~\unit{cm}~$\times$~10.2~\unit{cm}~$\times$~40.6~\unit{cm} prismatic NaI(Tl) scintillation crystals with the associated photomultiplier tubes (PMT), the electronic components, e.g. the multi-channel analyzers (MCA), embedded in a thermal-insulating and vibration-damping polyethylene (PE) foam protected by a rugged aluminum detector box. We inserted radiation sources consisting of a radionuclide carrying ion exchange sphere (diameter 1~\unit{mm}) embedded in a 25~\unit{mm}~$\times$~3~\unit{mm} solid plastic disc into a custom low absorption source holder made out of a polylactide polymer (PLA) and placed this holder on a tripod in a fixed distance of 1~\unit{m} to the detector front on the central detector $x$-axis. The mass model figures were created using the graphical interface \texttt{FLAIR} \citep{Vlachoudis2009}. For better visibility and interpretability, we applied false colors. \textbf{b}~Overview of the Bayesian inference framework highlighting the gamma-ray spectrometry based Compton edge probing measurements, the Monte Carlo simulations using the multi-purpose code \texttt{FLUKA} \citep{Ahdida2022NewCode} combined with the machine learning trained polynomial chaos expansion (PCE) emulator models supported by principal component analysis (PCA) as well as the Bayesian inference by Markov Chain Monte Carlo (MCMC) itself using \texttt{UQLab} \citep{Marelli2014UQLab:Matlab}. \textbf{c}~Radiation transport mechanisms inside the inorganic scintillation crystal, which is surrounded by a thin reflector layer and a rugged aluminum crystal casing. \textbf{d}~Schematic representation of an inorganic scintillation crystal lattice including the activator atoms and point defects. \textbf{e}~Mechanistic depictions of the various scintillation and quenching pathways for electron-hole pairs ($e^{-}/h$) as well as excitons within the inorganic scintillation crystal lattice.}\label{fig:scheme}
\end{figure}

To obtain the spectral Compton edge data, we performed gamma-ray spectrometry under controlled laboratory conditions \citep{Breitenmoser2022ExperimentalSpectrometry}. The adopted spectrometer consisted of four 10.2~\unit{cm}~$\times$~10.2~\unit{cm}~$\times$~40.6~\unit{cm} prismatic NaI(Tl) scintillation crystals with individual read-out. We used seven different calibrated radionuclide sources ($^{57}\text{Co}$, $^{60}\text{Co}$, $^{88}\text{Y}$, $^{109}\text{Cd}$, $^{133}\text{Ba}$, $^{137}\text{Cs}$ and $^{152}\text{Eu}$) for the radiation measurements. However, only $^{60}\text{Co}$, $^{88}\text{Y}$ and $^{137}\text{Cs}$ could be used for Compton edge probing. For the remaining sources, the Compton edges were obscured by additional full energy peaks (FEPs) and associated Compton continua. We used those remaining sources for energy and resolution calibrations. A schematic depiction of the measurement setup is shown in \hyperref[fig:scheme]{Fig.~\ref{fig:scheme}a}.

\subsection{Forward modeling}
\label{subsec:RESforward}

We simulated the detector response for the performed radiation measurements using the multi-purpose Monte Carlo code \texttt{FLUKA} \citep{Ahdida2022NewCode}. The performed simulations feature fully coupled photon, electron and positron radiation transport for
our source-detector configuration with a lower kinetic energy threshold of 1~\unit{keV}. As shown in \hyperref[fig:scheme]{Fig.~\ref{fig:scheme}a}, the applied mass model includes all relevant detector and source components in high detail. On the other hand, the laboratory room together with additional instruments and equipment are modelled in less detail. For this simplifications, care was taken to preserve the overall opacity as well as the mass density.

We used a mechanistic model recently published by Payne and his co-workers to include the non-proportional scintillation physics in our simulations \citep{Payne2009,Payne2011,Payne2014}. In general, the sequence of scintillation processes in inorganic scintillators can be qualitatively divided into five steps \citep{Rodnyi1997PhysicalScintillators,VasilEv2014MultiscaleCharacteristics,Lecoq2017}. After interaction of the ionizing radiation with the scintillator, the emitted high-energetic electrons are relaxed by the production of numerous secondary electrons, phonons and plasmons. The low energetic secondary electrons are then thermalized by a phonon coupling mechanism producing excitation carriers, i.e. electron-hole pairs ($e^{-}/h$) and excitons. These excitation carriers are then transferred to the luminescent centers within the scintillation crystal, where they recombine and induce radiative relaxation of the excited luminescent centers producing scintillation photons. The first two processes, i.e. the interaction of the ionizing radiation with the scintillator as well as the $e^{-}$--$e^{-}$ relaxation, are explicitly simulated by the Monte Carlo code. The creation and migration of the excitation carriers on the other hand is accounted for by Payne's mechanistic model.

In this mechanistic model it is assumed that only excitons are capable to radiatively recombine at the luminescent centers. Consequently, $e^{-}/h$ pairs need to convert to excitons by the classic Onsager mechanism \citep{Onsager1938InitialIons} in order to contribute to the scintillation emission. In addition, creation and migration of the excitation carriers compete with several quenching phenomena. The quenching mechanisms considered in Payne's model are the trapping of $e^{-}/h$ pairs at point defects \citep{VasilEv2014MultiscaleCharacteristics,Payne2014} as well as exciton--exciton annihilation described by the Birks mechanism \citep{Birks1951}. 

Using this NPSM, the non-proportional light yield $L$ as a function of the differential energy loss $dE$ per differential path length $ds$ for electrons is given by \citep{Payne2014}:

\begin{equation}
    L\left( dE/ds\right) \propto \frac{
    1-\eta_{e/h}\exp{ \left[ -\frac{dE/ds}{dE/ds\mid_{\text{Ons}}} 
    \exp{\left( - \frac{dE/ds\mid_{\text{Trap}}}{dE/ds} \right)}  
    \right] 
    }
    }{
    1+\frac{dE/ds}{dE/ds\mid_{\text{Birks}}}
    }  
\label{eq:StateEq}
\end{equation}

\noindent where $\eta_{e/h}$, $dE/ds\mid_{\text{Ons}}$, $dE/ds\mid_{\text{Trap}}$ and $dE/ds\mid_{\text{Birks}}$ are the model parameters characterizing the fraction of excitation carriers, which are created as $e^{-}/h$ pairs at the thermalization phase, as well as the stopping power related to the Onsager, trapping and Birks mechanisms, respectively. As a result, all the parameters of the NPSM reflect physical processes after thermalization of the secondary particles, i.e. generation and transport of excitation carriers. Consequently, these processes and thereby also the corresponding parameters can be regarded as statistically independent with respect to the energy of the secondary particles. From a physics perspective, it is important to note that the Onsager and trapping mechanisms are coupled in a nonlinear way, whereas the Birks mechanism can be regarded as independent of the other mechanisms. As discussed in detail by Vasil'ev and Gektin \citep{VasilEv2014MultiscaleCharacteristics}, we may therefore interpret the trapping of $e^{-}/h$ pairs as a screening mechanism on the Onsager term in \hyperref[eq:StateEq]{Eq.~\ref{eq:StateEq}}. A scheme highlighting the individual scintillation processes included in the present study is presented in the \hyperref[fig:scheme]{Figs.~\ref{fig:scheme}c--e}.
 
\begin{figure}[t!]
\centerline{
\includegraphics[]{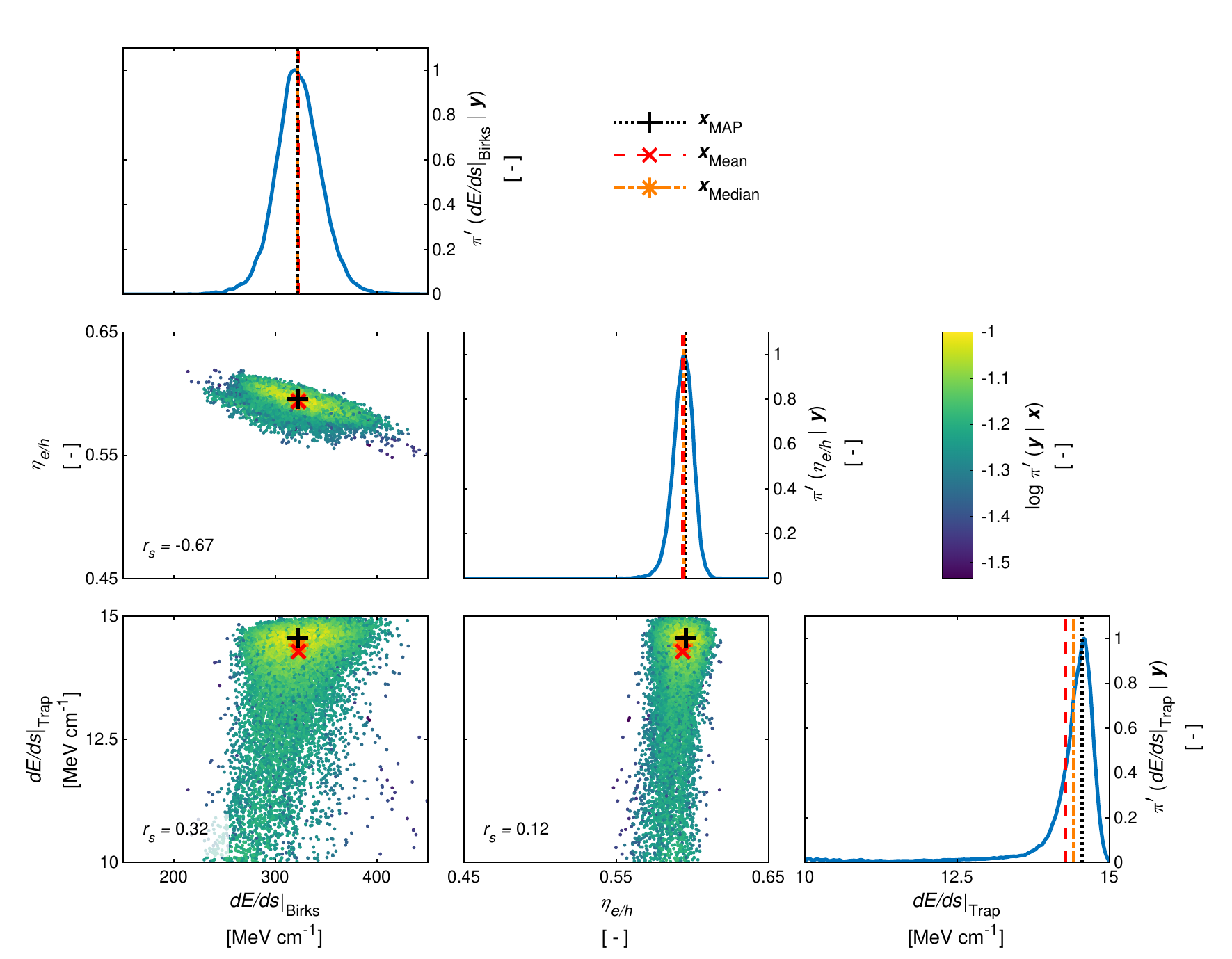}}
\caption{\textbf{ Posterior distribution estimate.} As a result of the sum mode inversion pipeline, the off-diagonal subfigures present samples from the multivariate posterior marginals given the experimental dataset $\boldsymbol{y}$ for the model parameters $\boldsymbol{x}\coloneqq\left( dE/ds\mid_{\text{Birks}},~\eta_{e/h},~dE/ds\mid_{\text{Trap}} \right)^{\intercal}$. We colored these samples by the corresponding normalized multivariate log-likelihood function values $\operatorname{log} \pi^{\prime}\left( \boldsymbol{y} \mid \boldsymbol{x} \right)$. In addition, the 
Spearman's rank correlation coefficient $r_s$ is provided for the model parameters in the corresponding off-diagonal subfigures. The subfigures on the diagonal axis highlight the normalized univariate marginal likelihood $\pi^{\prime}\left( x \mid \boldsymbol{y}\right)$ for the model parameter $x$. Both, the univariate and multivariate likelihood values, were normalized by their corresponding global maxima. Derived posterior point estimators, i.e. the maximum a posteriori (MAP) probability estimate $\boldsymbol{x}_{\text{MAP}}$, the posterior mean $\boldsymbol{x}_{\text{Mean}}$ and the posterior median $\boldsymbol{x}_{\text{Median}}$, are indicated as well in each subfigure.}\label{fig:scatter}
\end{figure}

\subsection{Bayesian inversion}
\label{subsec:RESinversion}

We applied Bayesian inversion using Markov Chain Monte Carlo \citep{Goodman2010EnsembleInvariance} to infer the NPSM parameters as well as to predict spectral and resolution scintillator properties from the measured Compton edge spectra and our forward model. To account for the sensitivity of the NPSMs on the activator concentration and other scintillation crystal specific properties, we developed two separate inversion pipelines. In the first approach, Bayesian inversion is carried out separately for each of the four crystals, using their individual pulse-height spectra. In the second pipeline, we consider all four scintillation crystals as one integrated scintillator and perform the Bayesian inversion on the combined pulse-height spectra (sum channel). Subsequently, we will refer to these two approaches as the \emph{single} and \emph{sum} mode inversion pipelines. For both pipelines, we performed the Bayesian inversion on the $^{60}\text{Co}$ (activity $A=3.08(5) \times 10^5$~\unit{Bq}) spectral dataset \citep{Breitenmoser2022ExperimentalSpectrometry} leaving the remaining measurements for validation. $^{60}\text{Co}$ possesses two main photon emission lines at 1173.228(3)~\unit{keV} and 1332.492(4)~\unit{keV} with corresponding Compton edges according to the Compton scattering theory (Methods) at 963.419(3)~\unit{keV} and 1118.101(4)~\unit{keV}, respectively. However, in this study, we will focus on the lower Compton edge at 963.419(3)~\unit{keV}, because the upper edge is heavily obscured by the FEP at 1173.228(3)~\unit{keV}. Furthermore, as suggested by previous investigators \citep{Payne2011,Payne2014}, we fixed the Onsager related stopping power parameter $dE/ds\mid_{\text{Ons}}$ to $36.4~\unit{MeV cm^{-1}}$ in both pipelines.

Because the high-fidelity radiation transport simulations described in the previous section are computationally intense, we emulated the detector response as a function of the NPSM parameters using a machine learning trained vector-valued PCE surrogate model \citep{Torre2019}. The surrogate model has excellent evaluation speed $\mathcal{O}(10^{-4}~\unit{s})$ on a local workstation compared to $\mathcal{O}(10^{3}~\unit{s})$ required for a single Monte Carlo simulation with sufficient precision on a computer cluster. Correspondingly, the surrogate model provides a significant acceleration of our Bayesian inversion computations, reducing their processing time by a factor of $10^7$. Considering the minimum number of forward model evaluations needed for a single Markov chain (Methods), the evaluation time can be reduced from $\mathcal{O}(10^2~\unit{d})$ to $\mathcal{O}(1~\unit{s})$. 

Following the sum mode inversion pipeline, we present the solution to our inversion problem as a multivariate posterior distribution estimate in \hyperref[fig:scatter]{Fig.~\ref{fig:scatter}}. We find a unimodal solution with a maximum a posteriori (MAP) probability estimate given by $\eta_{e/h}=5.96^{+0.10}_{-0.17}\times10^{-1}$, $dE/ds\mid_{\text{Trap}}=1.46^{+0.03}_{-0.31}\times10^{1}~\unit{MeV cm^{-1}}$ and $dE/ds\mid_{\text{Birks}}=3.22^{+0.46}_{-0.44}\times10^{2}~\unit{MeV cm^{-1}}$, where we used the central credible intervals with a probability mass of 95\% to estimate the associated uncertainties. Combining the individual multivariate posterior distribution estimates from the single mode inversion pipeline, we obtain statistically consistent estimates, i.e. $\eta_{e/h}=5.87^{+0.24}_{-0.20}\times10^{-1}$, $dE/ds\mid_{\text{Trap}}=1.41^{+0.17}_{-0.15}\times10^{1}~\unit{MeV cm^{-1}}$ and $dE/ds\mid_{\text{Birks}}=3.17^{+1.11}_{-0.82}\times10^{2}~\unit{MeV cm^{-1}}$. 

It is worth noting that, considering the uncertainty estimates, we observe only minor differences between the different posterior point estimators for both inversion pipelines (\hyperref[fig:scatter]{Fig.~\ref{fig:scatter}} and Supplementary Figs.~S11--S14). However, we find statistically significant differences between the posterior point estimators for the individual scintillation crystals (Supplementary Table~S2). Furthermore, our results significantly differ from best-estimate literature values, which we obtained using linear temperature interpolation on a dataset provided by Payne and his co-workers, i.e. $\eta_{e/h}=4.53\times10^{-1}$, $dE/ds\mid_{\text{Trap}}=1.2\times10^{1}~\unit{MeV cm^{-1}}$ and $dE/ds\mid_{\text{Birks}}=1.853\times10^{2}~\unit{MeV cm^{-1}}$ for an ambient temperature of 18.8~\unit{\degreeCelsius} \citep{Payne2014}.

\begin{figure}[t!]
\centerline{
\includegraphics[]{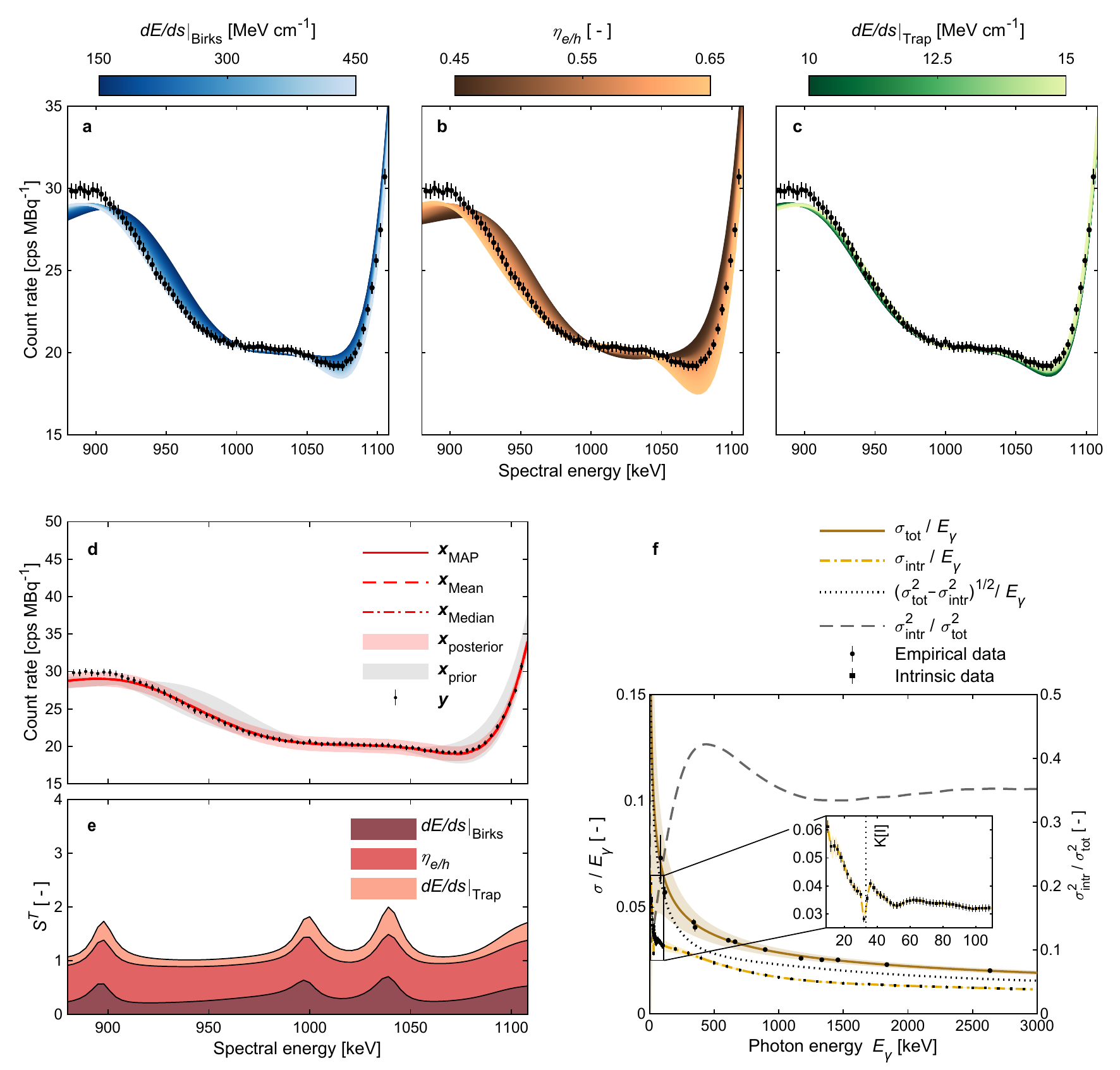}}
\caption{\textbf{ Compton edge and intrinsic resolution predictions.} \textbf{a--c}~Compton edge dynamics characterized by the trained polynomial chaos expansion emulator as a function of the individual non-proportional scintillation model parameters, i.e. the Birks related stopping power parameter $dE/ds\mid_{\text{Birks}}$, the free carrier fraction $\eta_{e/h}$ as well as the trapping related stopping power parameter $dE/ds\mid_{\text{Trap}}$, for the sum channel and the corresponding prior range. We fixed the remaining parameters at the corresponding maximum a posteriori (MAP) probability estimate values $\boldsymbol{x}_{\text{MAP}}$. The experimental data $\boldsymbol{y}$ from the measurement with a $^{60}\text{Co}$ source (activity $A=3.08(5) \times 10^5$~\unit{Bq}) is indicated as well \citep{Breitenmoser2022ExperimentalSpectrometry}. \textbf{d}~In this graph, we show the prior and posterior predictive distributions using the 99\% central credible interval obtained by the sum mode inversion pipeline. In addition, the experimental data $\boldsymbol{y}$ together with the derived posterior predictions using point estimators, i.e. the MAP probability estimate $\boldsymbol{x}_{\text{MAP}}$, the posterior mean $\boldsymbol{x}_{\text{Mean}}$ and the posterior median $\boldsymbol{x}_{\text{Median}}$, are indicated. \textbf{e}~We show the total Sobol indices $S^T$ computed by the polynomial chaos expansion emulator \citep{Sudret2008GlobalExpansions} as a function of the spectral energy for the individual model parameters on the sum channel. \textbf{f}~This graph presents the total ($\sigma_{\text{tot}}$) and the intrinsic ($\sigma_{\text{intr}}$) spectral resolution for the adopted detector system characterized by the standard deviation $\sigma$ as a function of the photon energy $E_{\gamma}$. The empirical resolution data as well as the corresponding resolution model were presented already elsewhere \citep{Breitenmoser2022ExperimentalSpectrometry}. For the zoomed inset with $E_{\gamma}<110$~\unit{keV}, the K-absorption edge for iodine K[I] is highlighted \citep{Bearden1967ReevaluationLevels}. For all graphs presented in this figure, uncertainties are provided as 1 standard deviation (SD) values (coverage factor $k=1$).}\label{fig:result}
\end{figure}

\subsection{Compton edge predictions}
\label{subsec:RESprediction}

We can use the trained PCE surrogate models to predict the spectral Compton edge as a function of the NPSM parameters and consequently the parameterized scintillation and quenching phenomena. In the \hyperref[fig:result]{Figs.~\ref{fig:result}a--c}, we present the spectral response of the PCE surrogate model for the sum channel as a function of the Birks related stopping power parameter $dE/ds\mid_{\text{Birks}}$, the free carrier fraction $\eta_{e/h}$ and the trapping related stopping power parameter $dE/ds\mid_{\text{Trap}}$. We observe a shift of the Compton edge toward smaller spectral energies for an increase in $dE/ds\mid_{\text{Birks}}$ and $\eta_{e/h}$ as well as a decrease in $dE/ds\mid_{\text{Trap}}$.

We leveraged the analytical relation between the polynomial chaos expansion and the Hoeffding-Sobol decomposition \citep{Sudret2008GlobalExpansions} to perform a global sensitivity analysis of the NPSM. Using the sum mode inversion pipeline, we present total Sobol indices $S^{T}$ for the model parameters $dE/ds\mid_{\text{Birks}}$, $\eta_{e/h}$ and $dE/ds\mid_{\text{Trap}}$ in \hyperref[fig:result]{Fig.~\ref{fig:result}e}. We find that the total Sobol indices can be ordered as $S^{T}(\eta_{e/h})>S^{T}(dE/ds\mid_{\text{Birks}})>S^{T}(dE/ds\mid_{\text{Trap}})$ over the entire spectral Compton edge domain indicating a corresponding contribution to the total model response variance. We get consistent results for a Hoeffding-Sobol sensitivity analysis of the individual scintillation crystals (Supplementary Fig.~S17).

In addition, we can also predict the spectral Compton edge using the prior and posterior predictive density estimates obtained by the two inversion pipelines. A comparison of these densities for the sum mode inversion pipeline indicates that our methodology successfully constrains the adopted NPSM (\hyperref[fig:result]{Fig.~\ref{fig:result}d}). However, we find also some model discrepancies, especially around the Compton continuum at the very low end of the investigated spectral range ($<920~\unit{keV}$). We get consistent results using the single mode inversion pipeline (Supplementary Fig.~S15). Furthermore, by comparing the posterior Compton edge predictions for the sum channel, we find no statistically significant difference between the two inversion pipelines (Supplementary Fig.~S16). From a modeling perspective, it is interesting to add that we observe no significant difference for Compton edge predictions using the various point estimators discussed in the previous section for both inversion pipelines.

\begin{figure}[t!]
\centerline{
\includegraphics[]{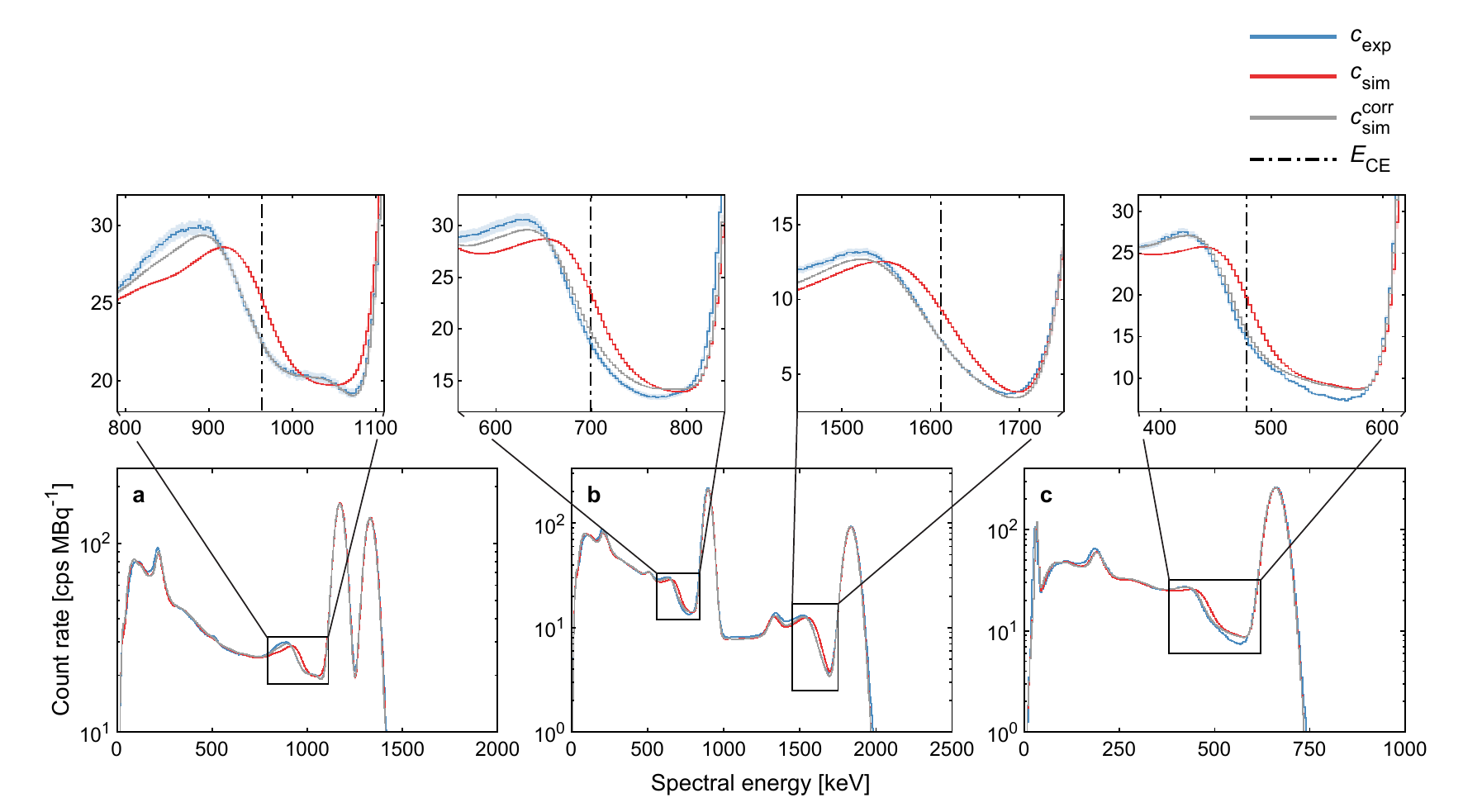}}
\caption{\textbf{ Simulated spectral detector response using a Bayesian calibrated non-proportional model.} The measured and simulated spectral detector responses are shown for three different calibrated radionuclide sources: \textbf{a}~$^{60}\text{Co}$ ($A=3.08(5) \times 10^5$~\unit{Bq}). \textbf{b}~$^{88}\text{Y}$ ($A=6.83(14) \times 10^5$~\unit{Bq}). \textbf{c}~$^{137}\text{Cs}$ ($A=2.266(34) \times 10^5$~\unit{Bq}). The zoomed-in subfigures highlight the Compton edge region and include also the Compton edge $E_{\text{CE}}$ predicted by the Compton scattering theory \citep{Knoll2010}, i.e. [477.334(3)\,,\,699.133(3)\,,\,963.419(3)\,,\,1611.77(1)]~\unit{keV} associated with the photon emission lines at [661.657(3)\,,\,898.042(3)\,,\,1173.228(3)\,,\,1836.063(3)]~\unit{keV}, respectively. The measured net count rate $c_{\text{exp}}$ as well as the simulated net count rate adopting a proportional scintillation model $c_{\text{sim}}$ were presented already elsewhere \citep{Breitenmoser2022ExperimentalSpectrometry}. We obtained the simulated net count rate $c_{\text{sim}}^{\text{corr}}$ the same way as $c_{\text{sim}}$ but accounted for the non-proportional scintillation effects by the Bayesian calibrated NPSM obtained by the sum mode inversion pipeline. For the calibration, we used the $^{60}\text{Co}$ dataset. For all graphs presented in this figure, uncertainties are provided as 1 standard deviation (SD) shaded areas (coverage factor $k=1$). These uncertainties are only visible for $c_{\text{exp}}$.}\label{fig:spec} 
\end{figure}

\subsection{Intrinsic resolution}
\label{subsec:RESintrinsic}

As already mentioned in the introduction, the scintillation non-proportionality not only distorts the spectral features in the pulse-height spectra but deteriorates also the spectral resolution of a scintillator detector. This contribution to the overall resolution due to the scintillation non-proportionality will be referred to as intrinsic resolution $\sigma_{\text{intr}}$ in accordance with previous studies \citep{Iredale1961,Zerby1961,Hill1966,Prescott1969ElectronNaITl,Dorenbos1995,Valentine1998a,Moszynski2002}. The intrinsic resolution is of great importance for two key reasons. 

First, it sets a fundamental lower limit on the achievable spectral resolution for a given scintillator material, making it a crucial factor in the development of new scintillators. As an example, in 1991, the scintillator $\text{Lu}_{2}\text{Si}\text{O}_{5}$ (LSO) was developed as an alternative to other available options at that time, such as $\text{Bi}_{4}\text{Ge}_{3}\text{O}_{12}$ (BGO). However, the performance of LSO led to considerable confusion within the research community as LSO exhibits a light yield more than four times greater than that of BGO, yet their energy resolutions are comparable \citep{Moses2008}. Consequently, the energy resolution for LSO was not dominated by counting statistics but some other factor. Thanks to the development of the Compton coincidence measurement technique in 1994 \citep{Valentine1994}, subsequent experimental studies have conclusively shown that the pronounced scintillation non-proportionality of the LSO scintillator was indeed the underlying factor responsible for the observed resolution degradation \citep{Rooney1997ScintillatorResponse,Valentine1998a}. This example showcases the need for a better understanding and prediction of the intrinsic resolution in the development of new scintillators \citep{Dujardin2018NeedsScintillators}.

Second, from a more technical perspective, the intrinsic resolution is a key component in the postprocessing pipeline for Monte Carlo simulations including NPSMs (Methods). In the forward model discussed before, the transport of scintillation photons, signal amplification by the photomultiplier tube and subsequent signal postprocessing in the multichannel analyzer are not included. As a result, to account for the additional resolution degradation by these processes, we need to perform a spectral broadening operation using a dedicated energy resolution model based on the measured total energy resolution as well as the intrinsic contribution \citep{Cano-Ott1999}.

Since our forward model explicitly accounts for the non-proportional scintillation physics by adopting a NPSM, we can use this numerical tool not only to predict pulse-height spectra but also to characterize the intrinsic resolution. We adopted a set of multiple monoenergetic Monte Carlo simulations to quantify the intrinsic resolution for different photon energies (Methods). Using this dataset, we then trained a Gaussian process (GP) regression model to predict the intrinsic resolution characterized by the standard deviation $\sigma$ for a given photon energy $E_{\gamma}$. The resulting GP model predictions together with the intrinsic data are highlighted in \hyperref[fig:result]{Fig.~\ref{fig:result}f}. In the same graph, we include also the empirical model to describe the overall energy resolution $\sigma_{\text{tot}}$ as well as the corresponding empirical dataset \citep{Breitenmoser2022ExperimentalSpectrometry}.

Comparing the intrinsic and overall spectral resolution, we find an almost constant ratio $\sigma_{\text{intr}}^2/\sigma_{\text{tot}}^2 \approx 0.35$ for $E_{\gamma}\gtrsim1500~\unit{keV}$. Around $E_{\gamma} \approx 440~\unit{keV}$, there is a pronounced peak with $\sigma_{\text{intr}}^2/\sigma_{\text{tot}}^2 \approx 0.42$ and for $E_{\gamma}\lesssim440~\unit{keV}$, we observe a significant decrease in $\sigma_{\text{intr}}^2/\sigma_{\text{tot}}^2$ with decreasing photon energy $E_{\gamma}$. Moreover, we find a more complex behaviour in $\sigma_{\text{intr}}$ for $E_{\gamma}\lesssim110~\unit{keV}$. For $28~\unit{keV}\lesssim E_{\gamma}\lesssim60~\unit{keV}$, the K-absorption edge for iodine K[I] at $E_{\gamma}=33.1694(4)~\unit{keV}$ \citep{Bearden1967ReevaluationLevels} alters the resolution significantly. On the other hand, at even smaller photon energies, there is again a pronounced increase in $\sigma_{\text{intr}}$ with decreasing energy compared to the mere moderate increase for $60~\unit{keV}\lesssim E_{\gamma}\lesssim110~\unit{keV}$.

\subsection{Bayesian calibrated NPSM simulations}
\label{subsec:REScalibration}

In addition to the insights into the Compton edge dynamics as well as the intrinsic resolution, the Bayesian inferred NPSM in combination with our forward model offers also the possibility to predict the full spectral detector response for new radiation sources accounting for non-proportional scintillation effects over the entire spectral range of our detector system. We used the $^{88}\text{Y}$ ($A=6.83(14) \times 10^5$~\unit{Bq}) and $^{137}\text{Cs}$ ($A=2.266(34) \times 10^5$~\unit{Bq}) radiation measurements to validate our calibrated NPSM. For the Monte Carlo simulations, we applied the posterior point estimators $\boldsymbol{x}_{\text{MAP}}$ obtained by the sum mode inversion pipeline in combination with the intrinsic and total resolution models discussed in the previous sections. 

In \hyperref[fig:spec]{Fig.~\ref{fig:spec}}, we present the measured and simulated spectral detector response for $^{88}\text{Y}$ and $^{137}\text{Cs}$ together with $^{60}\text{Co}$, whose Compton edge domain was used to perform the Bayesian inversion. For the simulations, we adopted a standard proportional scintillation model as well as the Bayesian inferred NPSM presented in this study. In line with the Compton scatter theory (Supplementary Methods~S1.4), we find an enhanced shift of the Compton edge toward smaller spectral energies as the photon energy increases. For all three measurements, we observe a significant improvement in the Compton edge prediction for the NPSM simulations compared to the standard proportional approach. However, there are still some discrepancies at the lower end of the Compton edge domain. Moreover, we find also some deviations between the Compton edge and the FEP for $^{88}\text{Y}$ and $^{137}\text{Cs}$. It is important to note that these discrepancies are smaller or at least of similar size for the NPSM simulations compared to the proportional approach indicating that the former performs statistically significantly better over the entire spectral domain. Additional validation results for $^{57}\text{Co}$, $^{109}\text{Cd}$, $^{133}\text{Ba}$ and $^{152}\text{Eu}$ together with a detailed uncertainty analysis for each source are attached in the Supplementary Information File for this study (Supplementary Figs.~S18--S25).


\section{Discussion}\label{sec:discuss}

Here, we demonstrated that Compton edge probing combined with Monte Carlo simulations and Bayesian inversion can successfully infer NPSMs for NaI(Tl) inorganic scintillators. A detailed Bayesian data analysis revealed no significant differences between standard posterior point estimators and the related spectral detector response predictions for both inversion pipelines. Consequently, the Bayesian inversion results indicate that our methodology successfully constrained the NPSM parameters to a unique solution. However, we found statistically significant differences between our results and best-estimate literature values as well as between the individual scintillation crystals themselves. These results corroborate the experimental findings of Hull and his co-workers \citep{Hull2009} and underscore the criticality of the NPSM calibration for every individual detector system.

Various studies reported a distortion of the Compton edge in gamma-ray spectrometry with inorganic scintillators \citep{Saito1981,Shi2002,Gardner2004,Cano-Ott1999,Breitenmoser2022ExperimentalSpectrometry}. In this study, we presented conclusive evidence that this shift is, at least partly, the result of the scintillation non-proportionality. Moreover, using our numerical models, we can predict the Compton edge shift as a function of the NPSM parameters. We observed a Compton edge shift toward smaller spectral energies for an increase in $dE/ds\mid_{\text{Birks}}$ and $\eta_{e/h}$ as well as a decrease in $dE/ds\mid_{\text{Trap}}$. These results imply that an enhanced scintillation non-proportionality promotes a Compton edge shift toward smaller spectral energies. In line with these observations, the non-proportionality is enhanced by a large $e^{-}/h$ fraction, an increased Birks mechanism as well as a reduction in the $e^{-}/h$ trapping rate \citep{Beck2015,VasilEv2014MultiscaleCharacteristics,Lecoq2017}.

Further, we quantified the sensitivity of the NPSM on the individual NPSM parameters using a PCE-based Sobol decomposition approach. The sensitivity results indicate that $\eta_{e/h}$ has the highest sensitivity on the Compton edge, followed by $dE/ds\mid_{\text{Birks}}$ and $dE/ds\mid_{\text{Trap}}$. However, previous studies showed a pronounced dependence of $dE/ds\mid_{\text{Trap}}$ on the ambient temperature \citep{Swiderski2006,Payne2014}. In addition, we expect also a substantial change of the crystal structure by radiation damage, i.e. the creation of new point defects in harsh radiation environments \citep{Zhu1998RadiationCrystals,Knoll2010}. Therefore, the obtained sensitivity results should be interpreted with care. $dE/ds\mid_{\text{Trap}}$ might be of significant importance to model the dynamics in the detector response with changing temperature or increase in radiation damage to the crystals, e.g. in deep space missions. 

Using the Bayesian calibrated NPSM, we are also able to numerically characterize the contribution of the scintillation non-proportionality to the overall energy resolution. This intrinsic resolution sets a fundamental lower limit on the achievable spectral resolution for a given scintillator material, making it a key factor in the development of new scintillators. At higher photon energies ($E_{\gamma}\gtrsim400~\unit{keV}$), we observed a significant contribution to the total spectral resolution ($\geq35\%$) with a maximum of $\approx42\%$ around 440~\unit{keV}. At lower energies ($10~\unit{keV}\lesssim E_{\gamma}\lesssim400~\unit{keV}$), the intrinsic contribution is reduced and shows substantial distortions around the K-absorption edge for iodine at about $33~\unit{keV}$. We conclude that the non-proportional scintillation is a significant contributor to the total energy resolution of NaI(Tl). These observations are in good agreement with previous results \citep{Narayan1968,Valentine1998a,Mengesha2002,Moszynski2002,Swiderski2013,Moszynski2016} and thereby substantiate not only the predictive power of our numerical model but showcase also its potential as a novel tool in the development of new scintillators.

Most of the theoretical studies focused on the prediction of NPSMs themselves. In contrast, available numerical models to predict the full detector response are scarce, computational intense and complex due to the adopted multi-step approaches with offline convolution computations \citep{Rooney1997ScintillatorResponse,Mengesha2002,Moszynski2002}. In this study, we present an alternative way to implement NPSMs and simulate the full spectral detector response to gamma-ray fields by directly evaluating the NPSM online during the Monte Carlo simulations. This approach saves considerable computation time and has the additional advantage of not having to store and analyze large files with secondary particle data. We have used this implementation to predict the full spectral detector response for additional radiation fields accounting for non-proportional scintillation effects. Validation measurements revealed a significant improvement in the simulated detector response compared to proportional scintillation models. However, there are still some model discrepancies, especially at the lower and higher end of the Compton edge domain. These discrepancies might be attributed to systematic uncertainties in the Monte Carlo mass model or deficiencies in the adopted NPSM. Sensitivity analysis performed in a previous study in conjunction with the prior prediction density results might indicate the latter \citep{Breitenmoser2022ExperimentalSpectrometry}. 

While we focused our work on NaI(Tl) in electron and gamma-ray fields, the presented methodology can easily be extended to a much broader range of applications. First, it is general consensus that the light yield as a function of the stopping power is, at least to a first approximation, independent of the ionizing particle type \citep{Murray1961b,Moses2008}. Second, the adopted NPSM was validated with an extensive database of measured scintillation light yields for inorganic scintillators, i.e. BGO, $\text{CaF}_{2}(\text{Eu})$, $\text{CeBr}_{3}$, Cs(Tl), Cs(Na), $\text{LaBr}_{3}(\text{Ce})$, LSO(Ce), NaI(Tl), $\text{SrI}_{2}$, $\text{SrI}_{2}(\text{Eu})$, YAP(Ce) and YAG(Ce), among others \citep{Payne2009,Payne2011,Payne2014}. From this, it follows that, given a gamma-ray field with resolvable Compton edges can be provided, our methodology may in principle be applied to any combination of inorganic scintillator and ionizing radiation field, including protons, $\alpha$-particles and heavy ions. However, it is important to note that our methodology relies on the observation of Compton edge shifts with a sufficient signal-to-noise ratio (SNR). We have shown that these shifts are influenced by the strength of scintillation non-proportionality of a given scintillator. As a result, scintillator materials that exhibit only a mild non-proportional scintillation response, e.g. $\text{LaBr}_{3}(\text{Ce})$ or $\text{YAP}(\text{Ce})$, may present challenges for the calibration of a NPSM due to the reduced SNR in the Compton edge shift. Further investigations are required to assess the applicability of our methodology to such scintillators. That said, the presented methodology can be readily adapted using Bayes's theorem to address low SNR cases more effectively by combining multiple Compton edge domains or by probing additional spectral features distorted by the non-proportional scintillation response.

In summary, we conclude that NPSMs are essential for accurate detector response simulations, especially for scintillators with large crystal volumes \citep{Murray1961b,Zerby1961,Valentine1998a}, e.g. in dark matter research, total absorption spectroscopy or remote sensing \citep{Cano-Ott1999,Bernabei2008,Adhikari2018AnDetectors,Paynter2021EvidenceBurst,Yang2022AOrigin,Lawrence1998,Trombka2000,Breitenmoser2022ExperimentalSpectrometry}. The novel methodology presented in this study offers a reliable and cost-effective alternative to existing experimental methods to investigate non-proportional scintillation physics phenomena and perform accurate full detector response predictions with Bayesian calibrated NPSM. Moreover, this new technique does not require any additional measurement equipment and can therefore be applied for any inorganic scintillator spectrometer, also during detector deployment. This is especially attractive for applications, where the scintillator properties change in operation, e.g. due to radiation damage or temperature changes, but also for detector design and the development of novel scintillator materials. Last but not least, we can use the derived numerical models not only for NPSM inference but also to investigate and predict various scintillator properties, e.g. intrinsic resolution or Compton edge dynamics, and thereby contribute to a better understanding of the complex scintillation physics in inorganic scintillators.


\section{Methods}\label{sec:method}

\subsection{Gamma-ray spectrometry}
\label{subsec:Mmeasur}

We performed gamma-ray spectrometric measurements in the calibration laboratory at the Paul Scherrer Institute (PSI) (inner room dimensions: 5.3~m~$\times$~4.5~m~$\times$~3~m). The adopted spectrometer consisted of four individual 10.2~\unit{cm}~$\times$~10.2~\unit{cm}~$\times$~40.6~\unit{cm} prismatic NaI(Tl) scintillation crystals with the associated photomultiplier tubes and the electronic components embedded in a thermal-insulating and vibration-damping polyethylene foam protected by a rugged aluminum detector box (outer dimensions: 86~\unit{cm}~$\times$~60~\unit{cm}~$\times$~30~\unit{cm}). The spectrometer features 1024 channels for an energy range between about 30 and 3000~\unit{keV} together with automatic linearization of the individual scintillation
crystal spectra \citep{Breitenmoser2022ExperimentalSpectrometry}. We used seven different calibrated radionuclide sources ($^{57}\text{Co}$, $^{60}\text{Co}$, $^{88}\text{Y}$, $^{109}\text{Cd}$, $^{133}\text{Ba}$, $^{137}\text{Cs}$ and $^{152}\text{Eu}$) from the Eckert~\&~Ziegler~Nuclitec~GmbH. We inserted these sources consisting of a radionuclide carrying ion exchange sphere (diameter 1~\unit{mm}) embedded in a 25~\unit{mm}~$\times$~3~\unit{mm} solid plastic disc into a custom low absorption source holder made out of a polylactide polymer (PLA) and placed this holder on a tripod in a fixed distance of 1~\unit{m} to the detector front on the central detector $x$-axis. To measure the source-detector distances and to position the sources accurately, distance as well as positioning laser systems were used. A schematic depiction of the measurement setup is shown in \hyperref[fig:scheme]{Fig.~\ref{fig:scheme}a}.

Between radiation measurements, background measurements were performed regularly for background correction and gain stability checks. For all measurements, the air temperature as well as the air humidity in the calibration laboratory was controlled by an air conditioning unit and logged by an external sensor. The air temperature was set at 18.8(4)~\unit{\degreeCelsius} and the relative air humidity at 42(3)\%. The ambient air pressure, which was also logged by the external sensor, fluctuated around 982(5)~\unit{hPa}.

During measurements, additional instruments and laboratory
equipment were located in the calibration laboratory,
e.g. shelves, a workbench, a source scanner or a boiler as
shown in \hyperref[fig:scheme]{Fig.~\ref{fig:scheme}a}. The effect of these features on the detector
response was carefully assessed in \citep{Breitenmoser2022ExperimentalSpectrometry}.

After postprocessing the spectral raw data according to the data reduction pipelines described in \citep{Breitenmoser2022ExperimentalSpectrometry}, we extracted the Compton edge spectral data from the net count rate spectra. The spectral domain of the Compton edge $\mathcal{D}_E$ was defined as $\mathcal{D}_E\coloneqq \left\{E:E_{\text{CE}}-3\cdot\sigma_{\text{tot}}\left(E_{\text{CE}}\right)\leq E \leq E_{\text{FEP}}-2\cdot\sigma_{\text{tot}}\left(E_{\text{FEP}}\right)\right\}$, where $E$ is the spectral energy, $\sigma_{\text{tot}}$ the energy dependent total resolution characterized by the standard deviation \citep{Breitenmoser2022ExperimentalSpectrometry} and $E_{\text{FEP}}$ the FEP associated with the Compton edge $E_{\text{CE}}$. Neglecting Doppler broadening and atomic shell effects, we compute $E_{\text{CE}}$ according to the Compton scattering theory \citep{Knoll2010} as follows: 

\begin{equation}
    E_{\text{CE}} \coloneqq E_{\text{FEP}} \left(
    1-\frac{1}{1+\frac{2E_{\text{FEP}}}{m_{e}c^2}}
    \right)
\label{eq:CE}
\end{equation}

\noindent where $m_{e}c^2$ is defined as the energy equivalent electron mass. In this study, we consulted the ENDF/B-VIII.0 nuclear data file library \citep{Brown2018} for nuclear decay related data as well as the Particle Data Group library \citep{Workman2022ReviewPhysics} for fundamental particle properties. 

To investigate the sensitivity of the selected Compton edge domain $\mathcal{D}_E$ on the Bayesian inversion results, we performed a sensitivity analysis on $\mathcal{D}_E$. Within the uncertainty bounds, the inversion results have proven to be insensitive to small alterations in $\mathcal{D}_E$ (Supplementary Table~S3).

It is important to note that, if not otherwise stated, uncertainties are provided as 1 standard deviation (SD) values in this study (coverage factor $k=1$). For more information about the radiation measurements and adopted data reduction pipelines, the reader is referred to the dedicated study \citep{Breitenmoser2022ExperimentalSpectrometry} as well as the Supporting Information File for this work (Supplementary Methods~S1.3).

\subsection{Monte Carlo simulations}
\label{subsec:Msim}

We performed all simulations with the multi-purpose Monte Carlo code \texttt{FLUKA} version 4.2.1 \citep{Ahdida2022NewCode} together with the graphical interface \texttt{FLAIR} version 3.1–15.1 \citep{Vlachoudis2009}. We used the most accurate physics settings (\texttt{precisio}) featuring a high-fidelity fully coupled photon, electron and positron radiation transport for our source-detector configuration. In addition, this module accounts for secondary electron production and transport, Landau fluctuations as well as X-ray fluorescence, all of which are essential for an accurate description of non-proportional scintillation effects \citep{Moszynski2002,Moses2008,Payne2011,Payne2015}. Motivated by the range of the transported particles, lower kinetic energy transport thresholds were set to 1~\unit{keV} for the scintillation crystals as well as the closest objects to the crystals, e.g. reflector, optical window and aluminum casing for the crystals. For the remaining model parts, the transport threshold was set to 10~\unit{keV} to decrease the computational load while maintaining the high-fidelity transport simulation in the scintillation crystals. All simulations were performed on a local computer cluster (7 nodes with a total number of 520 cores at a nominal clock speed of 2.6 GHz) at the Paul Scherrer Institute utilizing parallel computing.

We scored the energy deposition events in the scintillation crystals individually on an event-by-event basis using the custom user routine \texttt{usreou} together with the \texttt{detect} card. The number of primaries was set to $10^{7}$ for all simulations, which guarantees a maximum relative statistical standard deviation $\sigma_{\text{stat},\text{sim},k}/c_{\text{sim},k}<1\%$ and a maximum relative variance of the sample variance $\operatorname{VOV}_{k}<0.01\%$ for all detector channels $k$. More details on the simulation settings as well as on the postprocessing of the energy deposition data can be found in \citep{Breitenmoser2022ExperimentalSpectrometry}.

To implement the NPSM described by \hyperref[eq:StateEq]{Eq.~\ref{eq:StateEq}}, we developed an additional user routine \texttt{comscw}. Similar to \citep{Cano-Ott1999,Rasco2015}, we weight each individual energy deposition event in the scintillator, point-like or along the charged particle track, by the scintillation light yield given in \hyperref[eq:StateEq]{Eq.~\ref{eq:StateEq}} (Supplementary Algorithm~S1). The resulting simulated response is then rescaled to match the energy calibration models derived in \citep{Breitenmoser2022ExperimentalSpectrometry}. Using our methodology, we get simulated pulse-height spectra that incorporate non-proportional effects across the entire spectral range of our detector system.

\subsection{Surrogate modeling}
\label{subsec:Msurrogate}

We applied a custom machine learning trained vector-valued polynomial chaos expansion (PCE) surrogate model to emulate the spectral Compton edge detector response over $\mathcal{D}_E$ for both, the individual scintillation crystals as well as the sum channel. PCE models are ideal candidates to emulate expensive-to-evaluate vector-valued computational models \citep{Torre2019}. As shown by \citep{Xiu2003,Soize2005,Ernst2012}, any function $\boldsymbol{Y} = \mathcal{M}\left( \boldsymbol{X}\right)$ with the random input vector $\boldsymbol{X} \in \mathbb{R}^{M \times 1}$ and random response vector $\boldsymbol{Y} \in \mathbb{R}^{N \times 1}$ can be expanded as a so-called polynomial chaos expansion provided that $\mathbb{E}[ \left\| \boldsymbol{Y} \right\|^2]<\infty$:

\begin{equation}
    \boldsymbol{Y}=\mathcal{M}\left( \boldsymbol{X}\right)=\sum \limits_{\boldsymbol{\alpha} \in \mathbb{N}^{M}}
    \boldsymbol{a_{\alpha}}\Psi_{\boldsymbol{\alpha}}\left(\boldsymbol{X}\right) \label{eq:PCE_exact}
\end{equation}

\noindent where $\boldsymbol{a_{\alpha}}\coloneqq \left( a_{1, \boldsymbol{\alpha}},\dotsc,a_{N, \boldsymbol{\alpha}} \right)^{\intercal} \in \mathbb{R}^{N \times 1}$ are the deterministic expansion coefficients, $\boldsymbol{\alpha} \coloneqq \left( \alpha_{1},\dotsc,\alpha_{M} \right)^{\intercal} \in \mathbb{N}^{M \times 1}$ the multi-indices storing the degrees of the univariate polynomials $\psi_{\alpha}$ and $\Psi_{\boldsymbol{\alpha}}\left(\boldsymbol{X} \right) \coloneqq \prod_{i=1}^{M} \psi_{\alpha_{i}}^{i}\left(X_{i} \right)$ the multivariate polynomial basis functions, which are orthonormal with respect to the joint probability density function $f_{\boldsymbol{X}}$ of $\boldsymbol{X}$, i.e. $\langle \Psi_{\boldsymbol{\alpha}}, \Psi_{\boldsymbol{\beta}} \rangle_{f_{\boldsymbol{X}}}=\delta_{\boldsymbol{\alpha}, \boldsymbol{\beta}}$.

To reduce the computational burden, we combined the PCE model with principal component analysis (PCA) allowing us to characterize the main spectral Compton edge features of the response by means of a small number $N'$ of output variables compared to the original number $N$ of spectral variables, i.e. $N' \ll N$ \citep{Torre2019}. Similar to \citep{Wagner2020}, we computed the emulated computational model response $\hat{\mathcal{M}}_{\text{PCE}}\left( \boldsymbol{X}\right)$ in matrix form as:

\begin{equation}
    \boldsymbol{Y} \approx \hat{\mathcal{M}}_{\text{PCE}}\left( \boldsymbol{X}\right) =  \boldsymbol{\mu}_{\boldsymbol{Y}}+\operatorname{diag} \left( \boldsymbol{\sigma}_{\boldsymbol{Y}} \right) \boldsymbol{\Phi}' \boldsymbol{\mathrm{A}}\boldsymbol{\Psi}\left(\boldsymbol{X}\right)
    \label{eq:PCE_matrix}
\end{equation}

\noindent with $\boldsymbol{\mu}_{\boldsymbol{Y}}$ and $\boldsymbol{\sigma}_{\boldsymbol{Y}}$ being the mean and standard deviation of the random vector $\boldsymbol{Y}$ and $\boldsymbol{\Phi}'$ the matrix containing the retained eigenvectors $\boldsymbol{\phi}$ from the PCA, i.e. $\boldsymbol{\Phi}'\coloneqq \left( \boldsymbol{\phi}_{1},\dotsc, \boldsymbol{\phi}_{N'} \right) \in \mathbb{R}^{N \times N'}$. On the other hand, the vector $\boldsymbol{\Psi}\left(\boldsymbol{X}\right)\in \mathbb{R}^{\operatorname{card}\left( \mathcal{A}^{\star}\right) \times 1}$ and matrix $\boldsymbol{\mathrm{A}} \in \mathbb{R}^{N' \times \operatorname{card}\left( \mathcal{A}^{\star}\right)}$ store the multivariate orthonormal polynomials and corresponding PCE coefficients, respectively. The union set $\mathcal{A}^{\star} \coloneqq \bigcup_{j=1}^{N'} \mathcal{A}_{j}$ includes the finite sets of multi indices $\mathcal{A}_{j}$ for the $N'$ output variables following a specific truncation scheme.

We used a Latin hypercube experimental design $\boldsymbol{\mathcal{X}} \in \mathbb{R}^{M \times K}$ \citep{McKay1979,Choi2004} with $K=200$ instances sampled from a probabilistic model, which itself is defined by the model parameter priors described in the next subsection. The model response $\boldsymbol{\mathcal{Y}} \in \mathbb{R}^{N \times K}$ for this design was then evaluated using the forward model described in the previous subsection. We adopted a hyperbolic truncation scheme $\mathcal{A}_{j} \coloneqq \{ \boldsymbol{\alpha} \in \mathbb{N}^{M} : ( \sum_{i=1}^{M} \alpha_{i}^{q} )^{1/q} \leq p \}$ with $p$ and $q$ being hyperparameters defining the maximum degree for the associated polynomial and the q-norm, respectively. To compute the PCE coefficient matrix $\boldsymbol{\mathrm{A}}$, we applied adaptive least angle regression \citep{Blatman2011a} and optimized the hyperparameters $p\coloneqq\{1,2,\dotsc,7\}$ and $q\coloneqq\{0.5,0.6,\dotsc,1\}$ using machine learning with a holdout partition of 80\% and 20\% for the training and test set, respectively. For the PCA truncation, we adopted a relative PCA-induced error $\varepsilon_{\text{PCA}}$ of 0.1\%, i.e. $N'\coloneqq \min \{ S \in \left\{ 1,\dotsc,N \right\} : \sum_{j=1}^{S} \lambda_{j}/\sum_{j=1}^{N} \lambda_{j} \geq 1 - \varepsilon_{\text{PCA}} \}$ with $\lambda$ being the eigenvalues from the PCA. The resulting generalization error of the surrogate models, characterized by the relative mean squared error over the test sets, are $<1\%$ and $<2\%$ for the sum channel and the individual scintillation crystals, respectively. All PCE computations were performed with the \texttt{UQLab} code \citep{Marelli2014UQLab:Matlab} in combination with custom scripts to perform the PCA. More information about the PCE-PCA models as well as the PCE-PCA-based Sobol indices including detailed derivations are included in the Supplementary Information File for this study (Supplementary Methods~S1.1--S1.2).

\subsection{Bayesian inference}
\label{subsec:Minversion}

Following the Bayesian framework \citep{Kennedy2001BayesianModels}, we approximate the measured spectral detector response $\boldsymbol{y}\in \mathbb{R}^{N \times 1}$ with a probabilistic model combining the forward model $\mathcal{M}(\boldsymbol{x}_{\mathcal{M}})$ and model parameters $\boldsymbol{x}_{\mathcal{M}}\in \mathbb{R}^{M_{\mathcal{M}} \times 1}$ with an additive discrepancy term $\boldsymbol{\varepsilon}$, i.e. $\boldsymbol{y} \coloneqq \mathcal{M}(\boldsymbol{x}_{\mathcal{M}})+\boldsymbol{\varepsilon}$. For the discrepancy term $\boldsymbol{\varepsilon}$, which characterizes the measurement noise and prediction error, we assume a Gaussian model $\pi(\boldsymbol{\varepsilon} \mid \sigma^2_{\varepsilon})=\mathcal{N}(\boldsymbol{\varepsilon} \mid \boldsymbol{0},\sigma^2_{\varepsilon}\mathbb{I}_{N})$ with unknown discrepancy variance $\sigma^2_{\varepsilon}$. On the other hand, as discussed in the previous subsection, we emulate the forward model $\mathcal{M}(\boldsymbol{x}_{\mathcal{M}})$ with a PCE surrogate model $\hat{\mathcal{M}}_{\text{PCE}}(\boldsymbol{x}_{\mathcal{M}})$. Consequently, we can compute the likelihood function as follows:

\begin{equation}
    \pi\left(\boldsymbol{y} \mid \boldsymbol{x} \right) = \mathcal{N}\left(\boldsymbol{y} \mid \hat{\mathcal{M}}_{\text{PCE}}\left(\boldsymbol{x}_{\mathcal{M}}\right), \sigma^2_{\varepsilon}\mathbb{I}_{N}\right)
\label{eq:BI_likeli}
\end{equation}

\noindent with $\boldsymbol{x} \coloneqq [\,\boldsymbol{x}_{\mathcal{M}}\, , \, \sigma_{\varepsilon}^2\,]^{\intercal}$ and $\boldsymbol{x}_{\mathcal{M}} \coloneqq [\,dE/ds\mid_{\text{Birks}}\, , \,\eta_{e/h}\, , \, dE/ds\mid_{\text{Trap}}\,]^{\intercal}$. In combination with the prior density $\pi\left( \boldsymbol{x}\right)$, we can then compute the posterior distribution using Bayes' theorem \citep{Gelman2013BayesianAnalysis}:

\begin{equation}
    \pi\left( \boldsymbol{x} \mid \boldsymbol{y}\right) = \frac{\pi\left(\boldsymbol{y} \mid \boldsymbol{x} \right)\pi\left(\boldsymbol{x}\right)}{\int_{\mathcal{D}_{\boldsymbol{X}}}\pi\left(\boldsymbol{y} \mid \boldsymbol{x} \right)\pi\left(\boldsymbol{x}\right) \, \mathrm{d} \boldsymbol{x}}
\label{eq:BI_posterior}
\end{equation}

\noindent where we assume independent marginal priors, i.e. $\pi\left( \boldsymbol{x}\right) = \prod_{i=1}^{M}\pi\left( x_{i}\right)$ with $M=M_{\mathcal{M}}+1$. We defined the marginal priors based on the principle of maximum entropy \citep{Jaynes1957} as well as empirical data from previous studies \citep{Payne2009,Payne2011,Payne2014}. It should be emphasized that we applied the sum mode inversion pipeline first followed by the single mode inversion pipeline. In accordance with Bayes' theorem \citep{Gelman2013BayesianAnalysis}, we therefore incorporate the results obtained by the sum mode inversion pipeline in the marginal priors used for the single mode inversion pipeline. A full list of all adopted marginal priors for both pipelines is attached in the Supplementary Information File for this study (Supplementary Table~S1). Using the prior and posterior distributions, we can then also make predictions on future model response measurements $\boldsymbol{y}^{\ast}$ leveraging the prior and posterior predictive densities:

\begin{subequations}
\begin{align}
    \pi\left( \boldsymbol{y}^{\ast}\right) &= \int_{\mathcal{D}_{\boldsymbol{x}}}\pi\left(\boldsymbol{y}^{\ast} \mid \boldsymbol{x} \right)\pi\left(\boldsymbol{x}\right) \, \mathrm{d} \boldsymbol{x}
    \label{eq:BI_priorpred} \\
    \pi\left( \boldsymbol{y}^{\ast} \mid \boldsymbol{y} \right) &= \int_{\mathcal{D}_{\boldsymbol{x}}}\pi\left(\boldsymbol{y}^{\ast} \mid \boldsymbol{x} \right)\pi\left( \boldsymbol{x} \mid \boldsymbol{y}\right) \, \mathrm{d} \boldsymbol{x}
    \label{eq:BI_postpred}
\end{align}
\end{subequations}

All Bayesian computations were performed with the \texttt{UQLab} code \citep{Marelli2014UQLab:Matlab}. We applied an affine invariant ensemble algorithm \citep{Goodman2010EnsembleInvariance} to perform Markov Chain Monte Carlo (MCMC) and thereby estimate the posterior distribution $\pi\left( \boldsymbol{x} \mid \boldsymbol{y}\right)$. We used 10 parallel chains with $2\times10^4$ MCMC iterations per chain together with a 50\% burn-in. The convergence and precision of the MCMC simulations were carefully assessed using standard diagnostics tools \citep{Brooks1998GeneralSimulations,Gelman2013BayesianAnalysis}. We report a potential scale reduction factor $\hat{R}<1.02$ and an effective sample size $\text{ESS}\gg400$ for all performed MCMC simulations. Additional trace and convergence plots for the individual parameters $\boldsymbol{x}$ and point estimators (Supplementary Figs.~S1--S10), a full list of the Bayesian inversion results (Supplementary Table~S2) as well as a sensitivity analysis on the adopted Compton edge domain (Supplementary Table~S3) can be found in the attached Supplementary Information File for this study.

\subsection{Resolution modeling}
\label{subsec:Mres}

In this last section, we will discuss the derivation of the energy resolution models adopted in this study. We start with the model to characterize the overall or total energy resolution $\sigma_{\text{tot}}$ for our detector system and describe in a second step the derivation of the intrinsic resolution model $\sigma_{\text{intr}}$. It is important to note that in contrast to $\sigma_{\text{intr}}$, we provide here only a short summary of the key aspects involved in $\sigma_{\text{tot}}$. The entire postprocessing pipeline to derive $\sigma_{\text{tot}}$ was already thoroughly discussed in a previous study \citep{Breitenmoser2022ExperimentalSpectrometry}. For more details, we kindly refer the reader to the dedicated study. 

For each scintillation crystal, we quantified $\sigma_{\text{tot}}$ by characterizing the spectral dispersion of measured FEPs associated with known photon emission lines from specific radionuclides. The corresponding pipeline can be divided into three steps. In a first step, we extracted specific spectral domains containing a singlet or multiplet of FEPs from a set of measured count rate spectra covering a spectral range between 122 and 1836~\unit{keV} \citep{Breitenmoser2022LaboratoryRLL}. In a second step, we fitted a spectral peak model based on a sum of independent Gaussian peaks together with a numerical baseline \citep{Westmeier1992} to the selected singlets or multiplets using weighted non-linear least-squares (WNLLS) regression combined with the interior-reflective Newton method \citep{Coleman1996AnBounds}. In the third step, we extracted the Gaussian standard deviation parameters from the fitted FEPs as a characteristic measure for the spectral resolution. By combining these empirical resolution values with the known emission line energies, we derived an exponential model to describe $\sigma_{\text{tot}}$ as a function of the photon energy $E_{\gamma}$ adopting again WNLLS. The resulting relative generalization error, characterized by leave-one-out cross-validation, is $<0.2\%$ for all scintillation crystals.

To derive a model for $\sigma_{\text{intr}}$, we performed in a first step additional Monte Carlo simulations for an isotropic and uniform monoenergetic photon flux of energy $10~\unit{keV} \leq E_{\gamma} \leq 3200~\unit{keV}$. To account for the different spectral scales, we applied a non-uniform experimental design for the photon energy $E_{\gamma}$ with a 2~\unit{keV} spacing below 110~\unit{keV} and 100~\unit{keV} spacing above. Moreover, to account for the non-proportional scintillation physics, we ran all simulations with the Bayesian calibrated NPSM, i.e. the derived MAP point estimators. The mass model for those simulations features a 10.2~\unit{cm}~$\times$~10.2~\unit{cm}~$\times$~40.6~\unit{cm} prismatic NaI(Tl) scintillation crystal embedded in a vacuum environment. In a second step, we extracted the mean light yield values from the simulated FEPs (Supplementary Algorithm~S1). Similar to the measured spectra, we can then derive a simple polynomial energy calibration model using WNLLS to convert the simulated light yield to energy \citep{Breitenmoser2022ExperimentalSpectrometry}. In a third step, we adopted the extracted $\sigma_{\text{intr}}$ from the individual energy calibrated FEPs to train a Gaussian Process (GP) regression model with \citep{Rasmussen2006GaussianLearning}:

\begin{equation}
    \sigma_{\text{intr}}\left(E_{\gamma} \right) \sim \mathcal{GP}\left(\mathbf{f}\left( E_{\gamma} \right)^{\intercal}\boldsymbol{\beta},\kappa \left(E_{\gamma},E'_{\gamma}\right)+\sigma_{\mathcal{GP}}^{2}\delta_{E_{\gamma},E'_{\gamma}} \right)
\label{eq:res_intr}
\end{equation}

\noindent where we applied a polynomial trend function of the second order, i.e. $\mathbf{f}\left( E_{\gamma} \right) \coloneqq \left( 1,E_{\gamma},E^{2}_{\gamma} \right)^{\intercal}$ and $\boldsymbol{\beta} \coloneqq \left( \beta_{0},\beta_{1},\beta_{2} \right)^{\intercal}$, a homoscedastic noise model with the noise variance $\sigma_{\mathcal{GP}}^{2}$ and Kronecker delta $\delta_{E_{\gamma},E'_{\gamma}}$ as well as a Matérn-3/2 covariance function $\kappa \left(E_{\gamma},E'_{\gamma}\right) \coloneqq \left(1+\sqrt{3} \mid E_{\gamma} - E'_{\gamma} \mid / \theta \right) \exp{\left(-\sqrt{3}\mid E_{\gamma} - E'_{\gamma} \mid /\theta \right)}$ with the kernel scale $\theta$. It is worth noting that, due to the known asymmetry in the FEPs \citep{Zerby1961,Valentine1998a,Cano-Ott1999}, we adopted numerical estimates both for the mean and standard deviation parameters associated with the individual FEPs. With the $N$-dimensional intrinsic dataset $\left\{ \boldsymbol{E}_{\gamma},\boldsymbol{\sigma}_{\text{intr}} \right\}$, we can then predict the intrinsic resolution $\boldsymbol{\sigma}_{\text{intr}}^{\ast}$ for a new set of $N^{\ast}$ photon energies $\boldsymbol{E}^{\ast}_{\gamma}$ using the GP posterior predictive density as follows \citep{Rasmussen2006GaussianLearning}:

\begin{subequations}
\begin{align}
    \pi\left(\boldsymbol{\sigma}_{\text{intr}}^{\ast} \mid \boldsymbol{E}^{\ast}_{\gamma},\boldsymbol{E}_{\gamma},\boldsymbol{\sigma}_{\text{intr}}\right) &= \mathcal{N}\left( \boldsymbol{\sigma}_{\text{intr}}^{\ast} \mid \boldsymbol{\mu}_{\mathcal{GP}},\boldsymbol{\Sigma}_{\mathcal{GP}}\right)
    \label{eq:GP_postpred}\\
    \boldsymbol{\mu}_{\mathcal{GP}} &= \mathbf{F}^{\intercal}_{\ast} \hat{\boldsymbol{\beta}}+\mathbf{K}_{\ast}^{\intercal} \mathbf{K}^{-1} \left( \boldsymbol{\sigma}_{\text{intr}} - \mathbf{F}^{\intercal} \hat{\boldsymbol{\beta}} \right)
    \label{eq:GP_postmean}\\
    \boldsymbol{\Sigma}_{\mathcal{GP}} &=
    \mathbf{K}_{\ast\ast} - \mathbf{K}_{\ast}^{\intercal}\mathbf{K}^{-1}\mathbf{K}_{\ast}+
    \mathbf{U}^{\intercal}\left( \mathbf{F}\mathbf{K}^{-1} \mathbf{F}^{\intercal}\right)^{-1}\mathbf{U}
    \label{eq:GP_postcov}\\
    \hat{\boldsymbol{\beta}} &= \left(\mathbf{F}\mathbf{K}^{-1}\mathbf{F}^{\intercal}\right)^{-1}\mathbf{F}\mathbf{K}^{-1}\boldsymbol{\sigma}_{\text{intr}}
    \label{eq:GP_beta}\\
    \mathbf{U} &= \mathbf{F}_{\ast} - \mathbf{F} \mathbf{K}^{-1} \mathbf{K}_{\ast}
    \label{eq:GP_U}
\end{align}
\end{subequations}

\noindent with the matrices $\mathbf{F} = \mathbf{f}\left( \boldsymbol{E}_{\gamma} \right) \in \mathbb{R}^{3 \times N}$, $\mathbf{F}_{\ast} = \mathbf{f}\left( \boldsymbol{E}^{\ast}_{\gamma} \right) \in \mathbb{R}^{3 \times N^{\ast}}$, $\mathbf{K} = \kappa\left( \boldsymbol{E}_{\gamma},\boldsymbol{E}_{\gamma} \right) + \sigma_{\mathcal{GP}}^2 \mathbb{I}_{N} \in \mathbb{R}^{N \times N}$, $\mathbf{K}_{\ast} = \kappa\left( \boldsymbol{E}_{\gamma},\boldsymbol{E}^{\ast}_{\gamma} \right) \in \mathbb{R}^{N \times N^{\ast}}$ and $\mathbf{K}_{\ast\ast} = \kappa\left( \boldsymbol{E}^{\ast}_{\gamma},\boldsymbol{E}^{\ast}_{\gamma} \right) \in \mathbb{R}^{N^{\ast} \times N^{\ast}}$. To account for the different spectral scales, we trained two GP models, one for $10~\unit{keV}\leq E_{\gamma} \leq 90~\unit{keV}$ and the other one for $90~\unit{keV}\leq E_{\gamma} \leq 3200~\unit{keV}$, using the $\texttt{MATLAB}\textsuperscript{\textregistered}$ code. For both models, we applied 5-fold cross-validation in combination with Bayesian optimization to determine the GP hyperparameters $\sigma_{\mathcal{GP}}^{2}$ and $\theta$. It is important to add that in case of the experimental design $\boldsymbol{\mathcal{X}}$ adopted to train the surrogate model, we ran the pipeline for $\sigma_{\text{intr}}$ with the corresponding set of NPSM parameters defined by $\boldsymbol{\mathcal{X}}$.

As discussed already in the Results section, the intrinsic resolution is also a key component in the postprocessing pipeline for Monte Carlo simulations including NPSMs. Because the Monte Carlo simulations performed for the forward model only inherently include the intrinsic resolution, we need to perform a spectral broadening operation to account for the additional energy resolution degradation due to the transport of scintillation photons, signal amplification by the photomultiplier tube and subsequent signal postprocessing in the multichannel analyzer. Similar to \citep{Cano-Ott1999}, we assume statistical independence between the resolution degradation due to the scintillation non-proportionality and the aforementioned neglected processes in the Monte Carlo simulations. We can then perform the broadening operation as described in \citep{Breitenmoser2022ExperimentalSpectrometry} with an adapted dispersion $\sqrt{\sigma^2_{\text{tot}}-\sigma^2_{\text{intr}}}$. For completeness, we included this adapted dispersion model in \hyperref[fig:scheme]{Fig.~\ref{fig:result}f}.

\vspace{10mm}


\backmatter

\bmhead{Supplementary information}

The online version contains
supplementary information.

\bmhead{Acknowledgments}

We gratefully acknowledge
the technical support by Dominik Werthmüller for the
execution of the Monte Carlo simulations on the computer cluster
at the Paul Scherrer Institute. We also thank Eduardo Gardenali Yukihara for helpful discussions and advices. Further, we would like to express our gratitude to the Swiss Armed Forces and the National Emergency Operations Centre (NEOC) for providing the detector system. This research has been supported in part by the Swiss Federal Nuclear Safety Inspectorate (grant no. CTR00491).

\bmhead{Competing interests}

The authors declare no competing interests.

\bmhead{Author contributions}

D.B. designed the study, supervised the project, performed the measurements, simulations, data postprocessing and wrote the manuscript. F.C. significantly contributed to the implementation of the NPSM in \texttt{FLUKA}. G.B. supervised the project. S.M. acquired the project funding. All authors contributed to the completion of the manuscript.

\bmhead{Data availability}

The radiation measurement raw data presented herein are deposited on the ETH Research Collection repository: \url{https://doi.org/10.3929/ethz-b-000528920} \citep{Breitenmoser2022LaboratoryRLL}. Additional datasets related to this study are available from the corresponding author upon reasonable request.

\bmhead{Code availability}
The \texttt{FLUKA} code \citep{Ahdida2022NewCode} used for Monte Carlo radiation transport and detector response simulations is available at \url{https://fluka.cern/}. We adopted the graphical user interphase \texttt{FLAIR} \citep{Vlachoudis2009}, freely available at \url{https://flair.web.cern.ch/flair/}, to setup the \texttt{FLUKA} input files and create the mass model figures. The custom \texttt{FLUKA} user routines adopted in the Monte Carlo simulations are deposited on the ETH Research Collection repository: \url{https://doi.org/10.3929/ethz-b-000595727} \citep{Breitenmoser2023}. Data processing, machine learning computation and figure creation was performed by the $\texttt{MATLAB}\textsuperscript{\textregistered}$ code in combination with the open-source toolbox \texttt{UQLab} \citep{Marelli2014UQLab:Matlab} available at \url{https://www.uqlab.com/}.


\end{document}



\newpage
\noindent \Large \textbf{\textsf{Supplementary Materials for}}
\vspace{5mm}

\noindent \large \textbf{\textsf{Emulator-based Bayesian Inference on Non-Proportional Scintillation Models \\ by Compton-Edge Probing}}
\vspace{8mm}

\noindent \small \textsf{David Breitenmoser$^{\text{1,2,*}}$, Francesco Cerutti$^{\text{3}}$, Gernot Butterweck$^{\text{1}}$, Malgorzata Magdalena Kasprzak$^{\text{1}}$, Sabine Mayer$^{\text{1}}$}

\vspace{5mm}

\noindent \footnotesize \textsf{$^{\text{1}}$Department of Radiation Safety and Security, Paul~Scherrer~Institute (PSI), Forschungsstrasse~111, Villigen~PSI, 5232, Switzerland}

\vspace{0.5mm}

\noindent \footnotesize \textsf{$^{\text{2}}$Department of Physics, Swiss Federal Institute of Technology (ETH), Otto-Stern-Weg~5, Zurich, 8093, Switzerland}

\vspace{0.5mm}

\noindent \footnotesize \textsf{$^{\text{3}}$European Organization for Nuclear Research (CERN), Esplanade~des~Particules~1, Geneva, 1211, Switzerland}

\vspace{2mm}

\noindent \footnotesize \textsf{$^{*}$Corresponding author: David Breitenmoser (david.breitenmoser@psi.ch,~ORCID:~\href{0000-0003-0339-6592}{https://orcid.org/0000-0003-0339-6592})}

\vspace{20mm}

\noindent \normalsize \textbf{\textsf{The PDF includes:}}

\vspace{1mm}

\indent \textsf{ Materials and Methods }
\\
\indent \textsf{ Figs. S1--S29 }
\\
\indent \textsf{ Tables S1--S5 }
\\
\indent \textsf{ Algorithm S1 }
\\
\indent \textsf{ References }

\thispagestyle{empty}
\newpage


\setcounter{page}{1}
\renewcommand{\thepage}{S\Roman{page}} 
\tableofcontents

\newpage

\listoffigures
\listoftables
\cleardoublepage
\newpage

\setcounter{page}{1}
\renewcommand{\thepage}{S\arabic{page}} 


\section{Materials and Methods}\label{sec:M&M}

\subsection{Adaptive sparse PCE-PCA surrogate model}
\label{subsec:PCE_PCA}

\normalsize
Here, based on previous work \citep{Blatman2013,Nagel2020,Wagner2020}, we derive our custom vector-valued adaptive sparse polynomial chaos expansion surrogate model (PCE), which we combine with principal component analysis (PCA). 

We start with the PCA model part. Consider our vector-valued model response as a random vector $\boldsymbol{Y} \in \mathbb{R}^{N \times 1}$ with mean $\boldsymbol{\mu}_{\boldsymbol{Y}}$, standard deviation $\boldsymbol{\sigma}_{\boldsymbol{Y}}$ and correlation matrix $\boldsymbol{\Sigma}_{\boldsymbol{Y}} \coloneqq \operatorname{corr} \left( \boldsymbol{Y}\right) =\mathbb{E} \left[ \boldsymbol{Y}^{\ast} \left(\boldsymbol{Y}^{\ast}\right)^{\intercal} \right]$. Note that, in contrast to previous studies \citep{Blatman2013,Nagel2020,Wagner2020}, we standardize our model response $\boldsymbol{Y}$ with $\boldsymbol{Y}^{\ast} \coloneqq \operatorname{diag} \left(\boldsymbol{\sigma_{\boldsymbol{Y}}} \right)^{-1} \left(\boldsymbol{Y}-\boldsymbol{\mu_{\boldsymbol{Y}}} \right)$ to account for the differences in the variance of the individual response variables. We can then perform an eigenvalue decomposition of the correlation matrix $\boldsymbol{\Sigma}_{\boldsymbol{Y}}$ with the eigenvalues $\lambda_{j}$ and eigenvectors $\boldsymbol{\phi}_{j} \coloneqq \left( \phi_{1},\dotsc,\phi_{N}  \right)^{\intercal}$ satisfying $\boldsymbol{\Sigma}_{\boldsymbol{Y}}\boldsymbol{\phi}_{j}=\lambda_{j}\boldsymbol{\phi}_{j}$ for $j=1,\dotsc,N$. Since $\boldsymbol{\Sigma}_{\boldsymbol{Y}}$ is symmetric and positive definite, the eigenvectors define an  orthonormal basis $\mathbb{R}^{N}=\operatorname{span}(\{\boldsymbol{\phi}_{j}\}_{j=1}^{N})$ and we can perform an orthogonal transformation of our random vectors $\boldsymbol{Y}^{\ast}$ as follows:

\begin{equation}
    \boldsymbol{Z} = \boldsymbol{\Phi}^{\intercal} \boldsymbol{Y}^{\ast}
\label{eq:PCA0}
\end{equation}

\noindent with the orthonormal matrix $\boldsymbol{\Phi} \coloneqq \left( \boldsymbol{\phi}_{1},\dotsc, \boldsymbol{\phi}_{N}  \right) \in \mathbb{R}^{N \times N}$, where $\lambda_{1}\geq\lambda_{2}\geq\dotsc\geq\lambda_{N}$. We call the transformed vectors $\boldsymbol{Z} \coloneqq \left(Z_{1},\dotsc, Z_{N} \right)^{\intercal}$  the principal components of $\boldsymbol{Y}^{\ast}$. Once we get the principal components, we can transform them back to the original response variable space with

\begin{equation}
    \boldsymbol{Y} = \boldsymbol{\mu}_{\boldsymbol{Y}}+ \operatorname{diag} \left( \boldsymbol{\sigma}_{\boldsymbol{Y}} \right) \sum_{j=1}^{N} Z_{j} \boldsymbol{\phi}_{j}
    \label{eq:PCAa}
\end{equation}

\noindent To reduce the dimensions of our problem, we retain only $N'$ principal components with the highest variance and thereby approximate our random vector $\boldsymbol{Y}$ as

\begin{equation}
    \boldsymbol{Y} \approx \boldsymbol{\mu}_{\boldsymbol{Y}}+ \operatorname{diag} \left( \boldsymbol{\sigma}_{\boldsymbol{Y}} \right) \sum_{j=1}^{N'} Z_{j} \boldsymbol{\phi}_{j}
    \label{eq:PCAb}
\end{equation}

\noindent where we choose $N'\coloneqq \min \{ S \in \left\{ 1,\dotsc,N \right\} : \sum_{j=1}^{S} \lambda_{j}/\sum_{j=1}^{N} \lambda_{j} \geq 1 - \varepsilon_{PCA} \}$ with a prescribed approximation error $\varepsilon_{\text{PCA}}$. 

For the PCE model part, we start again with the polynomial chaos expansion of the model response $\mathcal{M}\left( \boldsymbol{X}\right)$ with the random input vector $\boldsymbol{X} \in \mathbb{R}^{M \times 1}$ as described in Eq.~3 in the main study:

\begin{equation}
    \boldsymbol{Y}=\sum \limits_{\boldsymbol{\alpha} \in \mathbb{N}^{M}}
    \boldsymbol{a_{\alpha}}\Psi_{\boldsymbol{\alpha}}\left(\boldsymbol{X}\right) \label{eq:PCEa}
\end{equation}

\noindent where  $\boldsymbol{a_{\alpha}}\coloneqq \left( a_{1, \boldsymbol{\alpha}},\dotsc,a_{N, \boldsymbol{\alpha}} \right)^{\intercal} \in \mathbb{R}^{N \times 1}$ are the deterministic expansion coefficients, $\boldsymbol{\alpha} \coloneqq \left( \alpha_{1},\dotsc,\alpha_{M} \right)^{\intercal} \in \mathbb{N}^{M \times 1}$ the multi-indices storing the degrees of the univariate polynomials $\psi_{\alpha}$ and $\Psi_{\boldsymbol{\alpha}}\left(\boldsymbol{X} \right) \coloneqq \prod_{i=1}^{M} \psi_{\alpha_{i}}^{i}\left(X_{i} \right)$ the multivariate polynomial basis functions, which are orthonormal with respect to the joint probability density function $f_{\boldsymbol{X}}$ of $\boldsymbol{X}$, i.e. $\langle \Psi_{\boldsymbol{\alpha}}, \Psi_{\boldsymbol{\beta}} \rangle_{f_{\boldsymbol{X}}}=\delta_{\boldsymbol{\alpha}, \boldsymbol{\beta}}$. For computational purposes, we truncate the PCE series by adopting a truncation set $\mathcal{A}_{j}$ for the multi-index $\boldsymbol{\alpha}$ of each individual response variable $j=1,\dotsc,N$ resulting in:

\begin{equation}
    Y_{j}\approx \sum \limits_{\boldsymbol{\alpha} \in \mathcal{A}_{j}}
    a_{j,\boldsymbol{\alpha}}\Psi_{\boldsymbol{\alpha}}\left(\boldsymbol{X}\right)
    \label{eq:PCEb}
\end{equation}

\noindent For the truncation, we can use a hyperbolic truncation scheme defining the multi-index set as $\mathcal{A}_{j} \coloneqq \{ \boldsymbol{\alpha} \in \mathbb{N}^{M} : ( \sum_{j=1}^{M} \alpha_{j}^{q} )^{1/q}  \leq p \}$  with $p$ and $q$ defining the maximum degree for the associated polynomial and the q-norm, respectively.

To reduce the computational burden, we can now combine these results and perform the PCE not in the original response variable space but in the truncated principal component space. For that, we insert \hyperref[eq:PCEb]{Eq.~\ref{eq:PCEb}} in \hyperref[eq:PCAb]{Eq.~\ref{eq:PCAb}}:

\begin{equation}
    \boldsymbol{Y} \approx \hat{\mathcal{M}}\left( \boldsymbol{X}\right) = \boldsymbol{\mu_{\boldsymbol{Y}}}+ \operatorname{diag} \left( \boldsymbol{\sigma_{\boldsymbol{Y}}} \right) \sum_{j=1}^{N'} \left( \sum_{\boldsymbol{\alpha} \in \mathcal{A}_j} a_{j, \boldsymbol{\alpha}} \Psi_{\boldsymbol{\alpha}} \left( \boldsymbol{X} \right) \right) \boldsymbol{\phi}_{j} 
    \label{eq:PCAEa}
\end{equation}

\noindent which we can rearrange by introducing the union set $\mathcal{A}^{\star} \coloneqq \bigcup_{j=1}^{N'} \mathcal{A}_{j}$ to:

\begin{equation}
    \boldsymbol{Y} \approx \hat{\mathcal{M}}\left( \boldsymbol{X}\right) = \boldsymbol{\mu_{\boldsymbol{Y}}}+ \operatorname{diag} \left( \boldsymbol{\sigma_{\boldsymbol{Y}}} \right) \sum_{\boldsymbol{\alpha} \in \mathcal{A}^{\star}} \sum_{j=1}^{N'}  a_{j, \boldsymbol{\alpha}} \Psi_{\boldsymbol{\alpha}} \left( \boldsymbol{X} \right) \boldsymbol{\phi}_{j}
    \label{eq:PCAEb}
\end{equation}

\noindent or expressed in a more compact matrix form:

\begin{equation}
    \boldsymbol{Y} \approx \hat{\mathcal{M}}\left( \boldsymbol{X}\right) = \boldsymbol{\mu_{\boldsymbol{Y}}}+\operatorname{diag} \left( \boldsymbol{\sigma_{\boldsymbol{Y}}} \right) \boldsymbol{\Phi}' \boldsymbol{\mathrm{A}}\boldsymbol{\Psi}\left( \boldsymbol{X}\right)
    \label{eq:PCAEc}
\end{equation}

\noindent with the vector $\boldsymbol{\Psi}\left(\boldsymbol{X}\right)\in \mathbb{R}^{\operatorname{card}\left( \mathcal{A}^{\star}\right) \times 1}$ as well as the two matrices $\boldsymbol{\Phi}'\in \mathbb{R}^{N \times N'}$ and $\boldsymbol{\mathrm{A}} \in \mathbb{R}^{N' \times \operatorname{card}\left( \mathcal{A}^{\star}\right)}$ storing the multivariate orthonormal polynomials $\Psi_{\boldsymbol{\alpha}}$, the retained eigenvectors $\boldsymbol{\phi}_{j}$ and the PCE coefficients $a_{j, \boldsymbol{\alpha}}$, respectively.

For model training, we introduce an experimental design with the input matrix $\boldsymbol{\mathcal{X}} \in \mathbb{R}^{M \times K}$ and response matrix $\boldsymbol{\mathcal{Y}} \in \mathbb{R}^{N \times K}$ for $K$ instances, $M$ input variables and $N$ response variables. For the PCA model, we can use the response matrix $\boldsymbol{\mathcal{Y}}$ to estimate $\boldsymbol{\mu}_{\boldsymbol{Y}}$, $\boldsymbol{\sigma}_{\boldsymbol{Y}}$ as well as $\boldsymbol{\Sigma}_{\boldsymbol{Y}}$:

\begin{subequations}
\begin{align}
    \hat{\boldsymbol{\mu}}_{\boldsymbol{Y}}&=\frac{1}{K} \sum_{k=1}^{K} \boldsymbol{y}^{ \left( k \right) } 
    \label{eq:PCAmean} \\
     \hat{\boldsymbol{\sigma}}_{\boldsymbol{Y}}&=\sqrt{\frac{1}{K-1}\sum_{k=1}^{K} \left( \boldsymbol{y}^{\left(k\right)} - \hat{\boldsymbol{\mu}}_{\boldsymbol{Y}}\right)^2}
    \label{eq:PCAstd}\\
    \hat{\boldsymbol{\Sigma}}_{\boldsymbol{Y}}&=\frac{1}{K-1}\boldsymbol{\mathcal{Y}}^{\ast}  \left(\boldsymbol{\mathcal{Y}}^{\ast} \right)^{\intercal}
    \label{eq:PCAcorr}
\end{align}
\end{subequations}

\noindent with $\boldsymbol{\mathcal{Y}}^{\ast}$ denoting the standardized response matrix storing the standardized response variables $\boldsymbol{y}^{\ast}\coloneqq \operatorname{diag}\left(\hat{\boldsymbol{\sigma}}_{\boldsymbol{Y}}\right)\left(\boldsymbol{y}-\hat{\boldsymbol{\mu}}_{\boldsymbol{Y}}\right)$, i.e. $\boldsymbol{\mathcal{Y}}^{\ast}\coloneqq\left(\boldsymbol{y}^{\ast(k)},\dotsc,\boldsymbol{y}^{\ast(k)},\dotsc,\boldsymbol{y}^{\ast(K)}\right)^{\intercal} \in \mathbb{R}^{N \times K}$. On the other hand, a rich variety of non-intrusive and sparse methods exist to estimate the PCE coefficient matrix $\boldsymbol{\mathrm{A}}$ using both, the input matrix $\boldsymbol{\mathcal{X}} \in \mathbb{R}^{M \times K}$ and response matrix $\boldsymbol{\mathcal{Y}}$ \citep{Luthen2021}. In the main study, we chose the least angle regression algorithm \citep{Blatman2011a} due to its high evaluation speed and its high accuracy even for very small experimental designs.

\subsection{PCA-PCE based Hoeffding-Sobol decomposition \& Sobol indices}
\label{subsec:Sobol}

One of the major advantages to use PCE emulators for computational intense simulations is the relation between PCE and the Hoeffding-Sobol decomposition and thereby Sobol indices \citep{Sudret2008GlobalExpansions}. For completeness, we repeat here some of the theory already discussed elsewhere \citep{Homma1996ImportanceModels,Sobol2001GlobalEstimates,Sudret2008GlobalExpansions,Wagner2020} and derive the PCA-PCE based Sobol indices accounting for the standardization in the PCA discussed in the previous subsection.

We start with the global variance decomposition theory derived by Sobol \citep{Sobol2001GlobalEstimates}. It can be shown that for any univariate integrable function $\mathcal{M}\left(\boldsymbol{X} \right)$ with $M$ mutually independent random input variables $X_i$ in $\mathcal{D}_{\boldsymbol{X}}$ and $i=\left\{1,2,\dotsc,M\right\}$, there exists a unique functional decomposition, which is often referred to as Hoeffding-Sobol decomposition \citep{Sobol2001GlobalEstimates}:

\begin{multline}
    \mathcal{M}\left(\boldsymbol{X} \right) = \mathcal{M}_{0} + \sum_{i=1}^{M} \mathcal{M}_{i}\left(X_{i} \right) + \sum_{1 \leq i < j \leq M} \mathcal{M}_{i,j}\left(X_{i},X_{j} \right) + \dotsc + \mathcal{M}_{1,2,\dotsc,M}\left(X_{1},\dotsc,X_{M} \right)
\label{eq:SOBdecomp1}
\end{multline}

\noindent where the following two conditions hold:

\begin{enumerate}
  \item The first term $\mathcal{M}_{0}$ is constant and equal to the expected value of $\mathcal{M}\left(\boldsymbol{x} \right)$:
  
  \begin{equation}
    \mathcal{M}_{0}=\mathbb{E}\left[\mathcal{M}\left(\boldsymbol{X} \right)\right]=\int_{\mathcal{D}_{\boldsymbol{X}}}\mathcal{M}\left(\boldsymbol{x} \right)\,\mathrm{d}\boldsymbol{x}
  \label{eq:SOBmean}
  \end{equation}

  \item All the terms in the functional decomposition are orthogonal:
  
  \begin{equation}
    \int_{\mathcal{D}_{\boldsymbol{X}_{u}}}\mathcal{M}_{u}\left(\boldsymbol{x}_{u} \right)\,dx_{i_{k}}=0 \ \ , \ \   1 \leq k \leq s
  \label{eq:SOBorth}
  \end{equation}
  
  

  \noindent with $u$ being defined as a subset of indices, i.e. $u \coloneqq \left\{ i_{1},\dotsc,i_{s} \right\} \subset \left\{ 1,\dotsc,M \right\}$
  
\end{enumerate}

\noindent Further assuming that the function $\mathcal{M}\left(\boldsymbol{X} \right)$ is square-integrable, the functional decomposition in \hyperref[eq:SOBdecomp1]{Eq.~\ref{eq:SOBdecomp1}} may be squared and integrated to provide the variance decomposition: 

\begin{equation}
    V = \sum_{i=1}^{M} V_{i} + \sum_{1 \leq i < j \leq M} V_{i,j} + \dotsc + V_{1,2,\dotsc,M} \label{eq:SOBdecomp2}
\end{equation}

\noindent with the total variance $V$ and the partial variances $V_{u}$ defined as:

\begin{subequations}
\begin{align}
    V &= \operatorname{Var}\left[ \mathcal{M}\left(\boldsymbol{X} \right)\right] = \int_{\mathcal{D}_{\boldsymbol{X}}}\mathcal{M}^{2}\left(\boldsymbol{x} \right)\,\mathrm{d}\boldsymbol{x} - \mathcal{M}_{0}^{2} \label{eq:SOBvartot} \\
    V_{u} &= \operatorname{Var}\left[ \mathcal{M}_u\left(\boldsymbol{X}_u \right)\right] =\int_{\mathcal{D}_{\boldsymbol{X}_{u}}}\mathcal{M}_{u}^{2}\left(\boldsymbol{x}_{u} \right)\,\mathrm{d}\boldsymbol{x}_{u} \label{eq:SOBvarpart}
\end{align}
\end{subequations}

\noindent Based on these results, Sobol indices $S_{u}$ can be defined as a natural global sensitivity measure of $\mathcal{M}\left(\boldsymbol{X} \right)$ on the input variables $\boldsymbol{X}_{u}$: 

\begin{equation}
   S_{u} \coloneqq \frac{V_u}{V} 
\label{eq:Sobol_tot}
\end{equation}

\noindent Consequently, $S_{u}$ represents the relative contribution of the set of variables $u$ to the total variance $V$. First order indices $S_i$ indicate the influence of $X_i$ alone, whereas the higher order indices quantify possible interactions or mixed influences between multiple variables. In addition, we can also define the total Sobol index $S_i^T$  to evaluate the total effect of an input parameter $X_i$ on $\mathcal{M}\left(\boldsymbol{X} \right)$:

\begin{equation}
   S_{i}^{T} \coloneqq \frac{1}{V} \sum_{u \supset i} V_{u}
\label{eq:SOBtot1}
\end{equation}

\noindent As a result, $S_i^T$ includes not only the effect of $X_i$ alone but in addition the effect induced by all interactions between $X_i$ and the other variables. This is also the reason, why the sum of the total Sobol indices $\sum_{i}S_i^T$ can in fact exceed 1. As an example, if we have an interaction between the variables $X_1$ and $X_2$, their interaction effect on $\mathcal{M}$ is counted twice, once in $S_1^T$ and another time in $S_2^T$. This example in mind, it is easy to see that the peaks in Fig.~3(e) in the main study highlight regions, where the interaction terms between the individual variables significantly contribute to the total effect on $\mathcal{M}$. We have to add that for our study, the absolute values of $S_i^T$ are of less importance. We are more interested in the relative size of $S_i^T$, because the comparison of these values allows us to draw conclusions about the relative importance of the corresponding variables $X_i$ for the model response $\mathcal{M}\left(\boldsymbol{X} \right)$. As shown by \citep{Sudret2008GlobalExpansions}, $S_i^T$ can also be computed as:

\begin{subequations}
\begin{align}
    S_{i}^{T} &= 1-S_{\sim i}
    \label{eq:SOBtot2a} \\
    &= 1-\frac{\operatorname{Var}_{X_{\sim i}}\left[ \mathbb{E}_{X_{i}}\left[\mathcal{M}\left(\boldsymbol{X}\right)\right]\right]}{\operatorname{Var}\left[\mathcal{M}\left(\boldsymbol{X}\right)\right]}
    \label{eq:SOBtot2b}
\end{align}
\end{subequations}

\noindent where we use ${\sim}i$ to denote a set of indices, which do not include $i$, i.e. $S_{\sim i}=S_{v}$ with $v=\{1,\dotsc,i-1,i+1,\dotsc,M\}$.

Suppose now that we have a PCA-PCE surrogate model to emulate the vector-valued model response $\boldsymbol{Y}=\mathcal{M}\left(\boldsymbol{X} \right)$ with the random input vector $\boldsymbol{X} \in \mathbb{R}^{M \times 1}$ and random response vector $\boldsymbol{Y} \in \mathbb{R}^{N \times 1}$. To derive the $S_{i,k}^{T}$ for each response variable $k \in\{1,2,\dotsc,N\}$, we start with $\operatorname{Var}_{X_{\sim i}}\left[ \mathbb{E}_{X_{i}}\left[Y_k\right]\right]$ from \hyperref[eq:SOBtot2b]{Eq.~\ref{eq:SOBtot2b}} by replacing $Y_k$ with the $k^{\text{th}}$ component of \hyperref[eq:PCAEc]{Eq.~\ref{eq:PCAEc}}:

\begin{subequations}
\begin{align}
    \operatorname{Var}_{X_{\sim i}}\left[ \mathbb{E}_{X_{i}}\left[Y_k\right]\right]&=\mathbb{E}_{X_{\sim i}}\left[ \left(\mathbb{E}_{X_{i}}\left[Y_k\right]\right)^2\right]-\left(\mathbb{E}_X\left[Y_k\right]\right)^2
    \label{eq:SOBder1a}\\
    &=\mathbb{E}_{X_{\sim i}}\left[ \left(\mathbb{E}_{X_{i}}\left[\mu_{Y_k}+\sigma_{Y_k}\boldsymbol{\phi}_k^{\text{row}}\mathbf{A}\boldsymbol{\Psi}\left(\boldsymbol{X}\right)\right]\right)^2\right]-\mu_{Y_k}^2
    \label{eq:SOBder1b}
\end{align}
\end{subequations}

\noindent where we used $\boldsymbol{\phi}_k^{\text{row}}\coloneqq \left(\phi_{k1},\dotsc,\phi_{kN'}\right)$. We can simplify this expression by expanding the first term and considering that the expectation vanishes for all principal components, i.e. $\mathbb{E}\left[\mathbf{A}\boldsymbol{\Psi}\left(\boldsymbol{X}\right)\right]=0$:

\begin{subequations}
\begin{align}
    \operatorname{Var}_{X_{\sim i}}\left[ \mathbb{E}_{X_{i}}\left[Y_k\right]\right]&=\mathbb{E}_{X_{\sim i}}\left[\left(\sigma_{Y_k}\boldsymbol{\phi}_k^{\text{row}}\mathbf{A}\mathbb{E}\left[\boldsymbol{\Psi}\left(\boldsymbol{X}\right)\right]\right)^2\right]
    \label{eq:SOBder2a}\\
    &=\mathbb{E}_{X_{\sim i}}\left[\left(\sum_{\boldsymbol{\alpha} \in \mathcal{A}^{\star}} \sum_{j=1}^{N'}  \sigma_{Y_k}\phi_{kj} a_{j, \boldsymbol{\alpha}} \mathbb{E}\left[\boldsymbol{\Psi}\left(\boldsymbol{X}\right)\right] \right)^2\right]
    \label{eq:SOBder2b}
\end{align}
\end{subequations}

\noindent As shown by \citep{Wagner2020}, due to the orthonormality of the polynomial basis $\left\{\Psi_{\boldsymbol{\alpha}}\right\}_{\boldsymbol{\alpha}\in\mathcal{A}^{\star}}$, we can further simplify \hyperref[eq:SOBder2b]{Eq.~\ref{eq:SOBder2b}} resulting in:

\begin{equation}
   \operatorname{Var}_{X_{\sim i}}\left[ \mathbb{E}_{X_{i}}\left[Y_k\right]\right]=\sigma_{Y_k}^2\sum_{\boldsymbol{\alpha} \in \mathcal{A}^{\star}_{i=0}} \left(\sum_{j=1}^{N'}  \phi_{kj} a_{j, \boldsymbol{\alpha}}\right)^2 
\label{eq:SOBder3}
\end{equation}

\noindent with the subset $\mathcal{A}^{\star}_{i=0} \coloneqq \left\{ \boldsymbol{\alpha}\in \mathcal{A}^{\star} \mid \alpha_i=0 \right\}$. Using these results, we can compute the total variance with:

\begin{equation}
   \operatorname{Var}\left[ Y_k\right]=\sigma_{Y_k}^2\sum_{\boldsymbol{\alpha} \in \mathcal{A}^{\star}} \left(\sum_{j=1}^{N'}  \phi_{kj} a_{j, \boldsymbol{\alpha}}\right)^2 
\label{eq:SOBder4}
\end{equation}

\noindent In the end, we get the total PCE-PCA based Sobol index $S_{i,k}^{T}$ for the input variable $i$ and the response variable $k$ by inserting \hyperref[eq:SOBder3]{Eq.~\ref{eq:SOBder3}} and \hyperref[eq:SOBder4]{Eq.~\ref{eq:SOBder4}} into \hyperref[eq:SOBtot2b]{Eq.~\ref{eq:SOBtot2b}}:

\begin{equation}
   S_{i,k}^{T} = 1-\frac{\sum_{\boldsymbol{\alpha}\in\mathcal{A}_{i=0}^{\star}}\left( \sum_{j=1}^{N'}\phi_{kj} \, a_{j,\boldsymbol{\alpha}}\right)^{2}}{\sum_{\boldsymbol{\alpha}\in\mathcal{A}^{\star}}\left( \sum_{j=1}^{N'}\phi_{kj} \, a_{j,\boldsymbol{\alpha}}\right)^{2}} 
\label{eq:SOBtotPCEPCA}
\end{equation}


%
%

\newpage

\subsection{Uncertainty analysis}
\label{subsec:Uncertainty}

For completeness, we repeat here the uncertainty analysis pipeline adopted for the measured and simulated pulse-height spectra and highlight some changes to \citep{Breitenmoser2022ExperimentalSpectrometry}. 

For the radiation measurements, the statistical uncertainty of the net count rate spectra $c_{\text{exp},k}$ characterized by the standard deviation was computed adopting a probabilistic Poisson model \citep{Knoll2010}:

\begin{equation}
    \sigma_{\text{pois},\text{exp},k} = \sqrt{ \frac{C_{\text{gr},k}}{t_{\text{gr}}^2} + \frac{C_{\text{bg},k}}{t_{\text{bg}}^2}}
\end{equation}

\noindent where $C_{\text{gr},k}$ and $C_{\text{bg},k}$ are the gross and background counts in channel $k$ together with the gross and background measurement live times~$t_{\text{gr}}$ and $t_{\text{bg}}$, respectively. The small statistical uncertainty in the live time measurement is neglected. To compute the source activity $A$ as a function of the measurement date $t$, we use the fundamental exponential law of decay, i.e. $A = A_{0} \cdot 2^{- \Delta t / t_{1/2}}$ \citep{Knoll2010}. The uncertainty induced by the source activity~$A$ normalization is quantified using the standard error propagation methodology for independent variables \citep{Abernethy1985ASMEUncertainty}: 


\begin{equation}
    \sigma_{A} = \sigma_{A_{0}} \cdot 2^{- \Delta t / t_{1/2}}
\end{equation}

\noindent with the reference activity~$A_{0}$ and associated uncertainty~$\sigma_{A_{0}}$ provided by the vendor, the source half life~$t_{1/2}$ \citep{Pearce2008} as well as the time difference~$\Delta t = t - t_{0}$ between the reference date~$t_{0}$ and the measurement date~$t$. Contributions of the uncertainties in $t_{1/2}$ and $\Delta t$ to $\sigma_{A}$ are found to be less than $1\%$ for all performed measurements and are therefore neglected. We then summarize the total experimental uncertainty as follows \citep{Abernethy1985ASMEUncertainty}:

\begin{equation}
    \sigma_{\text{tot},\text{exp},k} = \sqrt{ \left( \frac{\sigma_{\text{pois},\text{exp},k}}{A}\right)^{2} + \left( \frac{c_{\text{exp},k}}{A} \cdot \sigma_{A} \right)^{2}}
\end{equation}

For the simulations, we computed the statistical uncertainty of the net count rate spectrum $c_{\text{sim},k}$ characterized by the standard deviation as follows \citep{Knoll2010}:

\begin{equation}
    \sigma_{\text{stat},\text{sim},k} = \sqrt{ \frac{1}{N_{\text{pr}} \left( N_{\text{pr}}-1 \right)} \cdot \left[  \left(N_{\text{pr}} - N_{\text{dep}} \right) \cdot c_{\text{sim},k}^{2} + \sum_{l=1}^{N_{\text{dep}}} \left( c_{\text{sim},kl} - c_{\text{sim},k} \right)^{2} \right]}
\end{equation}

\noindent where $c_{\text{sim},kl}$ are the individual broadened energy deposition events in the detector channel $k$, $N_{\text{dep}}$ the number of recorded events and $N_{\text{pr}}$ the number of simulated primaries. It is good practice in Monte Carlo studies to report not only the estimated uncertainty in the sample mean $c_{\text{sim},k}$ using the sample standard deviation $\sigma_{\text{stat},\text{sim},k}$ but also the so called variance of the sample variance $\operatorname{VOV}_k$ for the detector channel $k$ to quantify the statistical uncertainty in $\sigma_{\text{stat},\text{sim},k}^2$ itself \citep{Forster1994TenMCNP}:

\begin{equation}
    \operatorname{VOV}_{k} = \frac{ \operatorname{Var} \left(\sigma_{\text{stat},\text{sim},k}^{2} \right)}{\sigma_{\text{stat},\text{sim},k}^{4}} = \frac{\left(N_{\text{pr}} - N_{\text{dep}} \right) \cdot c_{\text{sim},k}^{4} + \sum_{l=1}^{N_{\text{dep}}} \left( c_{\text{sim},kl} - c_{\text{sim},k} \right)^{4}}{ \left[ \left(N_{\text{pr}} - N_{\text{dep}} \right) \cdot c_{\text{sim},k}^{2} + \sum_{l=1}^{N_{\text{dep}}}  \left( c_{\text{sim},kl} - c_{\text{sim},k} \right)^{2} \right]^{2}} - \frac{1}{N_{\text{pr}}}
\end{equation}

\noindent The propagation of the systematic uncertainties for the simulated detector response was performed by the Monte Carlo sampling technique. We considered the same model parameters for the uncertainty propagation as in \citep{Breitenmoser2022ExperimentalSpectrometry}. These parameters are the energy calibration factor $D_{1}\left[\unit{keV^{-1}}\right]$ as well as the empirical resolution parameters $B_{1}\left[\unit{-}\right]$ and $B_{2}\left[\unit{-}\right]$. However, we adapted the marginal distributions by introducing truncated normal distributions as summarized in \hyperref[tab:marginal]{Table~\ref{tab:marginal}}. In addition, we accounted for the statistical dependence of the model parameters $B_{1}$ and $B_{2}$ by correlated sampling using the Gaussian copula $\mathcal{C}_{\mathcal{N}}$ \citep{Sklar1959FonctionsMarges}:

\begin{subequations}
\begin{align}
    \left\{ B_1^{\ast},B_2\right\} &\sim\mathcal{C}_{\mathcal{N}}\left(F_{B_1^{\ast}}\left(b_1^{\ast}\right),F_{B_2}\left(b_2\right);\mathbf{R}\right)
    \label{eq:COPa}  \\
    &\sim\Phi_2\left(\Phi^{-1}\left(F_{B_1^{\ast}}\left(b_1^{\ast}\right)\right),\Phi^{-1}\left(F_{B_2}\left(b_2\right)\right);\mathbf{R}\right)
    \label{eq:COPb}
\end{align}
\end{subequations}

\noindent with the log-transformed variable $B_1^{\ast}\coloneqq\log\left(B_1\right)$, the linear correlation matrix $\mathbf{R}$ obtained by the regression analysis, the marginal distribution functions $F$ provided in \hyperref[tab:marginal]{Table~\ref{tab:marginal}}, the bivariate Gaussian distribution function $\Phi_2$ associated with the Gaussian copula $\mathcal{C}_{\mathcal{N}}$ and the inverse cumulative distribution function of the standard normal distribution $\Phi^{-1}$, respectively. The energy calibration factor $D_1$ is sampled independently according to the corresponding marginal as in \citep{Breitenmoser2022ExperimentalSpectrometry}. For more details and relevant literature on the copula theory, the reader is referred to \citep{Joe2014DependenceCopulas,Nelsen2006AnCopulas}.

The $N_{\text{MC}} \in \mathbb{N}_{>1}$ independently drawn input samples $\boldsymbol{\mathcal{X}}_{\text{MC}}=(\boldsymbol{x}^{(1)},...,\boldsymbol{x}^{(m)},...\boldsymbol{x}^{(N_{\text{MC}})})^{\intercal}$ from the probabilistic input model with $\boldsymbol{X}\coloneqq\left(D_{1},B_{1},B_{2}\right)^{\intercal}$ are then propagated through the postprocessing pipeline described in \citep{Breitenmoser2022ExperimentalSpectrometry} to obtain the corresponding spectral count rate samples $\boldsymbol{\mathcal{Y}}_{\text{MC}}=(c_{\text{sim},k}^{(1)},...,c_{\text{sim},k}^{(m)},..., c_{\text{sim},k}^{(N_{\text{MC}})})^{\intercal}$ with $k \in \{1,...,1024\}$. These samples can then be used to compute the sample standard deviation $\sigma_{\text{sys},\text{sim},k}$ similar to \hyperref[eq:PCAstd]{Eq.~\ref{eq:PCAstd}} and thereby quantify the systematic uncertainty with respect to the empirical model parameters $D_{1}$, $B_{1}$ and $B_{2}$. The total uncertainty characterized by the sample standard deviation can be summarized in the same way as for the experimental uncertainty \citep{Abernethy1985ASMEUncertainty}:

\begin{equation}
    \sigma_{\text{tot},\text{sim},k} = \sqrt{ \sigma_{\text{stat},\text{sim},k}^2 + \sigma_{\text{sys},\text{sim},k}^2}
\end{equation}

\newpage

\subsection{Compton edge shift analysis}
\label{subsec:CEshift}

To better understand the nature of the Compton edge shift utilized in our study, we quantify here the spectral shift between the measured Compton edge energy and the theoretical value according to the Compton scattering theory (cf. Eq.~2 in the main study). Because the measured Compton edges are obscured by the finite spectral resolution, quantification of the exact position in the measured pulse-height spectrum would require additional coincidence measurements \citep{Cherubini1989GammaScintillators,Swiderski2010MeasurementScintillators}. Therefore, similar to a previous study \citep{Mauritzson2022GEANT4-basedScintillator}, we use an alternative approach by quantifying the spectral shift between the already available Monte Carlo simulations with a proportional scintillation response and the measured pulse-height spectra. It is important to add that, based on the findings reported in the main study as well as due to the improved signal-to-noise ratio, we focus our investigation here on the sum channel. 

In a first step, we determine the inflection points as a characteristic measure of the corresponding Compton edges using spline regression \citep{Reinsch1967SmoothingFunctions}. We then compute the Compton edge shift as the spectral difference between the determined inflection points for the measured and simulated spectra. We apply this method to different Compton edges, i.e. 477.334(3)~\unit{keV} associated with the $^{137}\text{Cs}$ emission line at 661.657(3)~\unit{keV}, 699.133(3)~\unit{keV} associated with the $^{88}\text{Y}$ emission line at 898.042(3)~\unit{keV}, 963.419(3)~\unit{keV} associated with the $^{60}\text{Co}$ emission line at 1173.228(3)~\unit{keV} and 1611.77(1)~\unit{keV} associated with the $^{88}\text{Y}$ emission line at 1836.063(3)~\unit{keV}. Moreover, we perform Monte Carlo based uncertainty quantification by systematically propagating the uncertainty in the hyperparamters for the individual spline regression models. 

In the \hyperref[fig:ECshiftsum]{Fig.~\ref{fig:ECshiftsum}}, we present the results of our Compton edge shift analysis for the sum channel. In general, we can identify a consistent trend in the shift, i.e. an enhanced Compton edge shift toward smaller spectral energies with increasing Compton edge energy. As discussed in the main study, our NPSM predicts the Compton edge shift for all analyzed Compton edges with high accuracy. However, because our approach is based on complex Monte Carlo simulations, interpretability of the results and their connection to the various underlying physical processes is challenging. Therefore, to improve our understanding of the relationship between the NPSM and the resulting Compton edge shift, we develop here a simplified semi-analytical model.

We start by deriving the light yield function $\mathcal{L}$ as a function of the initial electron kinetic energy $E_{\text{k}}$ by integrating Eq.~1 from the main study from $E_{\text{k}}$ down to the mean excitation energy $I$:

\begin{equation}
    \mathcal{L}\left( E_{\text{k}} \right) = \int_{I}^{E_{\text{k}}}L\left(dE/ds\right)dE_{\text{k}}'
\label{eq:LYintegral}
\end{equation}

\noindent To compute the integral in \hyperref[eq:LYintegral]{Eq.~\ref{eq:LYintegral}}, we first need a model to describe the differential energy loss $dE$ per differential path length $ds$ as a function of the kinetic electron energy $E_{\text{k}}$. Similar to Payne and his co-workers \citep{Payne2009,Payne2011}, we apply a modified Bethe-Bloch model derived by Joy and Luo \citep{Joy1989AnElectrons} to described $dE/ds$ as a function of $E_{\text{k}}$ in units of \unit{eV.\AA^{-1}}:

\begin{equation}
    dE/ds\left( E_{\text{k}} \right) = 785 \frac{\rho Z}{A E_{\text{k}}}\log{\left[\frac{1.166\left(E_{\text{k}}+cI\right)}{I}\right]}
\label{eq:dEds}
\end{equation}

\noindent with the scintillator related mass density $\rho$, the atomic number $Z$ and the molecular weight $A$. In accordance with the results obtained by Payne and his co-workers, we fix the stopping power correction factor $c$ in \hyperref[eq:dEds]{Eq.~\ref{eq:dEds}} to $c=2.8$. To account for radiative losses as well as relativistic effects at higher energies, we combine Joy's model with the \texttt{ESTAR} database \citep{Berger2017ESTAR2.0.1}. We also consulted the \texttt{ESTAR} database for all material related properties of NaI(Tl) (cf. \hyperref[tab:NaI]{Table~\ref{tab:NaI}}). The resulting total stopping power model together with the individual model components are shown in \hyperref[fig:LightYield]{Fig.~\ref{fig:LightYield}(a)} for NaI(Tl). 

We then combine the derived stopping power model with the light yield $L\left(dE/ds\right)$ to perform the integration in \hyperref[eq:LYintegral]{Eq.~\ref{eq:LYintegral}}. For the model parameters in $L\left(dE/ds\right)$, i.e. $\eta_{e/h}$, $dE/ds\mid_{\text{Ons}}$, $dE/ds\mid_{\text{Trap}}$ and $dE/ds\mid_{\text{Birks}}$, we applied the maximum a posteriori (MAP) probability point estimates to compute a mean light yield function as well as the full set of posterior samples to derive the corresponding credible intervals for both, the sum channel and the individual crystals associated with the sum and single mode inversion pipelines, respectively. The resulting relative light yield functions $\mathcal{L}\left( E_{\text{k}} \right)/E_{\text{k}}$ are shown in \hyperref[fig:LightYield]{Fig.~\ref{fig:LightYield}(b)}. The characteristic shape of these relative light yield curves in \hyperref[fig:LightYield]{Fig.~\ref{fig:LightYield}(b)} has been extensively documented by numerous previous empirical studies \citep{Wayne1998ResponseElectrons,Khodyuk2010,Rooney1997ScintillatorResponse,Hull2009,Payne2009,Payne2011}, illustrating an increase in light yield with increasing energy for $E_{\text{k}}\ll\qty{10}{keV}$, a prominent peak around \qty{10}{keV}, followed by a subsequent decrease in yield for higher energies.

After successfully deriving the light yield function $\mathcal{L}$ as a function of the kinetic electron energy $E_{\text{k}}$, we now continue by investigating the connection between the observed negative Compton edge shift
and the non-proportional nature of this light yield function. For an ideal detector with a proportional scintillation response, we would have a constant relative light yield function:

\begin{equation}
\frac{\mathcal{L}\left(E_{\text{k}}\right)}{E_{\text{k}}}=\text{const.}
\label{eq:rLYprop}
\end{equation}

\noindent So, we can easily see that for a detector with a non-proportional scintillation light yield function, we get a spectral shift $\Delta E$, if we convert the produced scintillation light of an electron with energy $E_{\text{k},1}$ at a different energy $E_{\text{k},2}$:

\begin{subequations}
\begin{align}
    \Delta E&=\frac{\mathcal{L}\left(E_{\text{k},1}\right)}{\mathcal{L}\left(E_{\text{k},2}\right)/E_{\text{k},2}}-\frac{\mathcal{L}\left(E_{\text{k},1}\right)}{\mathcal{L}\left(E_{\text{k},1}\right)/E_{\text{k},1}}
    \label{eq:CEshift1a}\\
    &=\frac{\mathcal{L}\left(E_{\text{k},1}\right)}{\mathcal{L}\left(E_{\text{k},2}\right)}E_{\text{k},2}-E_{\text{k},1}
    \label{eq:CEshift1b}
    \end{align}
\end{subequations}

\noindent It is important to add that in our analysis, we implicitly assume a continuous deceleration of the involved electrons, starting from their initial kinetic energy $E_{\text{k}}$ and progressing down to the scintillator specific excitation energy $I$. In particular. we neglect electrons escaping from the scintillator. From \hyperref[eq:CEshift1b]{Eq.~\ref{eq:CEshift1b}} we can conclude that this shift $\Delta E$ will be positive for $\mathcal{L}\left(E_{\text{k},1}\right)/E_{\text{k},1}>\mathcal{L}\left(E_{\text{k},2}\right)/E_{\text{k},2}$ and negative for $\mathcal{L}\left(E_{\text{k},1}\right)/E_{\text{k},1}<\mathcal{L}\left(E_{\text{k},2}\right)/E_{\text{k},2}$. Moreover, with an increase in the relative difference $\lvert 1 - \left[ \mathcal{L}\left(E_{\text{k},1}\right)/E_{\text{k},1} \right] / \left[ \mathcal{L}\left(E_{\text{k},2}\right)/E_{\text{k},2} \right] \rvert$, we expect a proportional increase in the magnitude of the spectral shift. 

We can now use these insights for our Compton edge shift analysis. In gamma-ray spectrometry with inorganic scintillators, the energy calibration is typically performed using full energy peaks (FEPs), sometimes also called photopeaks \citep{Knoll2010,Breitenmoser2022ExperimentalSpectrometry}. In other words, to relate the photon energy with the generated scintillation photons, we use the relative light yield function $\sum_{j=1}^{N_{e^{-}}} \mathcal{L}(E_{\text{k},j}^{\text{FEP}})/\sum_{j=1}^{N_{e^{-}}} E_{\text{k},j}^{\text{FEP}}$ for $j=\{1,\dotsc,N_{e^{-}}\}$ electrons with kinetic energy $E_{\text{k},j}$ generated in the scintillator leading subsequently to a FEP. Consequently, in order to compute the Compton edge shift for inorganic scintillators using our model in \hyperref[eq:CEshift1b]{Eq.~\ref{eq:CEshift1b}}, we need to investigate the light yield for both Compton edge (CE) and FEP events, more specifically the integrated light yield for all electrons generated during these events.

We start with the CE events: As already discussed in the main study, in a CE event, a photon enters the scintillator, undergoes a single Compton scattering (COM) event with a deflection angle of \ang{180}, i.e. full back-scattering, and subsequently escapes the scintillator. During this interaction, some of the photon's energy gets transferred to a single atomic electron. Neglecting Doppler broadening and atomic shell effects \citep{Ribberfors1975RelationshipStates,Brusa1996FastScattering}, the transferred energy $E_{\text{k}}^{\text{CE}}$ is equivalent to the CE energy discussed in Eq.~2 in the main study, i.e.:

\begin{equation}
    E_{\text{k}}^{\text{CE}} = E_{\gamma}^0 \left(
    1-\frac{1}{1+\frac{2E_{\gamma}^0}{m_{e}c^2}}
    \right)
\label{eq:CEshift2}
\end{equation}

\noindent with $E_{\gamma}^0$ being the initial photon energy and $m_{e}c^2$ the energy equivalent electron mass. Because only one COM event takes place with a deterministic energy transfer, the light yield for a CE event can easily be calculated as follows:

\begin{equation}
\mathcal{L}\left(E_{\text{k}}^{\text{CE}}\right)=\mathcal{L}\left[E_{\gamma}^0 \left(
    1-\frac{1}{1+\frac{2E_{\gamma}^0}{m_{e}c^2}}
    \right)\right]
\label{eq:CEshift3}
\end{equation}

FEP events on the other hand are more complex because they involve a variable number of COM events with a subsequent photoelectric absorption (PE) of the photon in the scintillator. For simplicity, we neglect again Doppler broadening as well as atomic shell effects and consider only secondary electrons generated during COM and PE events. In particular, we neglect fluorescence photons and Auger electrons. Using these simplifications, we can calculate the light yield for a FEP event involving $j=\{1,\dotsc,N_{e^{-}}\}$ electrons with kinetic energy $E_{\text{k},j}$ as a sequence of $N_{\text{COM}}$ COM events followed by a single PE event:

\begin{subequations}
\begin{align}
\sum_{j=1}^{N_{e^{-}}} \mathcal{L}\left(E_{\text{k},j}^{\text{FEP}}\right)&=\mathcal{L}\left(E_{\text{k}}^{\text{PE}}\right) + \sum_{i=1}^{N_{\text{COM}}}\mathcal{L}\left(E_{\text{k},i}^{\text{COM}} \right)
\label{eq:CEshift4a}\\
&=\mathcal{L}\left(E_{\gamma}^{N_{\text{COM}}}\right) + \sum_{i=1}^{N_{\text{COM}}}\mathcal{L}\left(E_{\gamma}^{i-1}-E_{\gamma}^{i} \right)
\label{eq:CEshift4b}
\end{align}
\end{subequations}

\noindent where we denote with $E_{\text{k},j}^{\text{PE}}$ and $E_{\text{k},j}^{\text{COM}}$ the kinetic energies of the electrons generated during PE and COM events, respectively, and with $E_{\gamma}^{i}$ the photon energy after $i$ subsequent COM events. Both, the number of COM events ($N_{\text{COM}}$) as well as the transferred energy in these COM events ($E_{\gamma}^{i-1}-E_{\gamma}^{i}$) are linked in a complex stochastic process and depend on the photon energy as well as the properties of the scintillator \citep{Klein1929UberDirac,Ribberfors1975RelationshipStates,Brusa1996FastScattering}. To estimate these variables for our specific detector system, we apply once again Monte Carlo methods. More specifically, we adopted the multi-purpose Monte Carlo code \texttt{FLUKA} with the same physics settings as described in the main study \citep{Ahdida2022NewCode}. For the semi-analytical model described here, the mass model only included the scintillation crystal embedded in a vacuum environment. To estimate the scintillator response, we irradiated the mass model with an isotropic and uniform monoenergetic photon flux of energy $E_{\gamma}^{0}$ using the \texttt{FLOOD} mode with the \texttt{BEAMPOSit} card. We repeated these simulations for 31 different photon energies $E_{\gamma}^{0}$ in the spectral range \qtyrange{500}{2000}{keV} with a spacing of \qty{50}{keV}. To score $N_{\text{COM}}$ as well as $E_{\gamma}^{i}$, we applied the user routine \texttt{mgdraw}. 

In \hyperref[fig:CEshiftPred]{Fig.~\ref{fig:CEshiftPred}(a)}, we present the probability density for the scored number of COM events before absorption ($N_{\text{COM}}$) as a function of the photon energy $E_{\gamma}^{0}$ in a prismatic NaI(Tl) scintillator with dimensions \qtyproduct[list-units = single]{10.2 x 10.2 x 40.6}{cm}, i.e. the same crystal size as for our detector system used in the main study. In line with previous results \citep{Cano-Ott1999}, we find a moderate increase of the mean number of COM events with increasing photon energy ranging from \num{1.4} at \qty{500}{keV} up to \num{2.3} at \qty{2000}{keV}. Combining these results together with \hyperref[eq:CEshift3]{Eq.~\ref{eq:CEshift3}} and \hyperref[eq:CEshift4b]{Eq.~\ref{eq:CEshift4b}}, we can now compute the spectral Compton edge shift according to \hyperref[eq:CEshift1b]{Eq.~\ref{eq:CEshift1b}} as follows:

\begin{equation}
    \Delta E=E_{\gamma}^0\left\{\frac{\mathcal{L}\left[E_{\gamma}^0 \left(
    1-\frac{1}{1+\frac{2E_{\gamma}^0}{m_{e}c^2}}
    \right)\right]}{\mathcal{L}\left(E_{\gamma}^{N_{\text{COM}}}\right) + \sum_{i=1}^{N_{\text{COM}}}\mathcal{L}\left(E_{\gamma}^{i-1}-E_{\gamma}^{i} \right)}- 
    1+\frac{1}{1+\frac{2E_{\gamma}^0}{m_{e}c^2}}
    \right\}
    \label{eq:CEshift5}
\end{equation}

\noindent where we used the fact that for a FEP event in our simplified framework, the following special relationship holds: $\sum_{j=1}^{N_{e^{-}}} E_{\text{k},j}^{\text{FEP}}=E_{\gamma}^0$. 

In \hyperref[fig:CEshiftPred]{Fig.~\ref{fig:CEshiftPred}(b)}, we show the resulting negative Compton edge shift $-\Delta E$ as a function of the photon energy $E_{\gamma}^{0}$ for the same prismatic NaI(Tl) scintillator. In general, we find a good agreement between the predictions of our simplified semi-analytical model and the experimental results. Deviations can be attributed to the various simplifications and assumptions made during model derivation, e.g. Landau fluctuations, Doppler broadening, atomic shell effects, detector cross-talk, neglected secondary particles such as fluorescence photons and Auger electrons or escaping electrons. 

By discriminating the predicted spectral shift for different number of COM events $N_{\text{COM}}$, we can gain also some further insights in the underlying physics. First, we find a pronounced increase in the Compton edge shift $\lvert\Delta E\rvert$, both for an increase in the photon energy $E_{\gamma}^{0}$ as well as for an increase in the number of COM events $N_{\text{COM}}$. This is in line with our predictions discussed above, i.e. we expect a proportional increase in the magnitude of the spectral shift for an increase in the relative difference $\lvert 1 - \mathcal{L}(\beta E_{\gamma}^0) / [\beta\sum_{j=1}^{N_{e^{-}}} \mathcal{L}(E_{\text{k},j}^{\text{FEP}})]\rvert$ with $\beta \coloneqq 1-1/[1+2E_{\gamma}^0/(m_{e}c^2)]$. It is now easy to see that, due to the decreasing trend in $\mathcal{L}$ for higher energies (cf. \hyperref[fig:LightYield]{Fig.~\ref{fig:LightYield}(b)}), the relative difference will increase for an increase in $E_{\text{k}}$ and subsequently $E_{\gamma}^0$. Furthermore, we find that the relative light yield for CE events is on average smaller than for FEP events with $E_{\gamma}^0\in\qtyrange{500}{2000}{keV}$, i.e. $\mathcal{L}(\beta E_{\gamma}^0)/\beta<\langle\sum_{j=1}^{N_{e^{-}}} \mathcal{L}(E_{\text{k},j}^{\text{FEP}})\rangle$. This explains the negative sign for the spectral shift observed by our NaI(Tl) detector system.


With this newly derived semi-analytical model, we have now also a tool to investigate the relation between the negative Compton edge shift and the size of a scintillation crystal. In \hyperref[fig:CEdetsize]{Fig.~\ref{fig:CEdetsize}(b)}, we present the predicted mean Compton edge shift as a function of the photon energy $E_{\gamma}^0$ alongside the relation between $N_{\text{COM}}$ and $E_{\gamma}^0$. We find a pronounced and consistent increase in the negative Compton edge shift for an increase in crystal size over the entire spectral domain \qtyrange{500}{2000}{keV}. From our semi-analytical model and the results in \hyperref[fig:CEdetsize]{Fig.~\ref{fig:CEdetsize}(a)}, it is evident that this trend can be explained by the increase in $N_{\text{COM}}$ for bigger scintillation crystals. 

In summary, the semi-analytical model derived in this section cannot only successfully predict the trends and sign of the Compton edge shift with increasing photon energy $E_{\gamma}^0$, but it can also be used to investigate the relation between crystal size and Compton edge shift and thereby supports the interpretation of the results and findings in the main study obtained by high-fidelity Monte Carlo simulations.

\newpage


\section{Supplementary Figures}\label{sec:Figures}

\vspace{20mm}

\begin{figure}[h!]%
\centerline{
\includegraphics[]{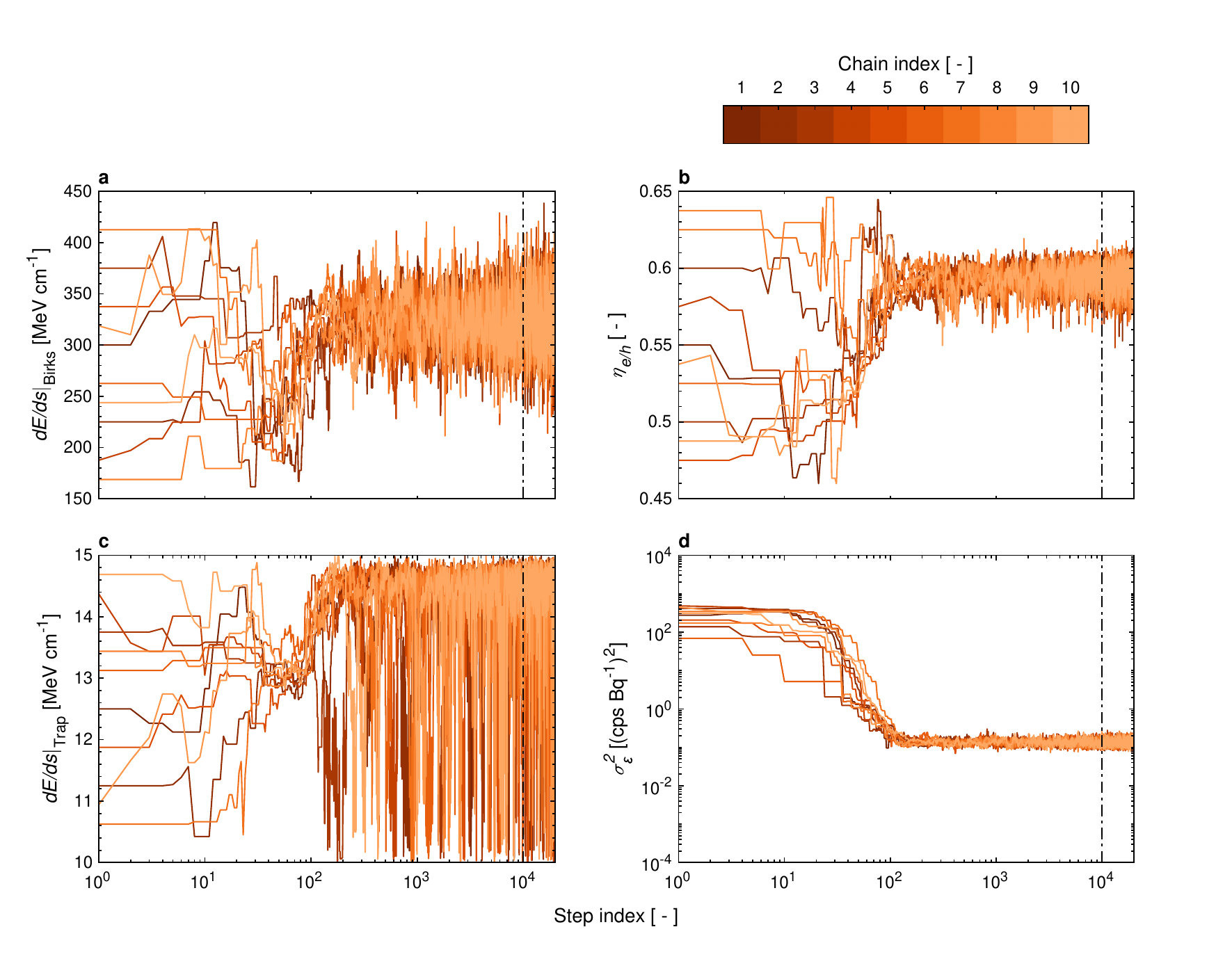}}
\caption[Markov Chain Monte Carlo trace plots for the sum mode]{\textbf{ Markov Chain Monte Carlo trace plots for the sum mode.} These graphs show the sample values of the Markov Chain Monte Carlo algorithm \citep{Goodman2010EnsembleInvariance} for each individual Markov chain and model parameter resulting from the sum mode inversion pipeline applied to the sum channel: \textbf{a}~The Birks related stopping power parameter $dE/ds\mid_{\text{Birks}}$. \textbf{b}~The free carrier fraction $\eta_{e/h}$. \textbf{c}~The trapping related stopping power parameter $dE/ds\mid_{\text{Trap}}$. \textbf{d}~The discrepancy model variance $\sigma^2_{\varepsilon}$. In addition, the burn-in threshold is highlighted as a dashed-dotted black line in each graph.}
\label{fig:tracesum}
\end{figure}

\newpage

\begin{figure}[h!]%
\centerline{
\includegraphics[]{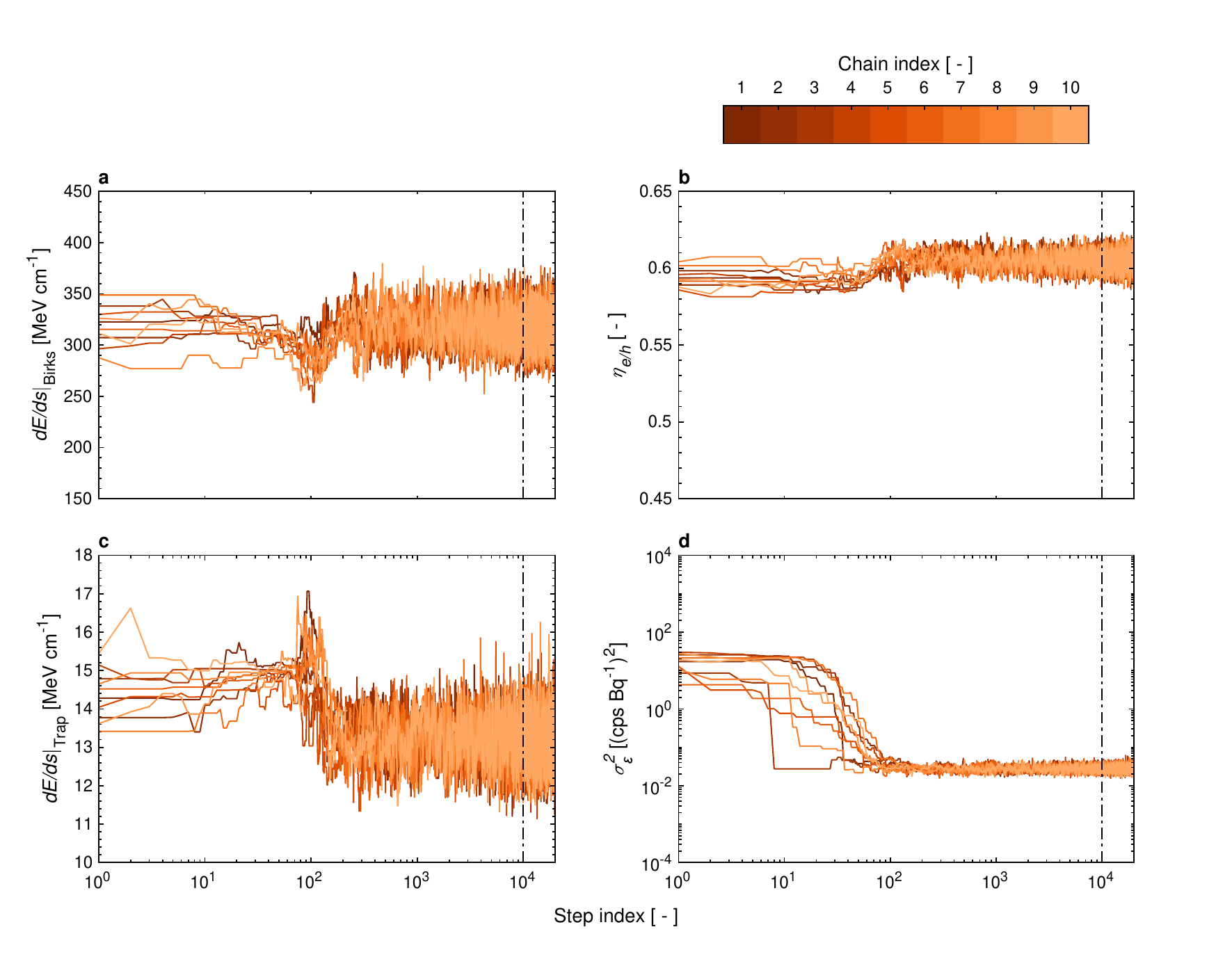}}
\caption[Markov Chain Monte Carlo trace plots for crystal 1]{\textbf{ Markov Chain Monte Carlo trace plots for crystal 1.} These graphs show the sample values of the Markov Chain Monte Carlo algorithm \citep{Goodman2010EnsembleInvariance} for each individual Markov chain and model parameter resulting from the single mode inversion pipeline applied to the scintillation crystal 1: \textbf{a}~The Birks related stopping power parameter $dE/ds\mid_{\text{Birks}}$. \textbf{b}~The free carrier fraction $\eta_{e/h}$. \textbf{c}~The trapping related stopping power parameter $dE/ds\mid_{\text{Trap}}$. \textbf{d}~The discrepancy model variance $\sigma^2_{\varepsilon}$. In addition, the burn-in threshold is highlighted as a dashed-dotted black line in each graph.}
\label{fig:trace1}
\end{figure}

\newpage

\begin{figure}[h!]%
\centerline{
\includegraphics[]{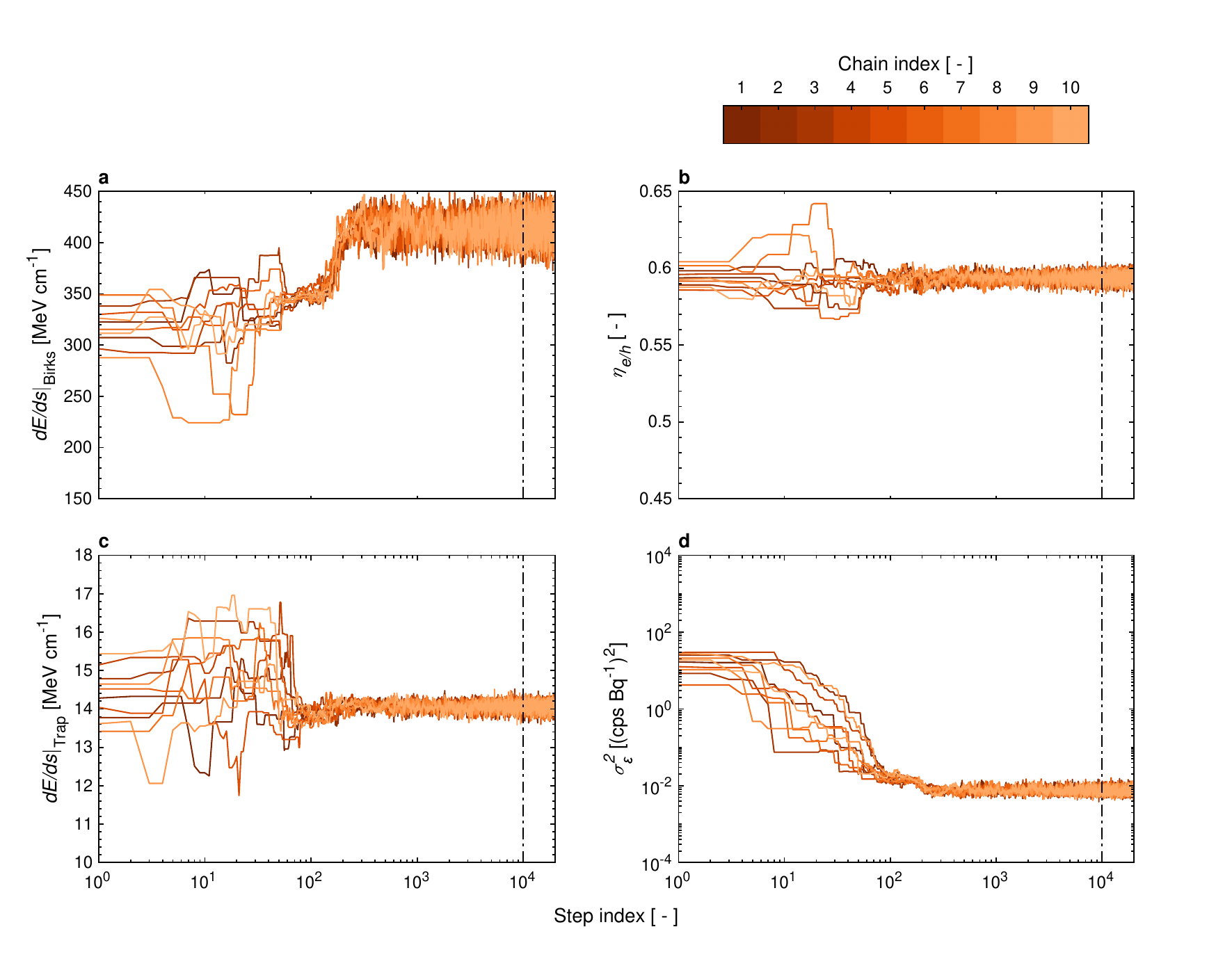}}
\caption[Markov Chain Monte Carlo trace plots for crystal 2]{\textbf{ Markov Chain Monte Carlo trace plots for crystal 2.} These graphs show the sample values of the Markov Chain Monte Carlo algorithm \citep{Goodman2010EnsembleInvariance} for each individual Markov chain and model parameter resulting from the single mode inversion pipeline applied to the scintillation crystal 2: \textbf{a}~The Birks related stopping power parameter $dE/ds\mid_{\text{Birks}}$. \textbf{b}~The free carrier fraction $\eta_{e/h}$. \textbf{c}~The trapping related stopping power parameter $dE/ds\mid_{\text{Trap}}$. \textbf{d}~The discrepancy model variance $\sigma^2_{\varepsilon}$. In addition, the burn-in threshold is highlighted as a dashed-dotted black line in each graph.}
\label{fig:trace2}
\end{figure}

\newpage

\begin{figure}[h!]%
\centerline{
\includegraphics[]{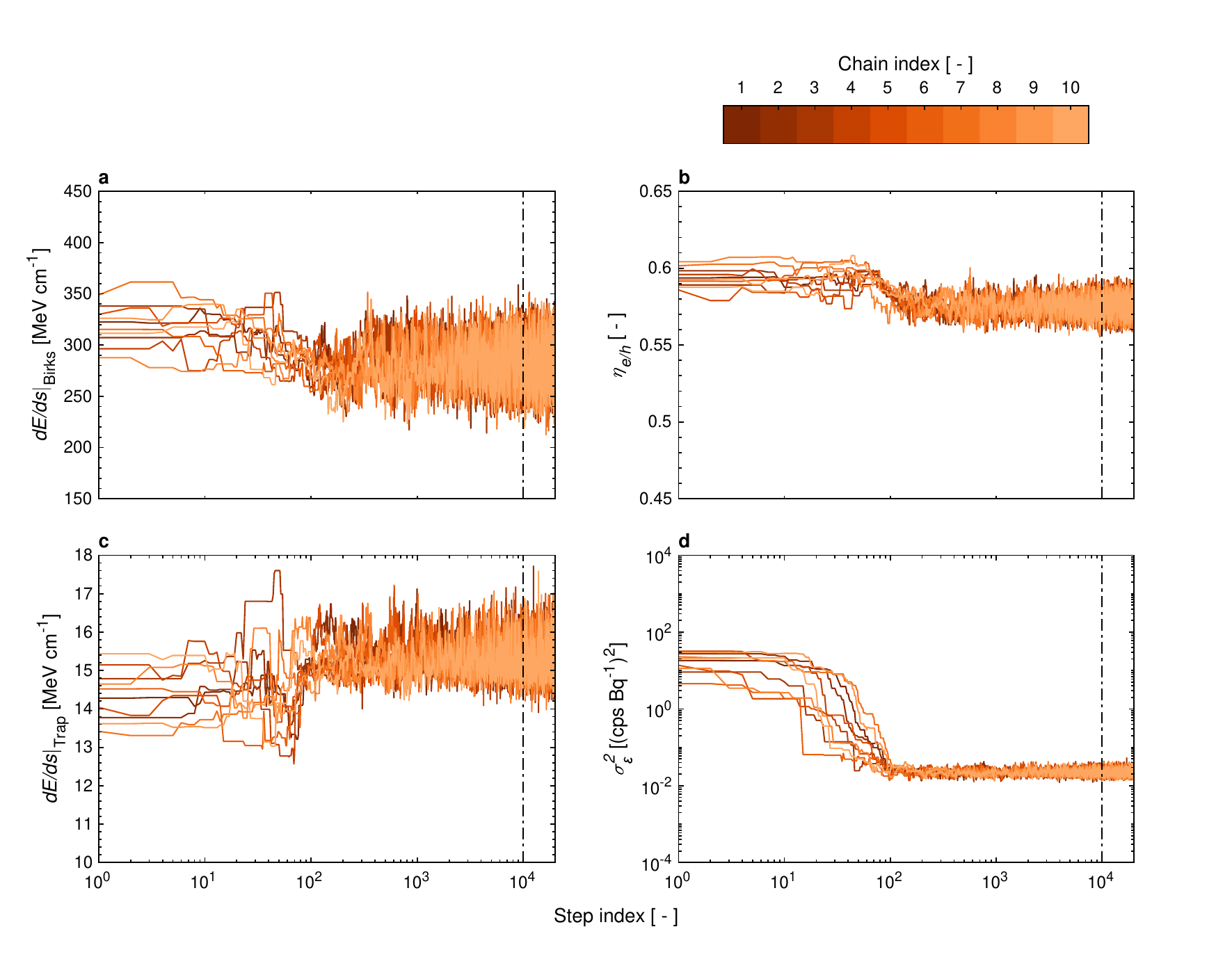}}
\caption[Markov Chain Monte Carlo trace plots for crystal 3]{\textbf{ Markov Chain Monte Carlo trace plots for crystal 3.} These graphs show the sample values of the Markov Chain Monte Carlo algorithm \citep{Goodman2010EnsembleInvariance} for each individual Markov chain and model parameter resulting from the single mode inversion pipeline applied to the scintillation crystal 3: \textbf{a}~The Birks related stopping power parameter $dE/ds\mid_{\text{Birks}}$. \textbf{b}~The free carrier fraction $\eta_{e/h}$. \textbf{c}~The trapping related stopping power parameter $dE/ds\mid_{\text{Trap}}$. \textbf{d}~The discrepancy model variance $\sigma^2_{\varepsilon}$. In addition, the burn-in threshold is highlighted as a dashed-dotted black line in each graph.}
\label{fig:trace3}
\end{figure}

\newpage

\begin{figure}[h!]%
\centerline{
\includegraphics[]{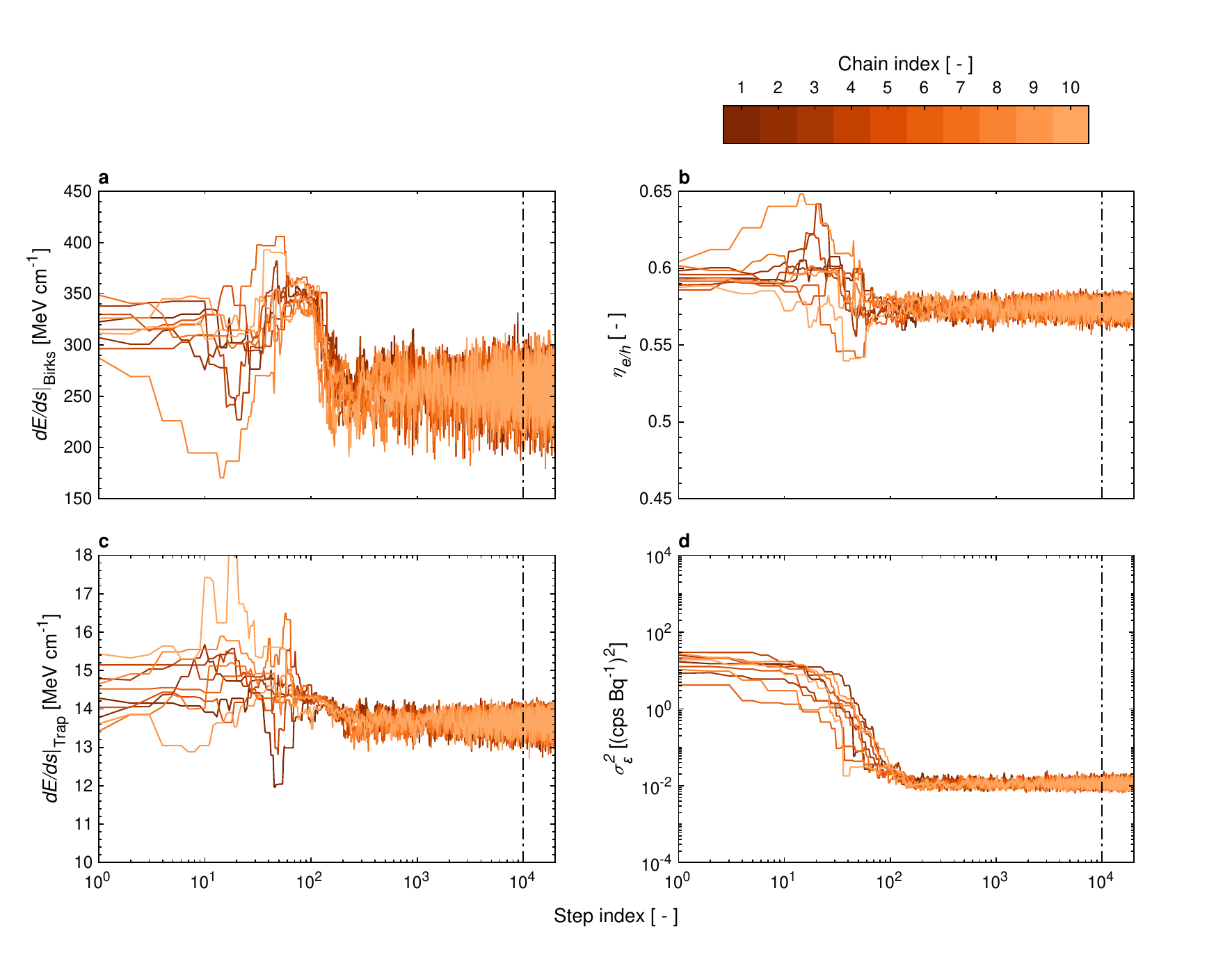}}
\caption[Markov Chain Monte Carlo trace plots for crystal 4]{\textbf{ Markov Chain Monte Carlo trace plots for crystal 4.} These graphs show the sample values of the Markov Chain Monte Carlo algorithm \citep{Goodman2010EnsembleInvariance} for each individual Markov chain and model parameter resulting from the single mode inversion pipeline applied to the scintillation crystal 4: \textbf{a}~The Birks related stopping power parameter $dE/ds\mid_{\text{Birks}}$. \textbf{b}~The free carrier fraction $\eta_{e/h}$. \textbf{c}~The trapping related stopping power parameter $dE/ds\mid_{\text{Trap}}$. \textbf{d}~The discrepancy model variance $\sigma^2_{\varepsilon}$. In addition, the burn-in threshold is highlighted as a dashed-dotted black line in each graph.}
\label{fig:trace4}
\end{figure}

\newpage

\begin{figure}[h!]%
\centerline{
\includegraphics[]{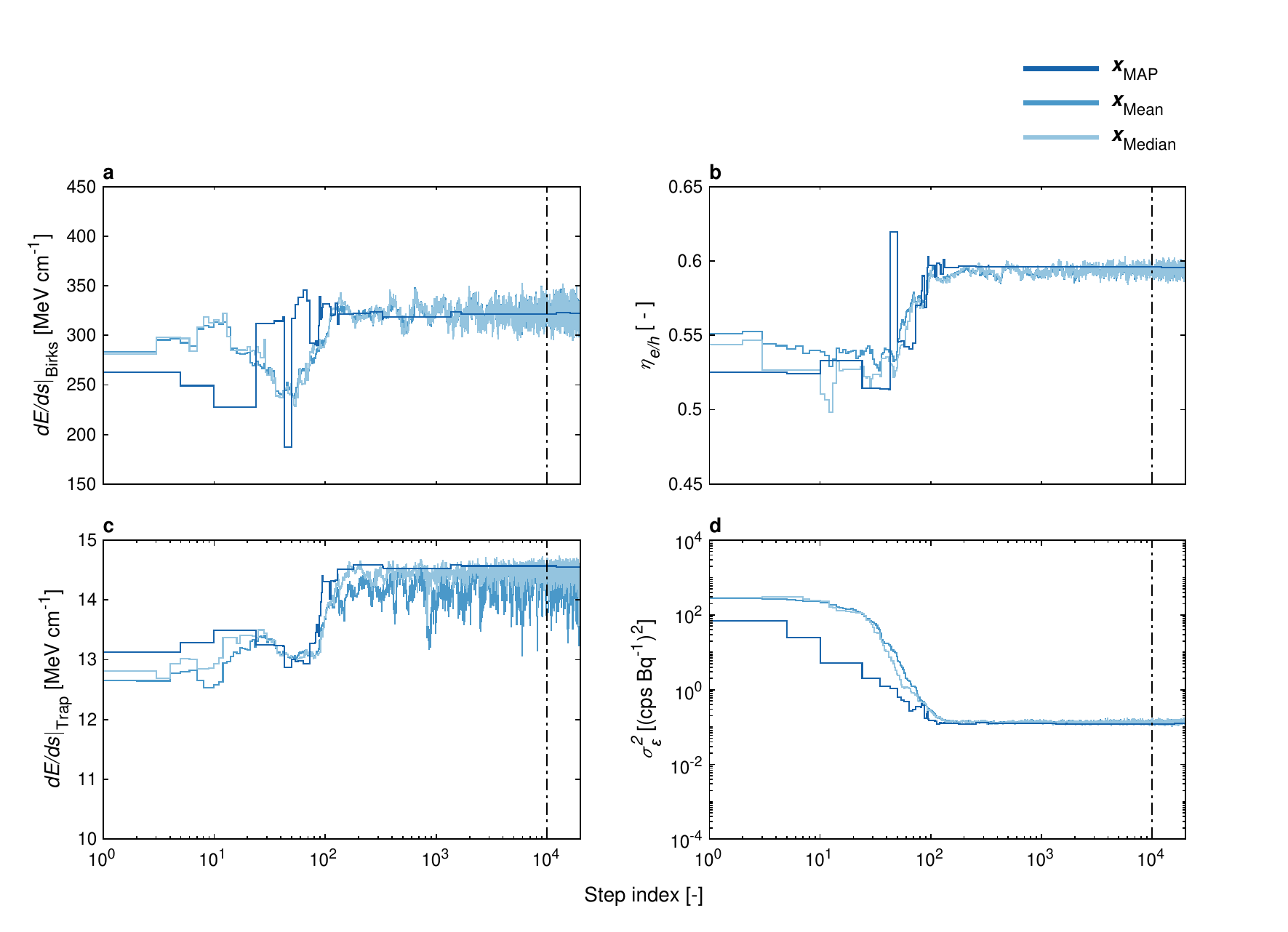}}
\caption[Posterior point estimator convergence for the sum mode]{\textbf{ Posterior point estimator convergence for the sum mode.} These graphs show the convergence of the posterior point estimators, i.e. the maximum a posteriori (MAP) probability estimate $\boldsymbol{x}_{\mathrm{MAP}}$, the posterior mean $\boldsymbol{x}_{\mathrm{Mean}}$ and the posterior median $\boldsymbol{x}_{\mathrm{Median}}$, as a function of the Markov Chain Monte Carlo steps and each individual model parameter resulting from the sum mode inversion pipeline applied to the sum channel: \textbf{a}~The Birks related stopping power parameter $dE/ds\mid_{\text{Birks}}$. \textbf{b}~The free carrier fraction $\eta_{e/h}$. \textbf{c}~The trapping related stopping power parameter $dE/ds\mid_{\text{Trap}}$. \textbf{d}~The discrepancy model variance $\sigma^2_{\varepsilon}$. In addition, the burn-in threshold is highlighted as a dashed-dotted black line in each graph.}
\label{fig:convsum}
\end{figure}

\newpage

\begin{figure}[h!]%
\centerline{
\includegraphics[]{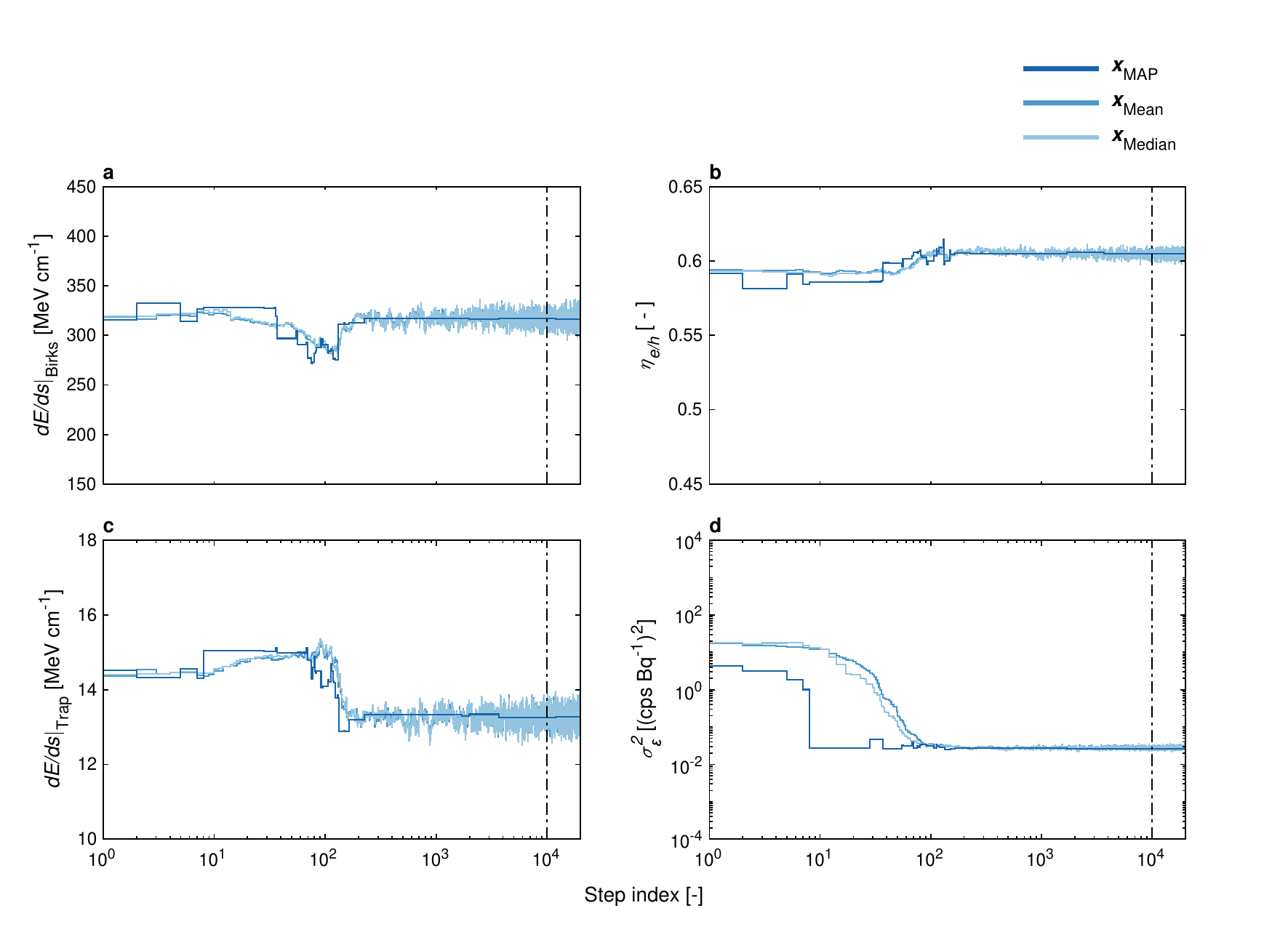}}
\caption[Posterior point estimator convergence for crystal 1]{\textbf{ Posterior point estimator convergence for crystal 1.} These graphs show the convergence of the posterior point estimators, i.e. the maximum a posteriori (MAP) probability estimate $\boldsymbol{x}_{\mathrm{MAP}}$, the posterior mean $\boldsymbol{x}_{\mathrm{Mean}}$ and the posterior median $\boldsymbol{x}_{\mathrm{Median}}$, as a function of the Markov Chain Monte Carlo steps and each individual model parameter resulting from the single mode inversion pipeline applied to the scintillation crystal 1: \textbf{a}~The Birks related stopping power parameter $dE/ds\mid_{\text{Birks}}$. \textbf{b}~The free carrier fraction $\eta_{e/h}$. \textbf{c}~The trapping related stopping power parameter $dE/ds\mid_{\text{Trap}}$. \textbf{d}~The discrepancy model variance $\sigma^2_{\varepsilon}$. In addition, the burn-in threshold is highlighted as a dashed-dotted black line in each graph.}
\label{fig:conv1}
\end{figure}

\newpage

\begin{figure}[h!]%
\centerline{
\includegraphics[]{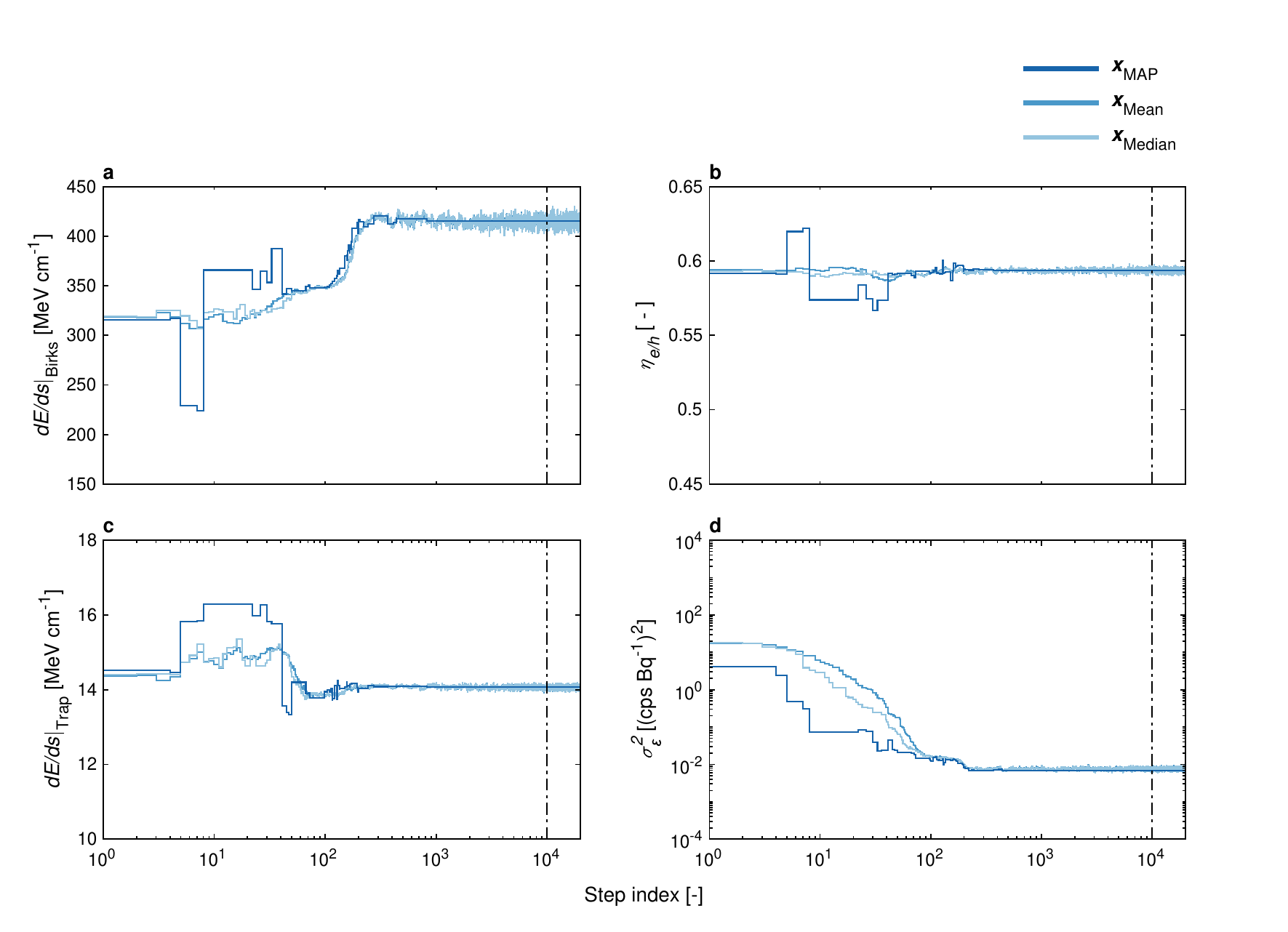}}
\caption[Posterior point estimator convergence for crystal 2]{\textbf{ Posterior point estimator convergence for crystal 2.} These graphs show the convergence of the posterior point estimators, i.e. the maximum a posteriori (MAP) probability estimate $\boldsymbol{x}_{\mathrm{MAP}}$, the posterior mean $\boldsymbol{x}_{\mathrm{Mean}}$ and the posterior median $\boldsymbol{x}_{\mathrm{Median}}$, as a function of the Markov Chain Monte Carlo steps and each individual model parameter resulting from the single mode inversion pipeline applied to the scintillation crystal 2: \textbf{a}~The Birks related stopping power parameter $dE/ds\mid_{\text{Birks}}$. \textbf{b}~The free carrier fraction $\eta_{e/h}$. \textbf{c}~The trapping related stopping power parameter $dE/ds\mid_{\text{Trap}}$. \textbf{d}~The discrepancy model variance $\sigma^2_{\varepsilon}$. In addition, the burn-in threshold is highlighted as a dashed-dotted black line in each graph.}
\label{fig:conv2}
\end{figure}

\newpage

\begin{figure}[h!]%
\centerline{
\includegraphics[]{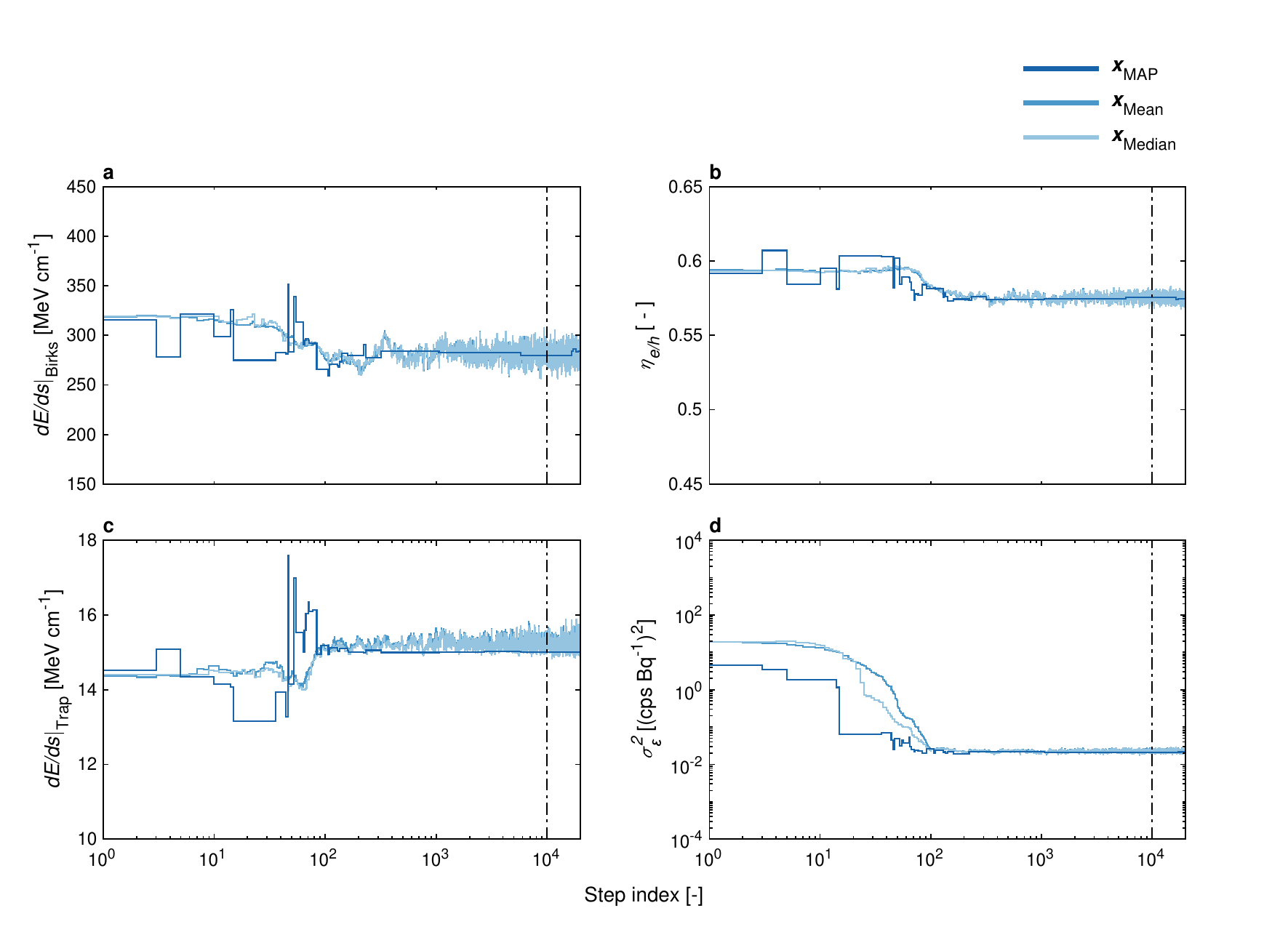}}
\caption[Posterior point estimator convergence for crystal 3]{\textbf{ Posterior point estimator convergence for crystal 3.} These graphs show the convergence of the posterior point estimators, i.e. the maximum a posteriori (MAP) probability estimate $\boldsymbol{x}_{\mathrm{MAP}}$, the posterior mean $\boldsymbol{x}_{\mathrm{Mean}}$ and the posterior median $\boldsymbol{x}_{\mathrm{Median}}$, as a function of the Markov Chain Monte Carlo steps and each individual model parameter resulting from the single mode inversion pipeline applied to the scintillation crystal 3: \textbf{a}~The Birks related stopping power parameter $dE/ds\mid_{\text{Birks}}$. \textbf{b}~The free carrier fraction $\eta_{e/h}$. \textbf{c}~The trapping related stopping power parameter $dE/ds\mid_{\text{Trap}}$. \textbf{d}~The discrepancy model variance $\sigma^2_{\varepsilon}$. In addition, the burn-in threshold is highlighted as a dashed-dotted black line in each graph.}
\label{fig:conv3}
\end{figure}

\newpage

\begin{figure}[h!]%
\centerline{
\includegraphics[]{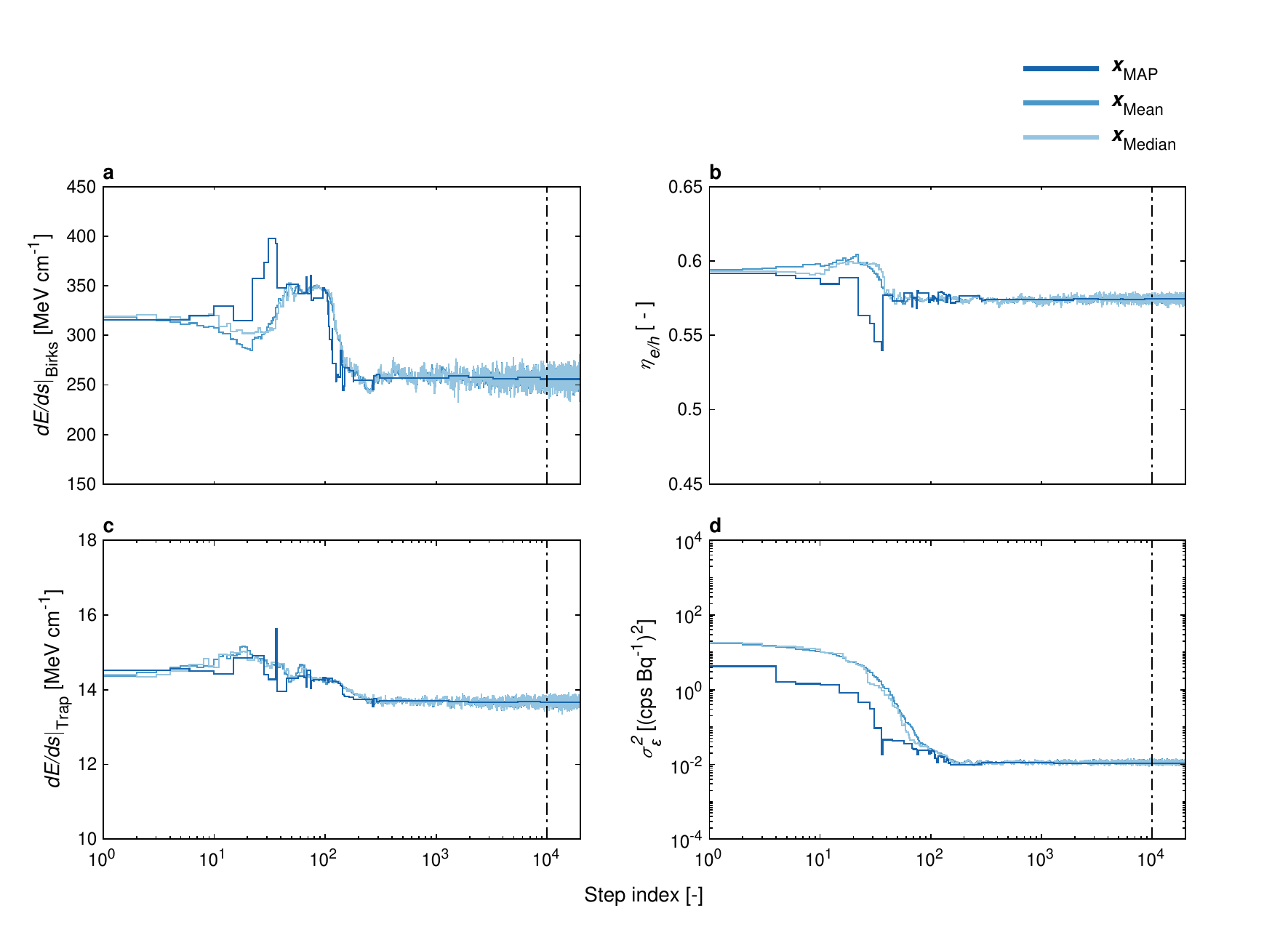}}
\caption[Posterior point estimator convergence for crystal 4]{\textbf{ Posterior point estimator convergence for crystal 4.} These graphs show the convergence of the posterior point estimators, i.e. the maximum a posteriori (MAP) probability estimate $\boldsymbol{x}_{\mathrm{MAP}}$, the posterior mean $\boldsymbol{x}_{\mathrm{Mean}}$ and the posterior median $\boldsymbol{x}_{\mathrm{Median}}$, as a function of the Markov Chain Monte Carlo steps and each individual model parameter resulting from the single mode inversion pipeline applied to the scintillation crystal 4: \textbf{a}~The Birks related stopping power parameter $dE/ds\mid_{\text{Birks}}$. \textbf{b}~The free carrier fraction $\eta_{e/h}$. \textbf{c}~The trapping related stopping power parameter $dE/ds\mid_{\text{Trap}}$. \textbf{d}~The discrepancy model variance $\sigma^2_{\varepsilon}$. In addition, the burn-in threshold is highlighted as a dashed-dotted black line in each graph.}
\label{fig:conv4}
\end{figure}

\newpage

\begin{figure}[h!]%
\centerline{
\includegraphics[]{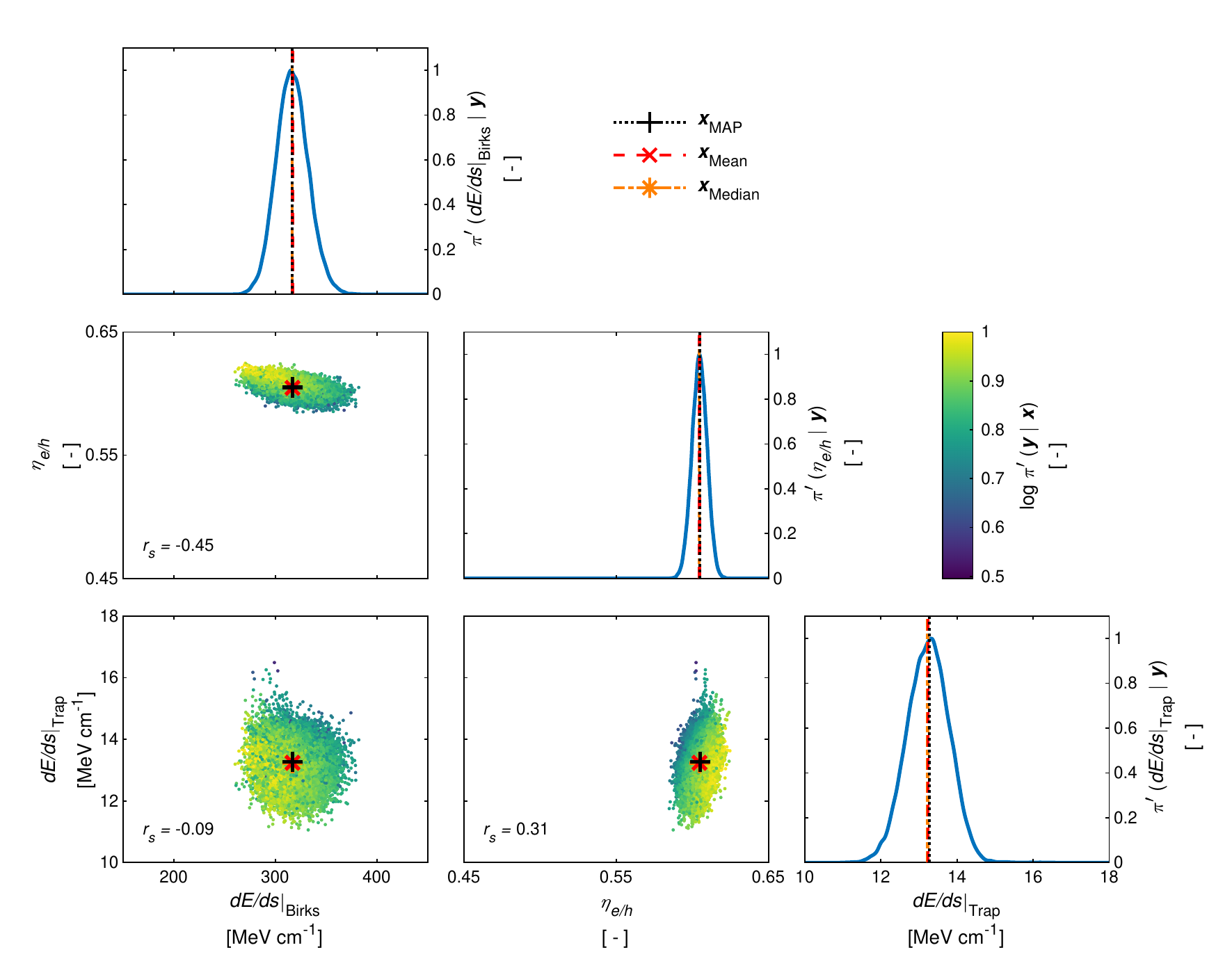}}
\caption[Posterior distribution estimate for crystal 1]{\textbf{ Posterior distribution estimate for crystal 1.} As a result of the single mode inversion pipeline applied to the scintillation crystal 1, the off-diagonal subfigures present samples from the multivariate posterior marginals given the experimental dataset $\boldsymbol{y}$ for the model parameters $\boldsymbol{x}\coloneqq\left( dE/ds\mid_{\text{Birks}},~\eta_{e/h},~dE/ds\mid_{\text{Trap}} \right)^{\intercal}$. We colored these samples by the corresponding normalized multivariate log-likelihood function values $\operatorname{log} \pi^{\prime}\left( \boldsymbol{y} \mid \boldsymbol{x} \right)$. In addition, the 
Spearman's rank correlation coefficient $r_s$ is provided for the model parameters in the corresponding off-diagonal subfigures. The subfigures on the diagonal axis highlight the normalized univariate marginal likelihood $\pi^{\prime}\left( x \mid \boldsymbol{y}\right)$ for the model parameter $x$. Both, the univariate and multivariate likelihood values, were normalized by their corresponding global maxima. Derived posterior point estimators, i.e. the maximum a posteriori (MAP) probability estimate $\boldsymbol{x}_{\text{MAP}}$, the posterior mean $\boldsymbol{x}_{\text{Mean}}$ and the posterior median $\boldsymbol{x}_{\text{Median}}$, are indicated as well in each subfigure.}\label{fig:scatter1}
\end{figure}

\newpage

\begin{figure}[h!]%
\centerline{
\includegraphics[]{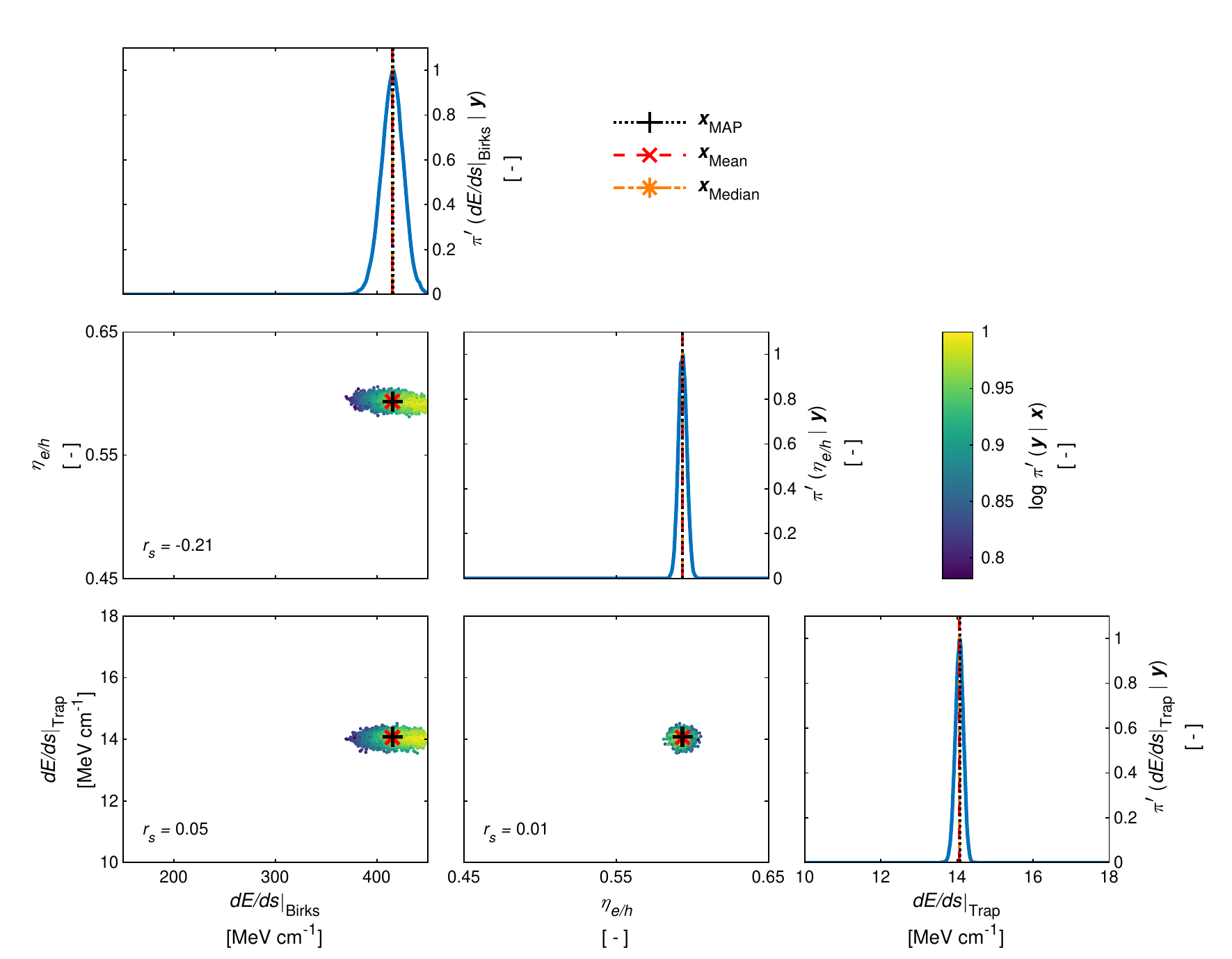}}
\caption[Posterior distribution estimate for crystal 2]{\textbf{ Posterior distribution estimate for crystal 2.} As a result of the single mode inversion pipeline applied to the scintillation crystal 2, the off-diagonal subfigures present samples from the multivariate posterior marginals given the experimental dataset $\boldsymbol{y}$ for the model parameters $\boldsymbol{x}\coloneqq\left( dE/ds\mid_{\text{Birks}},~\eta_{e/h},~dE/ds\mid_{\text{Trap}} \right)^{\intercal}$. We colored these samples by the corresponding normalized multivariate log-likelihood function values $\operatorname{log} \pi^{\prime}\left( \boldsymbol{y} \mid \boldsymbol{x} \right)$. In addition, the 
Spearman's rank correlation coefficient $r_s$ is provided for the model parameters in the corresponding off-diagonal subfigures. The subfigures on the diagonal axis highlight the normalized univariate marginal likelihood $\pi^{\prime}\left( x \mid \boldsymbol{y}\right)$ for the model parameter $x$. Both, the univariate and multivariate likelihood values, were normalized by their corresponding global maxima. Derived posterior point estimators, i.e. the maximum a posteriori (MAP) probability estimate $\boldsymbol{x}_{\text{MAP}}$, the posterior mean $\boldsymbol{x}_{\text{Mean}}$ and the posterior median $\boldsymbol{x}_{\text{Median}}$, are indicated as well in each subfigure.}\label{fig:scatter2}
\end{figure}

\newpage

\begin{figure}[h!]%
\centerline{
\includegraphics[]{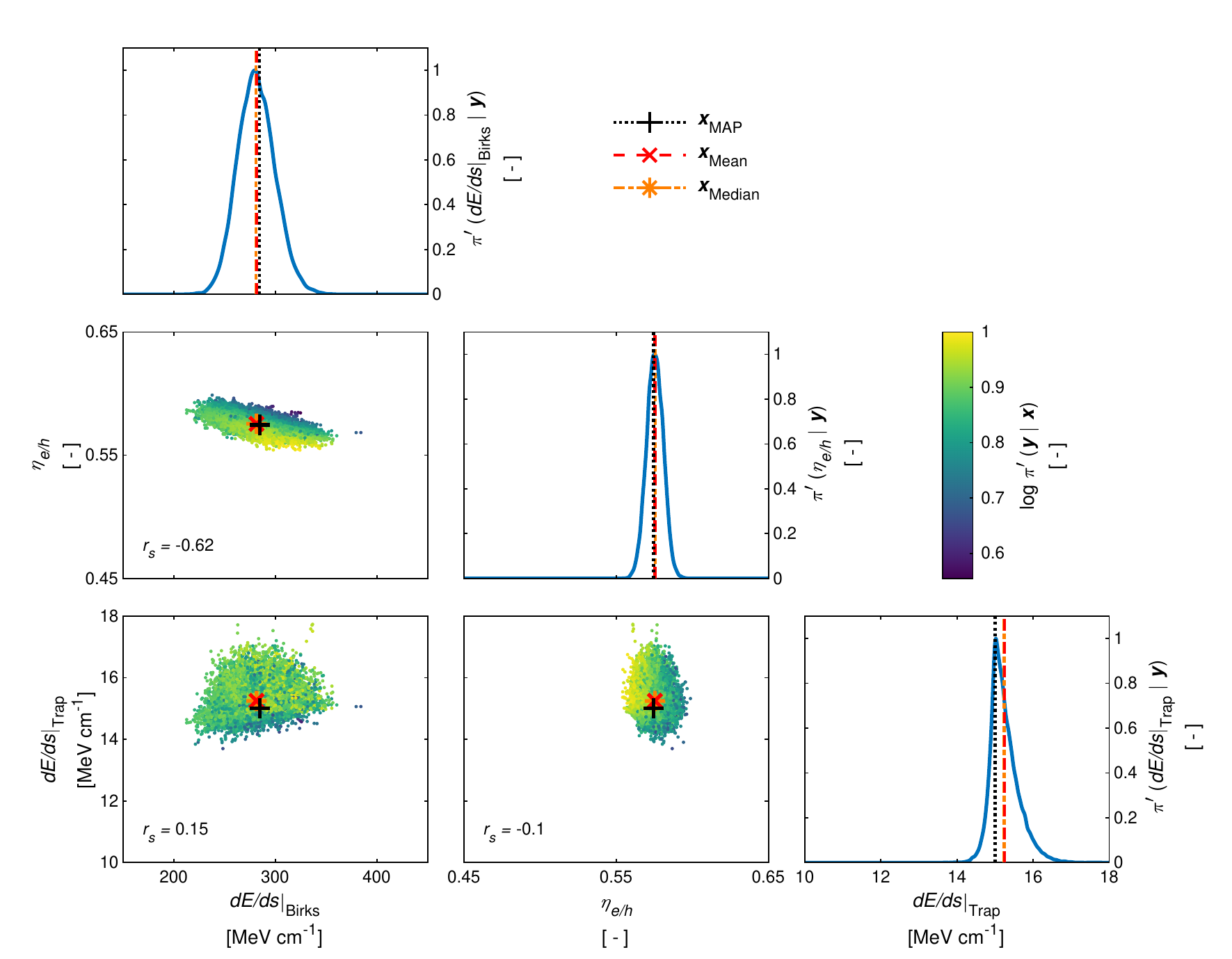}}
\caption[Posterior distribution estimate for crystal 3]{\textbf{ Posterior distribution estimate for crystal 3.} As a result of the single mode inversion pipeline applied to the scintillation crystal 3, the off-diagonal subfigures present samples from the multivariate posterior marginals given the experimental dataset $\boldsymbol{y}$ for the model parameters $\boldsymbol{x}\coloneqq\left( dE/ds\mid_{\text{Birks}},~\eta_{e/h},~dE/ds\mid_{\text{Trap}} \right)^{\intercal}$. We colored these samples by the corresponding normalized multivariate log-likelihood function values $\operatorname{log} \pi^{\prime}\left( \boldsymbol{y} \mid \boldsymbol{x} \right)$. In addition, the 
Spearman's rank correlation coefficient $r_s$ is provided for the model parameters in the corresponding off-diagonal subfigures. The subfigures on the diagonal axis highlight the normalized univariate marginal likelihood $\pi^{\prime}\left( x \mid \boldsymbol{y}\right)$ for the model parameter $x$. Both, the univariate and multivariate likelihood values, were normalized by their corresponding global maxima. Derived posterior point estimators, i.e. the maximum a posteriori (MAP) probability estimate $\boldsymbol{x}_{\text{MAP}}$, the posterior mean $\boldsymbol{x}_{\text{Mean}}$ and the posterior median $\boldsymbol{x}_{\text{Median}}$, are indicated as well in each subfigure.}\label{fig:scatter3}
\end{figure}

\newpage

\begin{figure}[h!]%
\centerline{
\includegraphics[]{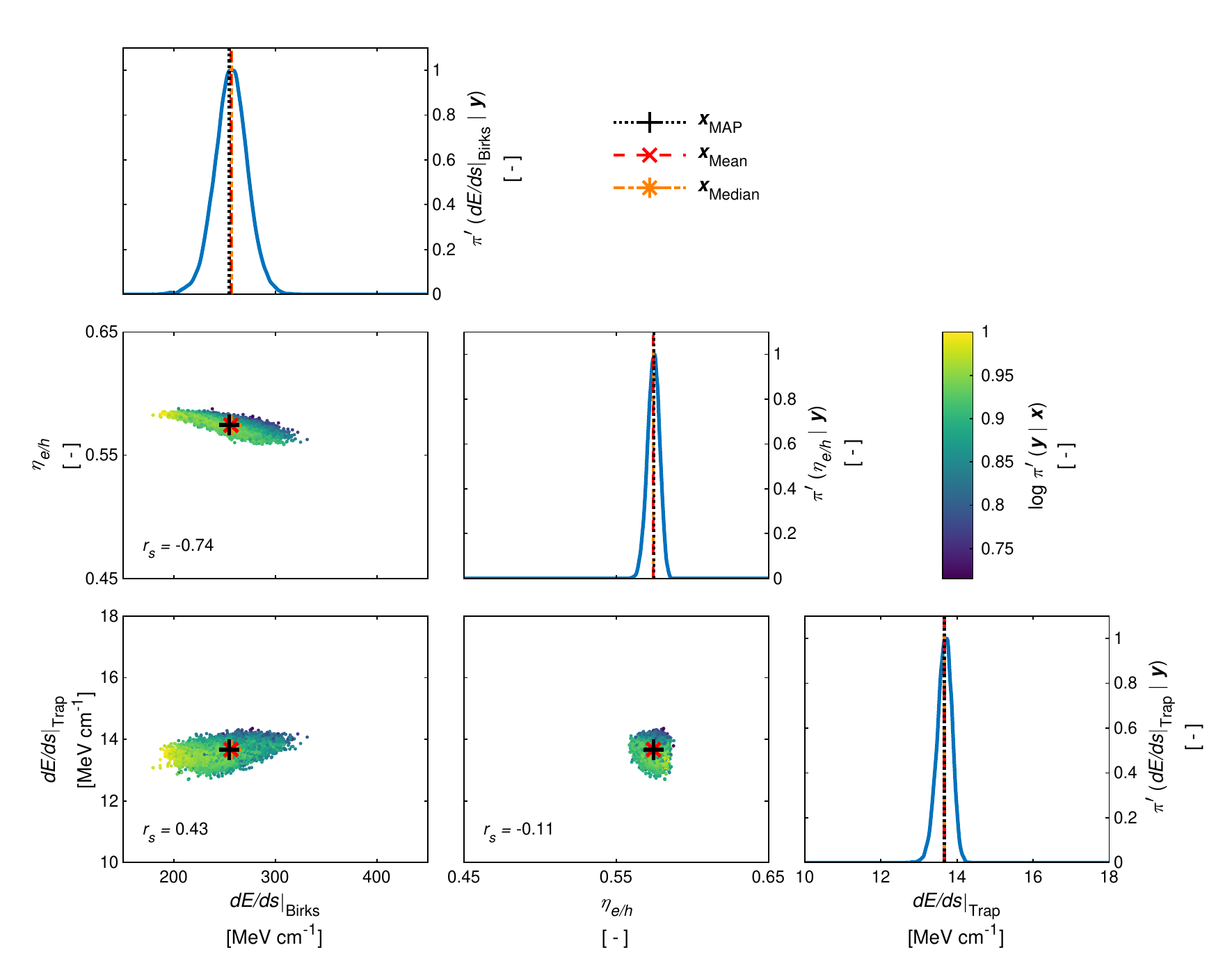}}
\caption[Posterior distribution estimate for crystal 4]{\textbf{ Posterior distribution estimate for crystal 4.} As a result of the single mode inversion pipeline applied to the scintillation crystal 4, the off-diagonal subfigures present samples from the multivariate posterior marginals given the experimental dataset $\boldsymbol{y}$ for the model parameters $\boldsymbol{x}\coloneqq\left( dE/ds\mid_{\text{Birks}},~\eta_{e/h},~dE/ds\mid_{\text{Trap}} \right)^{\intercal}$. We colored these samples by the corresponding normalized multivariate log-likelihood function values $\operatorname{log} \pi^{\prime}\left( \boldsymbol{y} \mid \boldsymbol{x} \right)$. In addition, the 
Spearman's rank correlation coefficient $r_s$ is provided for the model parameters in the corresponding off-diagonal subfigures. The subfigures on the diagonal axis highlight the normalized univariate marginal likelihood $\pi^{\prime}\left( x \mid \boldsymbol{y}\right)$ for the model parameter $x$. Both, the univariate and multivariate likelihood values, were normalized by their corresponding global maxima. Derived posterior point estimators, i.e. the maximum a posteriori (MAP) probability estimate $\boldsymbol{x}_{\text{MAP}}$, the posterior mean $\boldsymbol{x}_{\text{Mean}}$ and the posterior median $\boldsymbol{x}_{\text{Median}}$, are indicated as well in each subfigure.}\label{fig:scatter4}
\end{figure}

\newpage

\begin{figure}[h!]%
\centerline{
\includegraphics[]{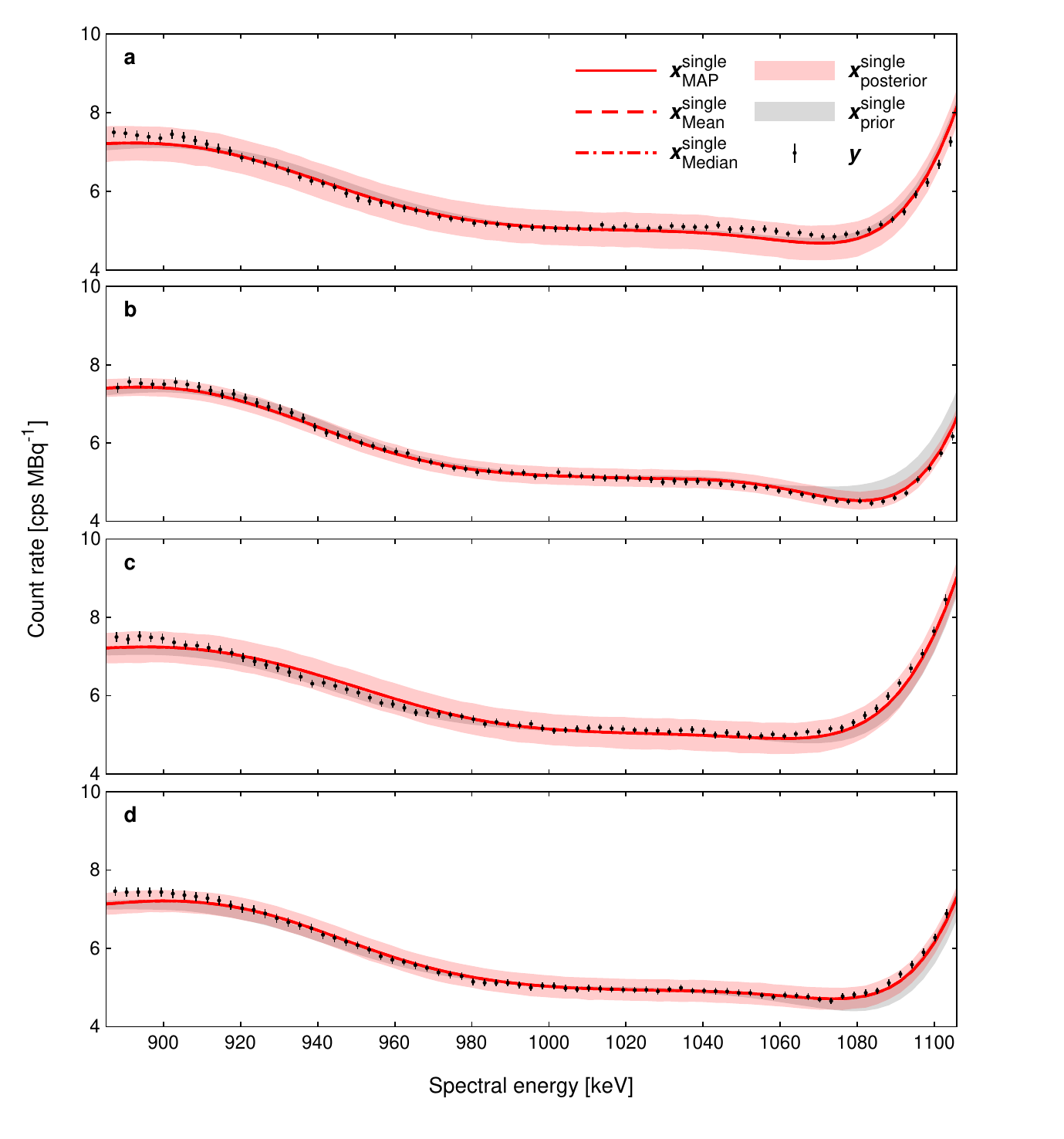}}
\caption[Compton edge predictions for the individual crystals]{\textbf{ Compton edge predictions for the individual crystals.} Here, we show the prior and posterior predictive distributions using the 99\% central credible interval for the individual scintillation crystals 1--4 (\textbf{a--d}) obtained by the single mode inversion pipeline applied to the corresponding crystals and the spectral Compton edge domain $\mathcal{D}_E\coloneqq \left\{E:E_{\text{CE}}-3\cdot\sigma_{\text{tot}}\left(E_{\text{CE}}\right)\leq E \leq E_{\text{FEP}}-2\cdot\sigma_{\text{tot}}\left(E_{\text{FEP}}\right)\right\}$ (cf. Methods in the main study). In addition, the experimental data $\boldsymbol{y}$ together with the posterior predictions using point estimators, i.e. the maximum a posteriori (MAP) probability estimate $\boldsymbol{x}_{\text{MAP}}^{\text{single}}$, the posterior mean $\boldsymbol{x}_{\text{Mean}}^{\text{single}}$ and the posterior median $\boldsymbol{x}_{\text{Median}}^{\text{single}}$, are indicated in each subfigure. Experimental uncertainties are provided as 1 standard deviation (SD) values (coverage factor $k=1$).}\label{fig:PostPred_det}
\end{figure}

\newpage

\begin{figure}[h!]%
\centerline{
\includegraphics[]{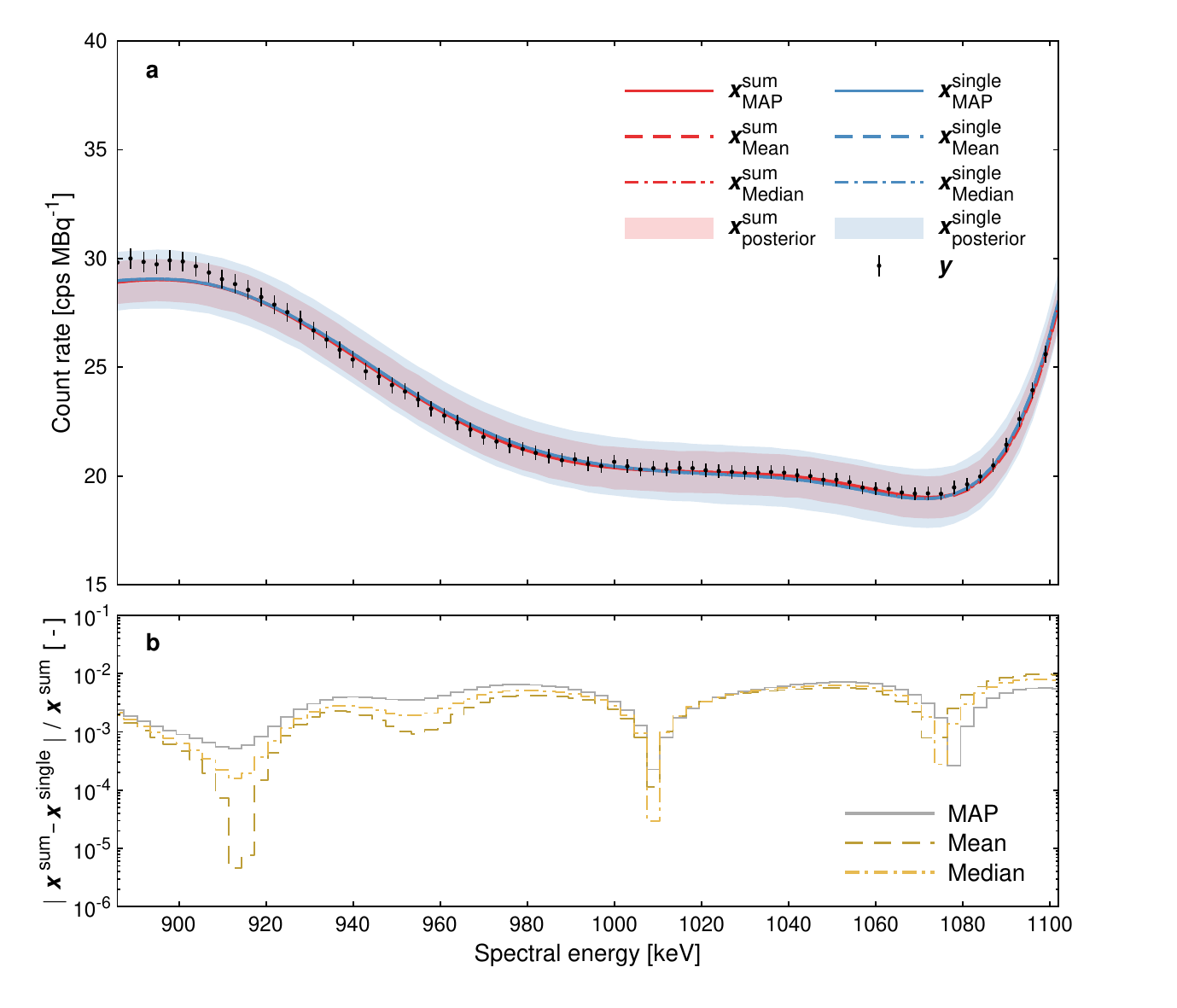}}
\caption[Spectral comparison of the sum and single mode inversion pipelines]{\textbf{ Spectral comparison of the sum and single mode inversion pipelines.} Here, we quantitatively compare the two inversion pipelines, sum and single, for the spectral Compton edge domain $\mathcal{D}_E\coloneqq \left\{E:E_{\text{CE}}-3\cdot\sigma_{\text{tot}}\left(E_{\text{CE}}\right)\leq E \leq E_{\text{FEP}}-2\cdot\sigma_{\text{tot}}\left(E_{\text{FEP}}\right)\right\}$ (cf. Methods in the main study). For that purpose, the posterior predictions for the individual scintillation crystals obtained by the single mode inversion pipeline are summed together and compared to the predictions for the sum channel obtained by the sum mode inversion pipeline. \textbf{a}~In this graph, we show the prior and posterior predictive distributions using the 99\% central credible interval for the two inversion pipelines, sum and single. In addition, the experimental data $\boldsymbol{y}$ together with the derived posterior predictions using point estimators, i.e. the maximum a posteriori (MAP) probability estimate $\boldsymbol{x}_{\text{MAP}}$, the posterior mean $\boldsymbol{x}_{\text{Mean}}$ and the posterior median $\boldsymbol{x}_{\text{Median}}$, are indicated for each pipeline. Experimental uncertainties are provided as 1 standard deviation (SD) values (coverage factor $k=1$). \textbf{b}~In this subfigure, we present the relative difference between the two inversion pipelines, sum and single, for the posterior predictions shown in subfigure~a, i.e. predictions using the three point estimators $\boldsymbol{x}_{\text{MAP}}$, $\boldsymbol{x}_{\text{Mean}}$ and $\boldsymbol{x}_{\text{Median}}$.}\label{fig:PostPred_comparison}
\end{figure}

\newpage

\begin{figure}[h!]%
\centerline{
\includegraphics[]{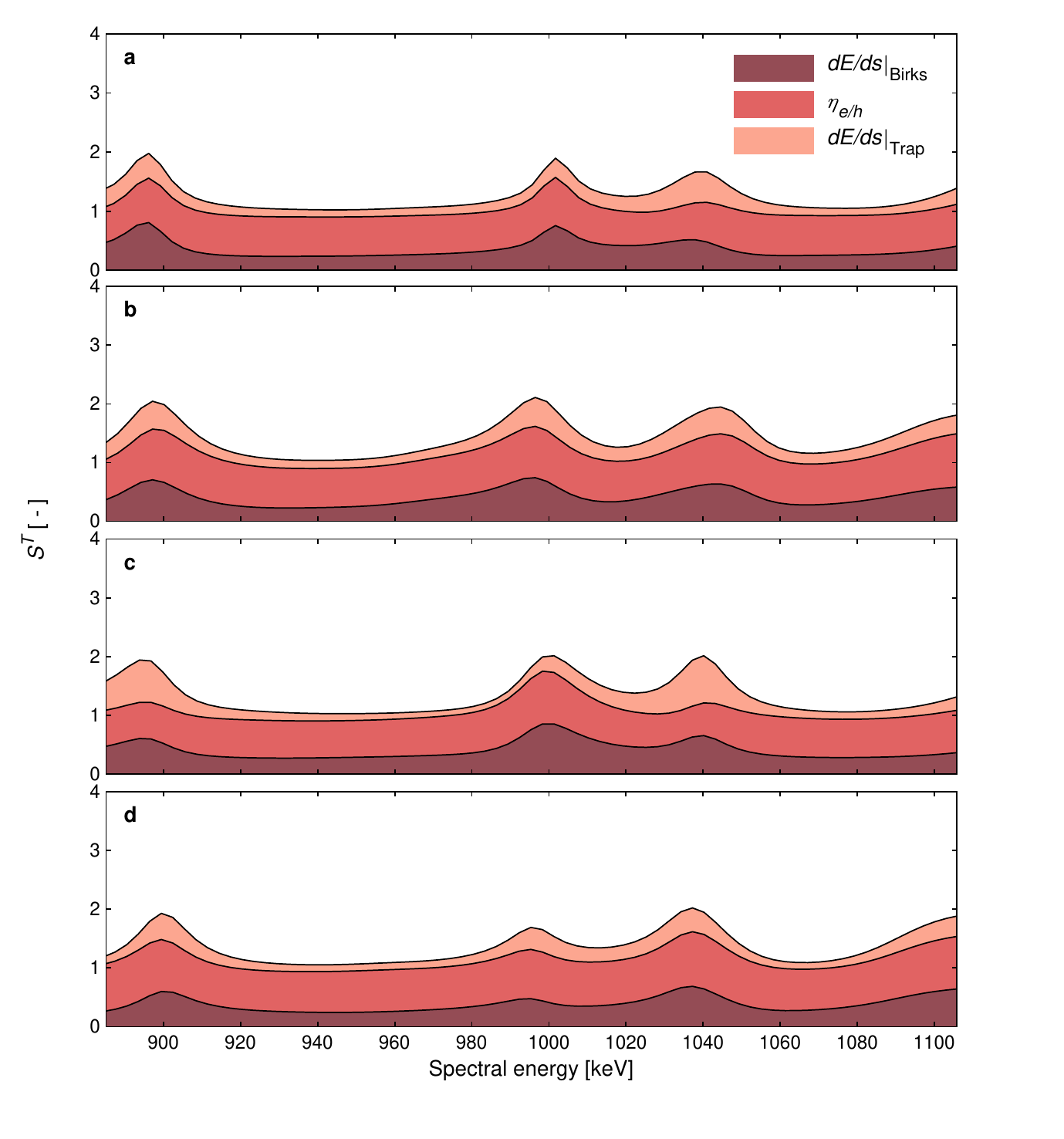}}
\caption[Hoeffding-Sobol decomposition for the individual crystals]{\textbf{ Hoeffding-Sobol decomposition for the individual crystals.} In this graph, we present the total Sobol indices $S^T$ computed by the polynomial chaos expansion emulators \citep{Sudret2008GlobalExpansions} for the individual scintillation crystals 1--4 (\textbf{a--d}) following the single mode inversion pipeline. The total Sobol indices are computed for the individual non-proportional scintillation model parameters, i.e. the Birks related stopping power parameter $dE/ds\mid_{\text{Birks}}$, the free carrier fraction $\eta_{e/h}$ as well as the trapping related stopping power parameter $dE/ds\mid_{\text{Trap}}$, on the spectral Compton edge domain $\mathcal{D}_E\coloneqq \left\{E:E_{\text{CE}}-3\cdot\sigma_{\text{tot}}\left(E_{\text{CE}}\right)\leq E \leq E_{\text{FEP}}-2\cdot\sigma_{\text{tot}}\left(E_{\text{FEP}}\right)\right\}$ (cf. Methods in the main study).}\label{fig:Sobol}
\end{figure}

\newpage

\begin{figure}[h!]%
\centerline{
\includegraphics[]{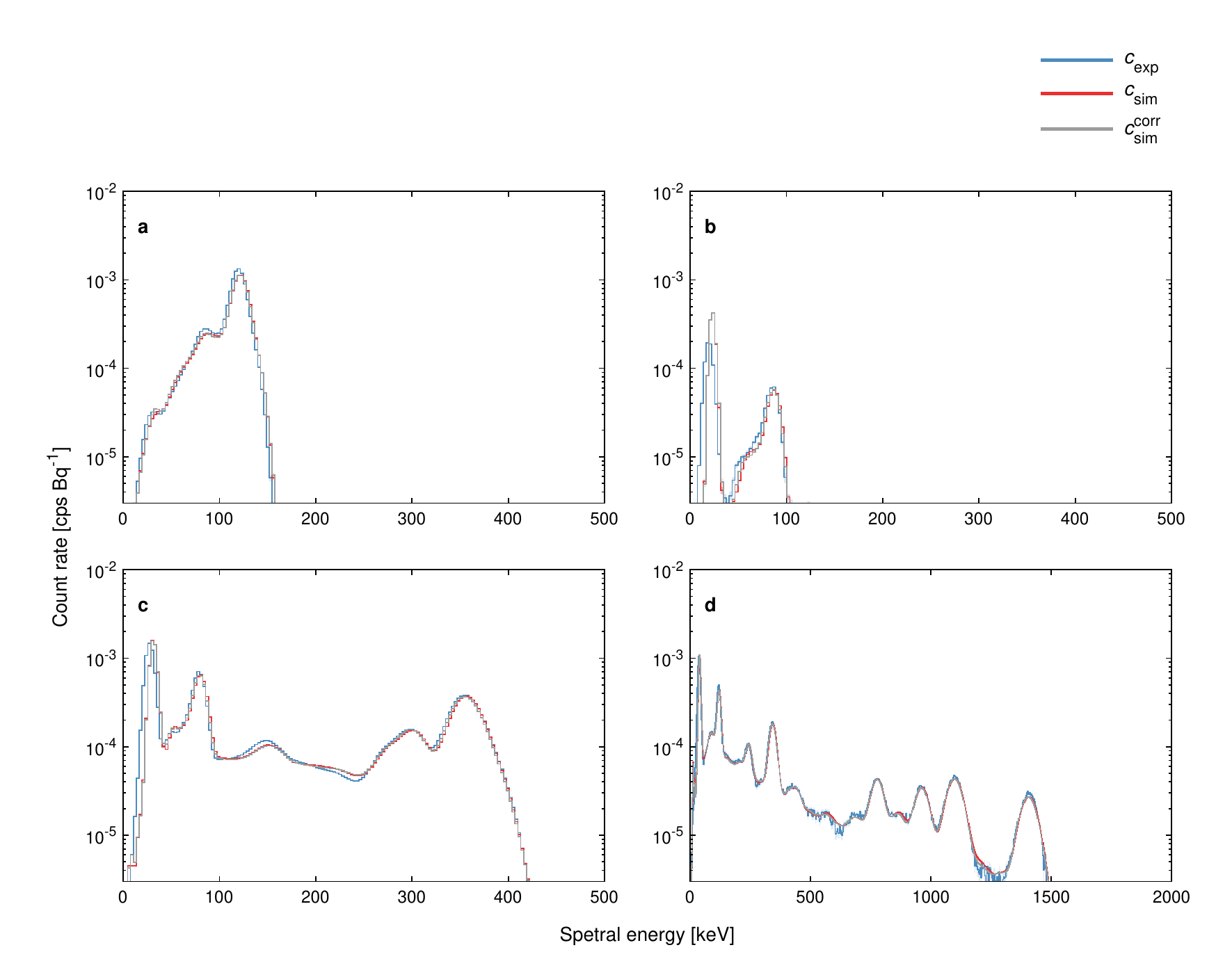}}
\caption[Spectral detector response for $^{57}\text{Co}$, $^{109}\text{Cd}$, $^{133}\text{Ba}$ and $^{152}\text{Eu}$]{\textbf{ Spectral detector response for $^{57}\text{Co}$, $^{109}\text{Cd}$, $^{133}\text{Ba}$ and $^{152}\text{Eu}$.} The measured and simulated spectral detector responses are shown for the sum channel using the four different calibrated radionuclide sources: \textbf{a}~$^{57}\text{Co}$, $^{109}\text{Cd}$ ($A=1.113(18) \times 10^5$~\unit{Bq}). \textbf{b}~$^{109}\text{Cd}$ ($A=7.38(15) \times 10^4$~\unit{Bq}). \textbf{c}~$^{133}\text{Ba}$ ($A=2.152(32) \times 10^5$~\unit{Bq}). \textbf{d}~$^{152}\text{Eu}$ ( $A=1.973(30) \times 10^4$~\unit{Bq}). The measured net count rate $c_{\text{exp}}$ as well as the simulated net count rate adopting a proportional scintillation model $c_{\text{sim}}$ were presented already elsewhere \citep{Breitenmoser2022ExperimentalSpectrometry}. We obtained the simulated net count rate $c_{\text{sim}}^{\text{corr}}$ the same way as $c_{\text{sim}}$ but accounted for the non-proportional scintillation effects by the sum mode inversion pipeline presented in this study. For the calibration, we used the $^{60}\text{Co}$ dataset \citep{Breitenmoser2022ExperimentalSpectrometry}. For all graphs presented in this figure, uncertainties are provided as 1 standard deviation (SD) shaded areas (coverage factor $k=1$). These uncertainties are only visible for $c_{\text{exp}}$.}
\label{fig:spec}
\end{figure}

\newpage

\begin{figure}[h!]%
\centerline{
\includegraphics[]{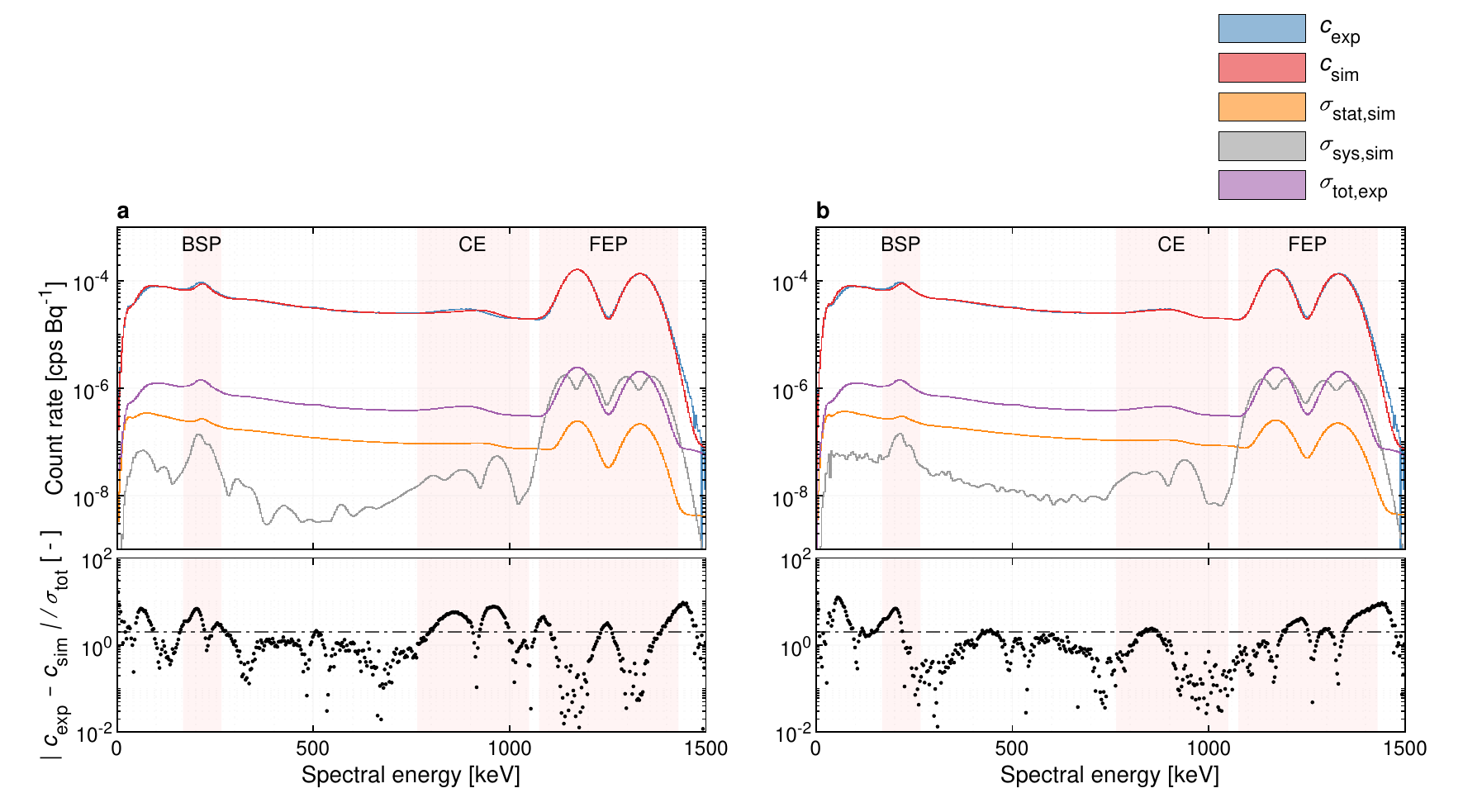}}
\caption[Uncertainty quantification for the $^{60}\text{Co}$ spectral detector response]{\textbf{ Uncertainty quantification for the $^{60}\text{Co}$ spectral detector response.} The measured and simulated mean net count rates $c_{\text{exp}}$ and $c_{\text{sim}}$ are shown for the sum channel using a $^{60}\text{Co}$ calibrated radionuclide source ($A=3.08(5) \times 10^5$~\unit{Bq}) together with the corresponding uncertainty estimates, i.e. the combined statistical and systematic measured uncertainty $\sigma_{\text{tot,exp}}$, the simulated statistical uncertainty $\sigma_{\text{stat,\text{sim}}}$ as well as the simulated systematic uncertainty $\sigma_{\text{sys,\text{sim}}}$, using 1~standard~deviation values. The measurement results were presented already elsewhere \citep{Breitenmoser2022ExperimentalSpectrometry}. Two different scintillation models have been used for the simulations: \textbf{a}~Proportional scintillation model published in \citep{Breitenmoser2022ExperimentalSpectrometry}. \textbf{b}~Bayesian calibrated non-proportional scintillation model obtained by the sum mode inversion pipeline presented in this study. Distinct spectral regions, i.e. the backscatter peak (BSP), the Compton edge (CE) as well as the full energy peaks (FEP) are highlighted for both graphs. Note that the highlighted CE region refers to the lower Compton edge at 963.419(3)~\unit{keV} associated with the photon emission line at 1173.228(3)~\unit{keV}. The normalized residual level $\mid c_{\text{exp}}-c_{\text{sim}} \mid / \sigma_{\text{tot}}$ with $\sigma_{\text{tot}} \coloneqq \sqrt{\sigma_{\text{tot,exp}}^{2}+\sigma_{\text{tot,sim}}^{2}}$ for a coverage~factor~of~2 is marked with the horizontal dash-dotted black line in the lower subfigures. More information on the numerical computation of the uncertainty estimates can be found in \hyperref[subsec:Uncertainty]{Section~\ref{subsec:Uncertainty}} and \citep{Breitenmoser2022ExperimentalSpectrometry}.}\label{fig:Co60Unc}
\end{figure}

\newpage

\begin{figure}[h!]%
\centerline{
\includegraphics[]{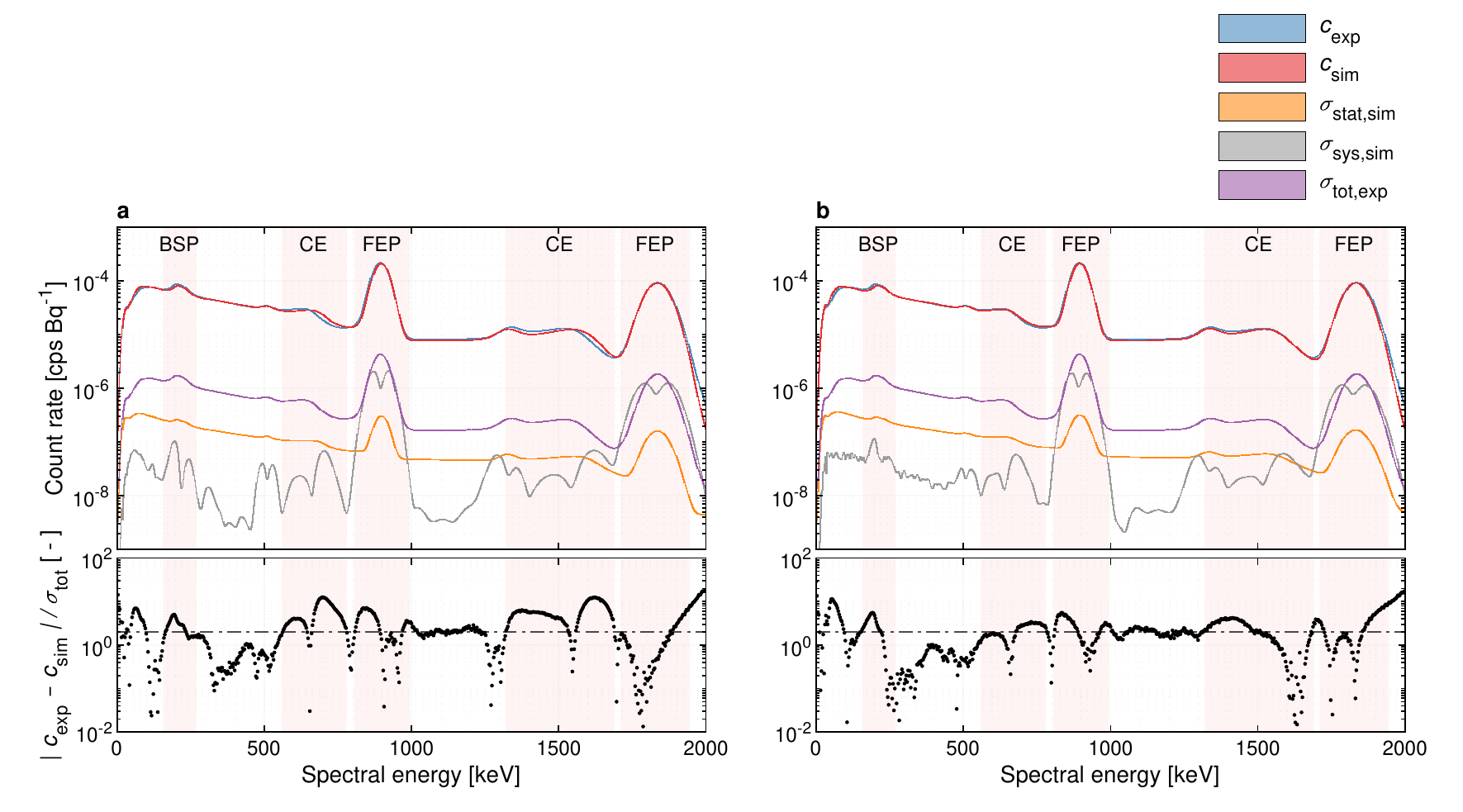}}
\caption[Uncertainty quantification for the $^{88}\text{Y}$ spectral detector response]{\textbf{ Uncertainty quantification for the $^{88}\text{Y}$ spectral detector response.} The measured and simulated mean net count rates $c_{\text{exp}}$ and $c_{\text{sim}}$ are shown for the sum channel using a $^{88}\text{Y}$ calibrated radionuclide source ($A=6.83(14) \times 10^5$~\unit{Bq}) together with the corresponding uncertainty estimates, i.e. the combined statistical and systematic measured uncertainty $\sigma_{\text{tot,exp}}$, the simulated statistical uncertainty $\sigma_{\text{stat,sim}}$ as well as the simulated systematic uncertainty $\sigma_{\text{sys,sim}}$, using 1~standard~deviation values. The measurement results were presented already elsewhere \citep{Breitenmoser2022ExperimentalSpectrometry}. Two different scintillation models have been used for the simulations: \textbf{a}~Proportional scintillation model published in \citep{Breitenmoser2022ExperimentalSpectrometry}. \textbf{b}~Bayesian calibrated non-proportional scintillation model obtained by the sum mode inversion pipeline presented in this study. Distinct spectral regions, i.e. the backscatter peak (BSP), the Compton edges (CE) as well as the full energy peaks (FEP) are highlighted for both graphs. The normalized residual level $\mid c_{\text{exp}}-c_{\text{sim}} \mid / \sigma_{\text{tot}}$ with $\sigma_{\text{tot}} \coloneqq \sqrt{\sigma_{\text{tot,exp}}^{2}+\sigma_{\text{tot,sim}}^{2}}$ for a coverage~factor~of~2 is marked with the horizontal dash-dotted black line in the lower subfigures. More information on the numerical computation of the uncertainty estimates can be found in \hyperref[subsec:Uncertainty]{Section~\ref{subsec:Uncertainty}} and \citep{Breitenmoser2022ExperimentalSpectrometry}.}\label{fig:Y88Unc}
\end{figure}

\newpage

\begin{figure}[h!]%
\centerline{
\includegraphics[]{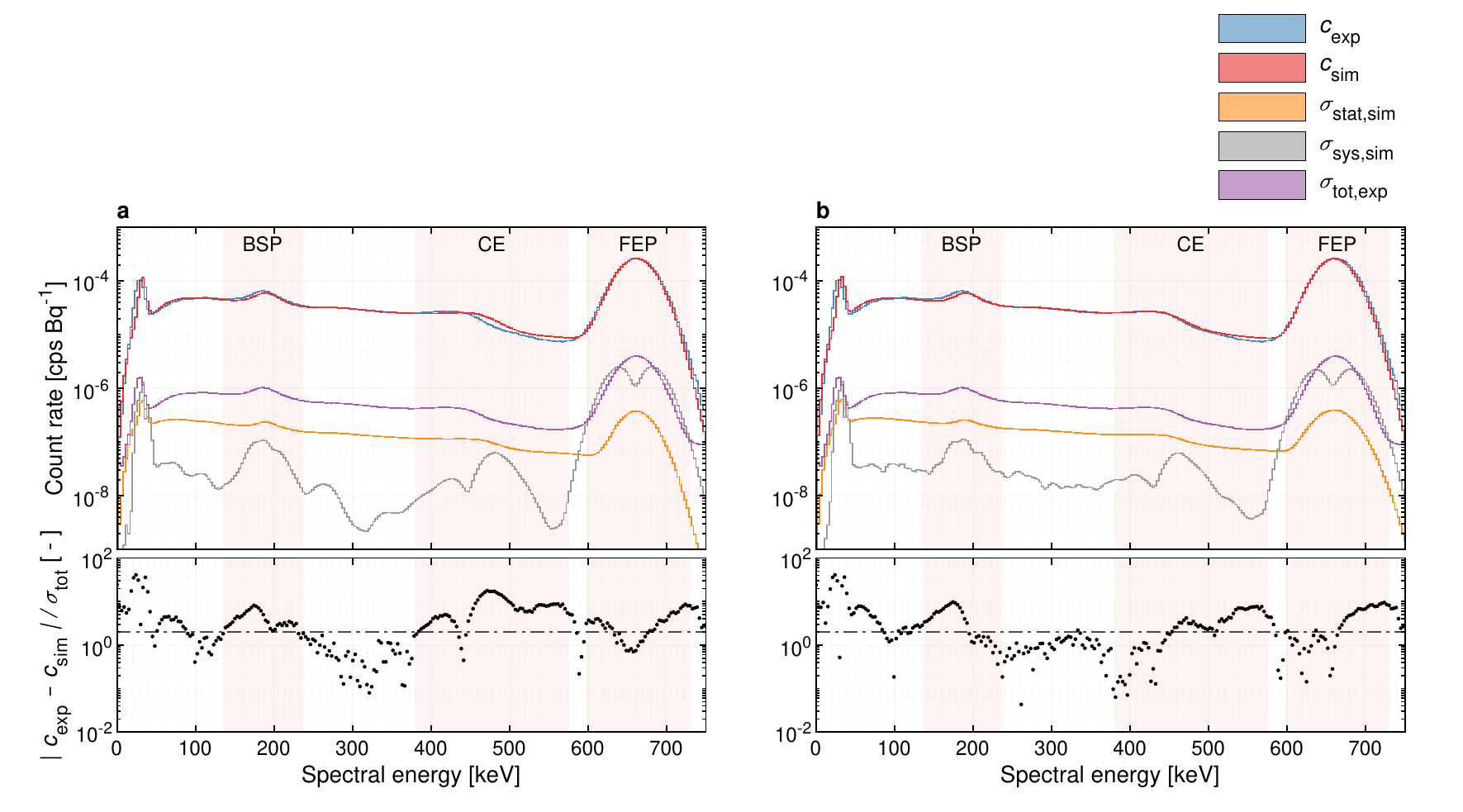}}
\caption[Uncertainty quantification for the $^{137}\text{Cs}$ spectral detector response]{\textbf{ Uncertainty quantification for the $^{137}\text{Cs}$ spectral detector response.} The measured and simulated mean net count rates $c_{\text{exp}}$ and $c_{\text{sim}}$ are shown for the sum channel using a $^{137}\text{Cs}$ calibrated radionuclide source ($A=2.266(34) \times 10^5$~\unit{Bq}) together with the corresponding uncertainty estimates, i.e. the combined statistical and systematic measured uncertainty $\sigma_{\text{tot,exp}}$, the simulated statistical uncertainty $\sigma_{\text{stat,sim}}$ as well as the simulated systematic uncertainty $\sigma_{\text{sys,sim}}$, using 1~standard~deviation values. The measurement results were presented already elsewhere \citep{Breitenmoser2022ExperimentalSpectrometry}. Two different scintillation models have been used for the simulations: \textbf{a}~Proportional scintillation model published in \citep{Breitenmoser2022ExperimentalSpectrometry}. \textbf{b}~Bayesian calibrated non-proportional scintillation model obtained by the sum mode inversion pipeline presented in this study. Distinct spectral regions, i.e. the backscatter peak (BSP), the Compton edge (CE) as well as the full energy peak (FEP) are highlighted for both graphs. The normalized residual level $\mid c_{\text{exp}}-c_{\text{sim}} \mid / \sigma_{\text{tot}}$ with $\sigma_{\text{tot}} \coloneqq \sqrt{\sigma_{\text{tot,exp}}^{2}+\sigma_{\text{tot,sim}}^{2}}$ for a coverage~factor~of~2 is marked with the horizontal dash-dotted black line in the lower subfigures. More information on the numerical computation of the uncertainty estimates can be found in \hyperref[subsec:Uncertainty]{Section~\ref{subsec:Uncertainty}} and \citep{Breitenmoser2022ExperimentalSpectrometry}.}\label{fig:Cs137Unc}
\end{figure}

\newpage

\begin{figure}[h!]%
\centerline{
\includegraphics[]{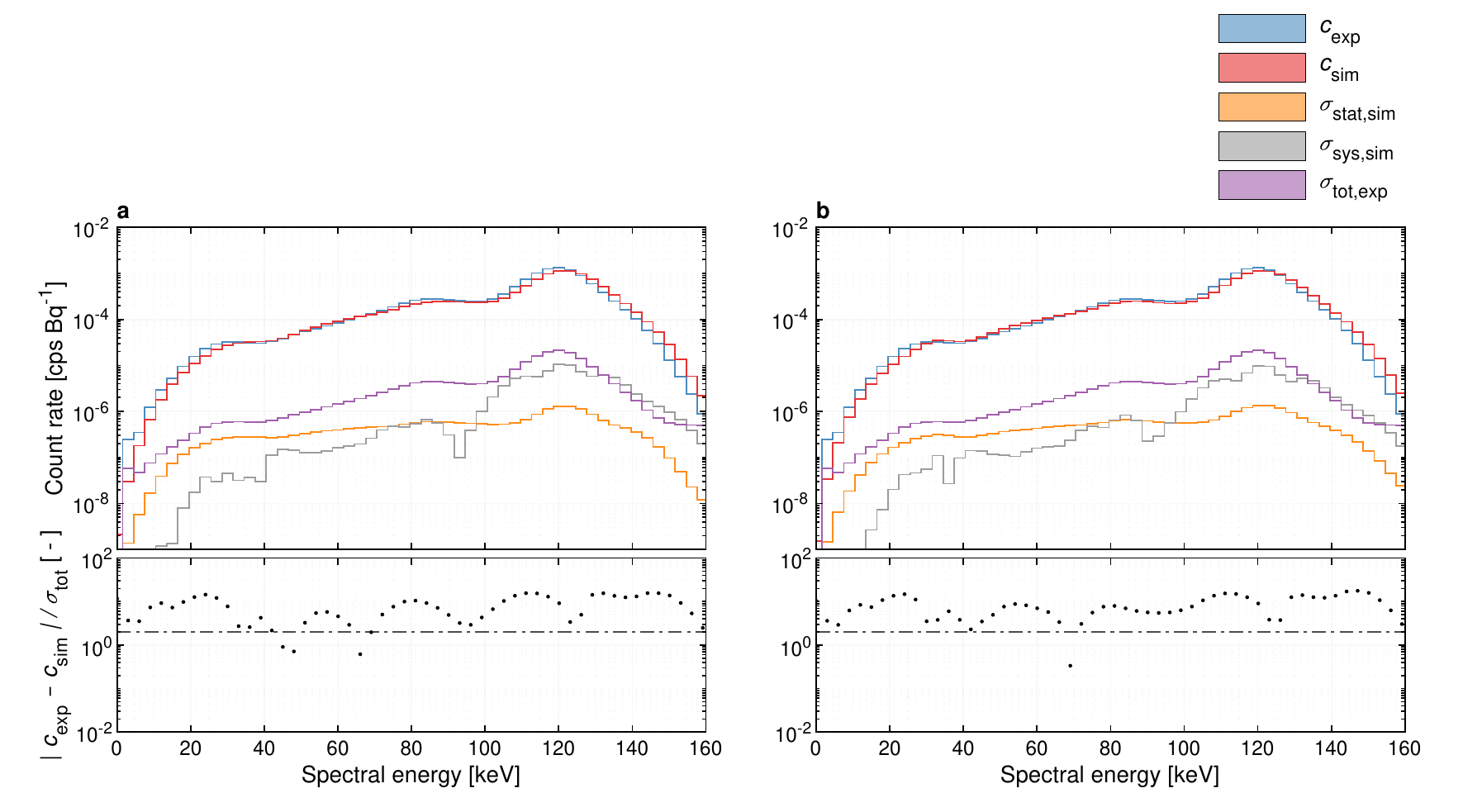}}
\caption[Uncertainty quantification for the $^{57}\text{Co}$ spectral detector response]{\textbf{ Uncertainty quantification for the $^{57}\text{Co}$ spectral detector response.} The measured and simulated mean net count rates $c_{\text{exp}}$ and $c_{\text{sim}}$ are shown for the sum channel using a $^{57}\text{Co}$ calibrated radionuclide source ($A=1.113(18) \times 10^5$~\unit{Bq}) together with the corresponding uncertainty estimates, i.e. the combined statistical and systematic measured uncertainty $\sigma_{\text{tot,exp}}$, the simulated statistical uncertainty $\sigma_{\text{stat,sim}}$ as well as the simulated systematic uncertainty $\sigma_{\text{sys,sim}}$, using 1~standard~deviation values. The measurement results were presented already elsewhere \citep{Breitenmoser2022ExperimentalSpectrometry}. Two different scintillation models have been used for the simulations: \textbf{a}~Proportional scintillation model published in \citep{Breitenmoser2022ExperimentalSpectrometry}. \textbf{b}~Bayesian calibrated non-proportional scintillation model obtained by the sum mode inversion pipeline presented in this study. The normalized residual level $\mid c_{\text{exp}}-c_{\text{sim}} \mid / \sigma_{\text{tot}}$ with $\sigma_{\text{tot}} \coloneqq \sqrt{\sigma_{\text{tot,exp}}^{2}+\sigma_{\text{tot,sim}}^{2}}$ for a coverage~factor~of~2 is marked with the horizontal dash-dotted black line in the lower subfigures. More information on the numerical computation of the uncertainty estimates can be found in \hyperref[subsec:Uncertainty]{Section~\ref{subsec:Uncertainty}} and \citep{Breitenmoser2022ExperimentalSpectrometry}.}\label{fig:Co57Unc}
\end{figure}

\newpage

\begin{figure}[h!]%
\centerline{
\includegraphics[]{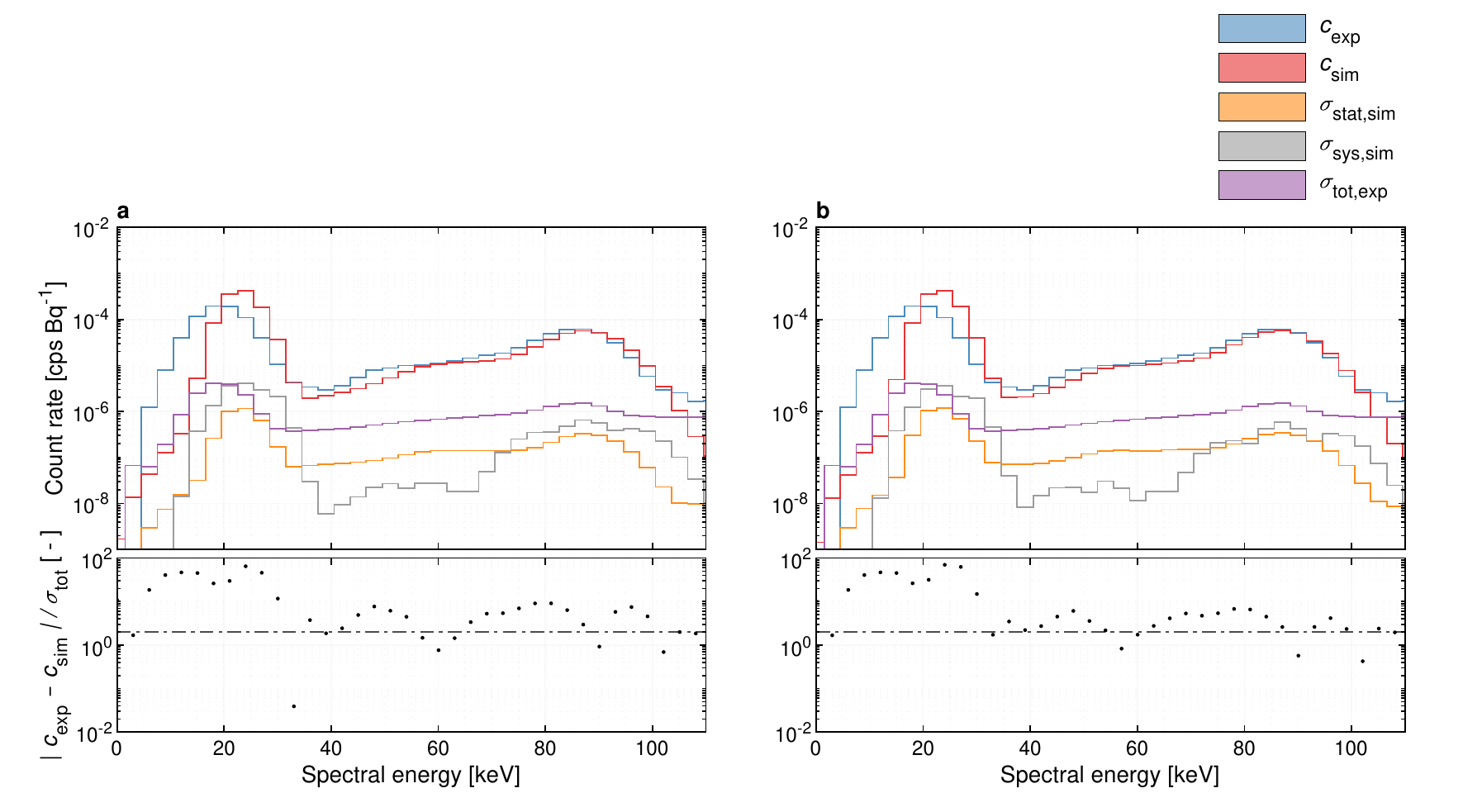}}
\caption[Uncertainty quantification for the $^{109}\text{Cd}$ spectral detector response]{\textbf{ Uncertainty quantification for the $^{109}\text{Cd}$ spectral detector response.} The measured and simulated mean net count rates $c_{\text{exp}}$ and $c_{\text{sim}}$ are shown for the sum channel using a $^{109}\text{Cd}$ calibrated radionuclide source ($A=7.38(15) \times 10^4$~\unit{Bq}) together with the corresponding uncertainty estimates, i.e. the combined statistical and systematic measured uncertainty $\sigma_{\text{tot,exp}}$, the simulated statistical uncertainty $\sigma_{\text{stat,sim}}$ as well as the simulated systematic uncertainty $\sigma_{\text{sys,sim}}$, using 1~standard~deviation values. The measurement results were presented already elsewhere \citep{Breitenmoser2022ExperimentalSpectrometry}. Two different scintillation models have been used for the simulations: \textbf{a}~Proportional scintillation model published in \citep{Breitenmoser2022ExperimentalSpectrometry}. \textbf{b}~Bayesian calibrated non-proportional scintillation model obtained by the sum mode inversion pipeline presented in this study. The normalized residual level $\mid c_{\text{exp}}-c_{\text{sim}} \mid / \sigma_{\text{tot}}$ with $\sigma_{\text{tot}} \coloneqq \sqrt{\sigma_{\text{tot,exp}}^{2}+\sigma_{\text{tot,sim}}^{2}}$ for a coverage~factor~of~2 is marked with the horizontal dash-dotted black line in the lower subfigures. More information on the numerical computation of the uncertainty estimates can be found in \hyperref[subsec:Uncertainty]{Section~\ref{subsec:Uncertainty}} and \citep{Breitenmoser2022ExperimentalSpectrometry}.}\label{fig:Cd109Unc}
\end{figure}

\newpage

\begin{figure}[h!]%
\centerline{
\includegraphics[]{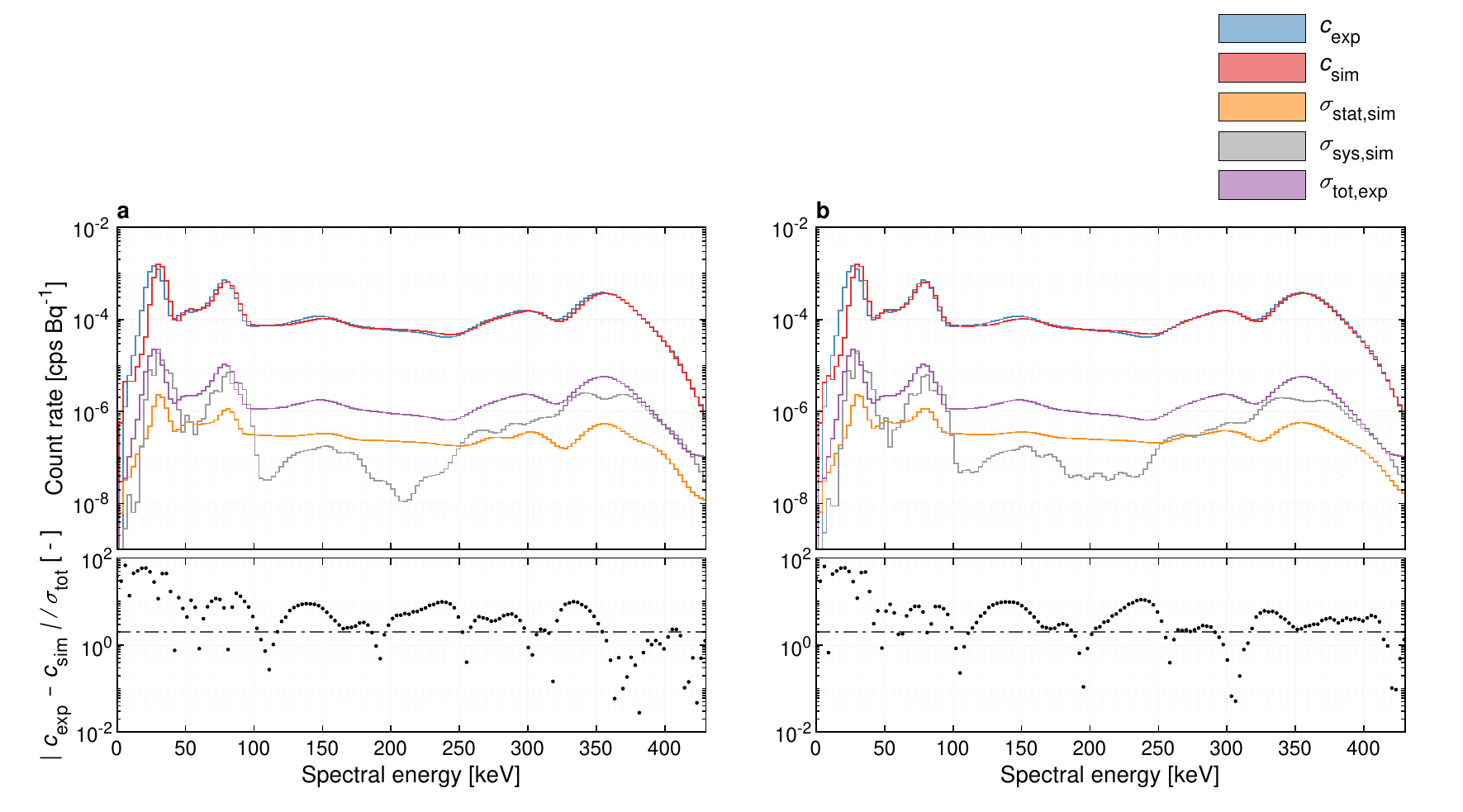}}
\caption[Uncertainty quantification for the $^{133}\text{Ba}$ spectral detector response]{\textbf{ Uncertainty quantification for the $^{133}\text{Ba}$ spectral detector response.} The measured and simulated mean net count rates $c_{\text{exp}}$ and $c_{\text{sim}}$ are shown for the sum channel using a $^{133}\text{Ba}$ calibrated radionuclide source ($A=2.152(32) \times 10^5$~\unit{Bq}) together with the corresponding uncertainty estimates, i.e. the combined statistical and systematic measured uncertainty $\sigma_{\text{tot,exp}}$, the simulated statistical uncertainty $\sigma_{\text{stat,sim}}$ as well as the simulated systematic uncertainty $\sigma_{\text{sys,sim}}$, using 1~standard~deviation values. The measurement results were presented already elsewhere \citep{Breitenmoser2022ExperimentalSpectrometry}. Two different scintillation models have been used for the simulations: \textbf{a}~Proportional scintillation model published in \citep{Breitenmoser2022ExperimentalSpectrometry}. \textbf{b}~Bayesian calibrated non-proportional scintillation model obtained by the sum mode inversion pipeline presented in this study. The normalized residual level $\mid c_{\text{exp}}-c_{\text{sim}} \mid / \sigma_{\text{tot}}$ with $\sigma_{\text{tot}} \coloneqq \sqrt{\sigma_{\text{tot,exp}}^{2}+\sigma_{\text{tot,sim}}^{2}}$ for a coverage~factor~of~2 is marked with the horizontal dash-dotted black line in the lower subfigures. More information on the numerical computation of the uncertainty estimates can be found in \hyperref[subsec:Uncertainty]{Section~\ref{subsec:Uncertainty}} and \citep{Breitenmoser2022ExperimentalSpectrometry}.}\label{fig:Ba133Unc}
\end{figure}

\newpage

\begin{figure}[h!]%
\centerline{
\includegraphics[]{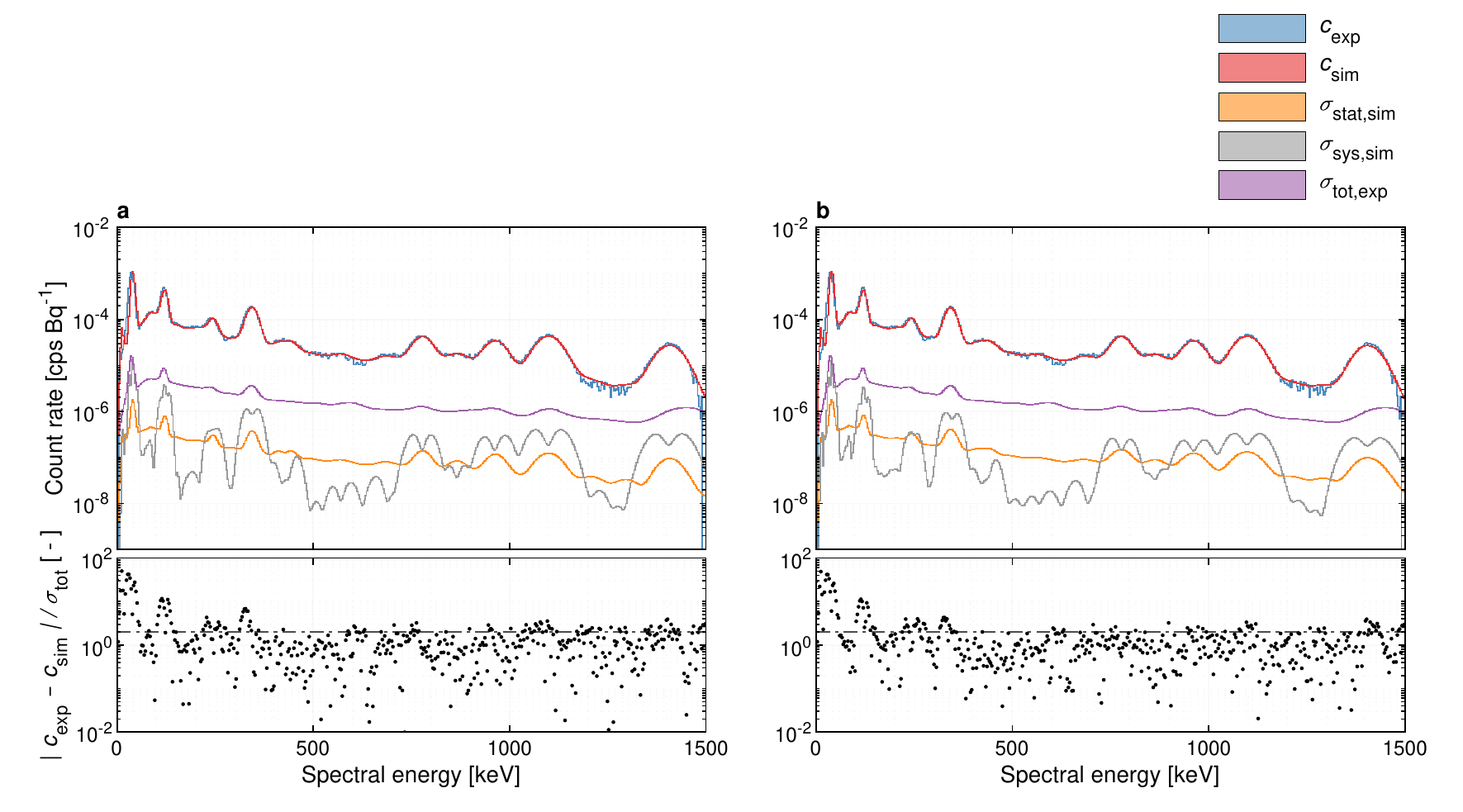}}
\caption[Uncertainty quantification for the $^{152}\text{Eu}$ spectral detector response]{\textbf{ Uncertainty quantification for the $^{152}\text{Eu}$ spectral detector response.} The measured and simulated mean net count rates $c_{\text{exp}}$ and $c_{\text{sim}}$ are shown for the sum channel using a $^{152}\text{Eu}$ calibrated radionuclide source ($A=1.973(30) \times 10^4$~\unit{Bq}) together with the corresponding uncertainty estimates, i.e. the combined statistical and systematic measured uncertainty $\sigma_{\text{tot,exp}}$, the simulated statistical uncertainty $\sigma_{\text{stat,sim}}$ as well as the simulated systematic uncertainty $\sigma_{\text{sys,sim}}$, using 1~standard~deviation values. The measurement results were presented already elsewhere \citep{Breitenmoser2022ExperimentalSpectrometry}. Two different scintillation models have been used for the simulations: \textbf{a}~Proportional scintillation model published in \citep{Breitenmoser2022ExperimentalSpectrometry}. \textbf{b}~Bayesian calibrated non-proportional scintillation model obtained by the sum mode inversion pipeline presented in this study. The normalized residual level $\mid c_{\text{exp}}-c_{\text{sim}} \mid / \sigma_{\text{tot}}$ with $\sigma_{\text{tot}} \coloneqq \sqrt{\sigma_{\text{tot,exp}}^{2}+\sigma_{\text{tot,sim}}^{2}}$ for a coverage~factor~of~2 is marked with the horizontal dash-dotted black line in the lower subfigures. More information on the numerical computation of the uncertainty estimates can be found in \hyperref[subsec:Uncertainty]{Section~\ref{subsec:Uncertainty}} and \citep{Breitenmoser2022ExperimentalSpectrometry}.}\label{fig:Eu152Unc}
\end{figure}

\newpage

\begin{figure}[h!]%
\centerline{
\includegraphics[]{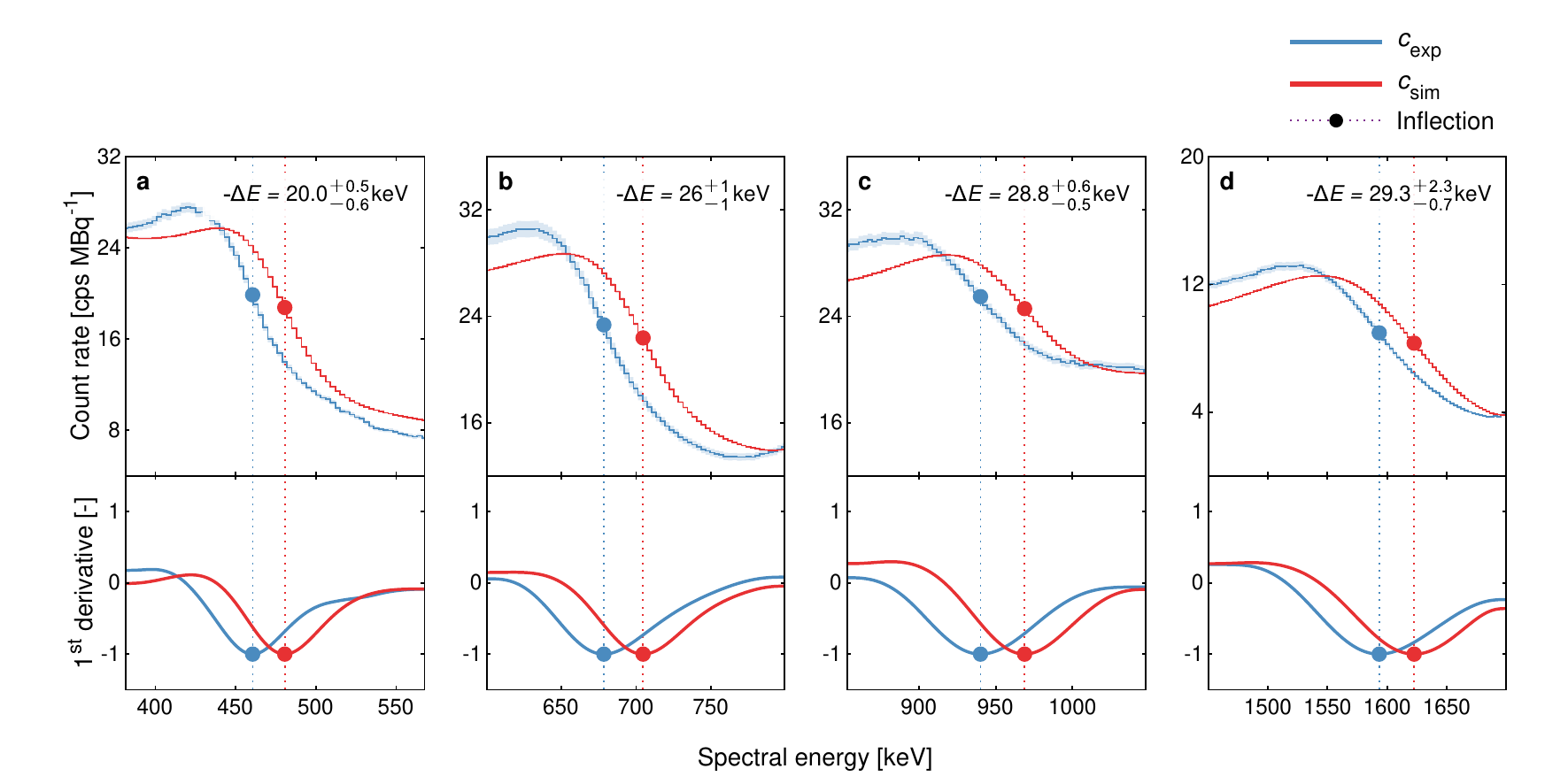}}
\caption[Compton edge shift analysis for the sum channel]{\textbf{ Compton edge shift analysis for the sum channel.} Here, we present the results from our Compton edge shift analysis for the sum channel. We characterize this negative spectral shift $-\Delta E$ for four different Compton edges: \textbf{a}~477.334(3)~\unit{keV} associated with the $^{137}\text{Cs}$ emission line at 661.657(3)~\unit{keV}. \textbf{b}~699.133(3)~\unit{keV} associated with the $^{88}\text{Y}$ emission line at 898.042(3)~\unit{keV}. \textbf{c}~963.419(3)~\unit{keV} associated with the $^{60}\text{Co}$ emission line at 1173.228(3)~\unit{keV}. \textbf{d}~1611.77(1)~\unit{keV} associated with the $^{88}\text{Y}$ emission line at 1836.063(3)~\unit{keV}. First, we determine the inflection points at the individual Compton edges for the measured net spectra $c_{\text{exp}}$ and the simulated net spectra $c_{\text{sim}}$ (proportional scintillation model) by computing the $1^{\text{st}}$ derivative of the corresponding spectra using spline regression \citep{Reinsch1967SmoothingFunctions} (bottom panels). Note that for visualization purposes, we have normalized the first derivatives of the net count rate spectra shown in the bottom panels by their corresponding global minima. In a second step, we compute the Compton edge shift as the spectral difference between the determined inflection points for $c_{\text{exp}}$ and $c_{\text{sim}}$, i.e. $-\Delta E \coloneqq c_{\text{sim}}^{\text{infl}}- c_{\text{exp}}^{\text{infl}}$ (top panels).}\label{fig:ECshiftsum}
\end{figure}

\newpage

\begin{figure}[h!]%
\centerline{
\includegraphics[]{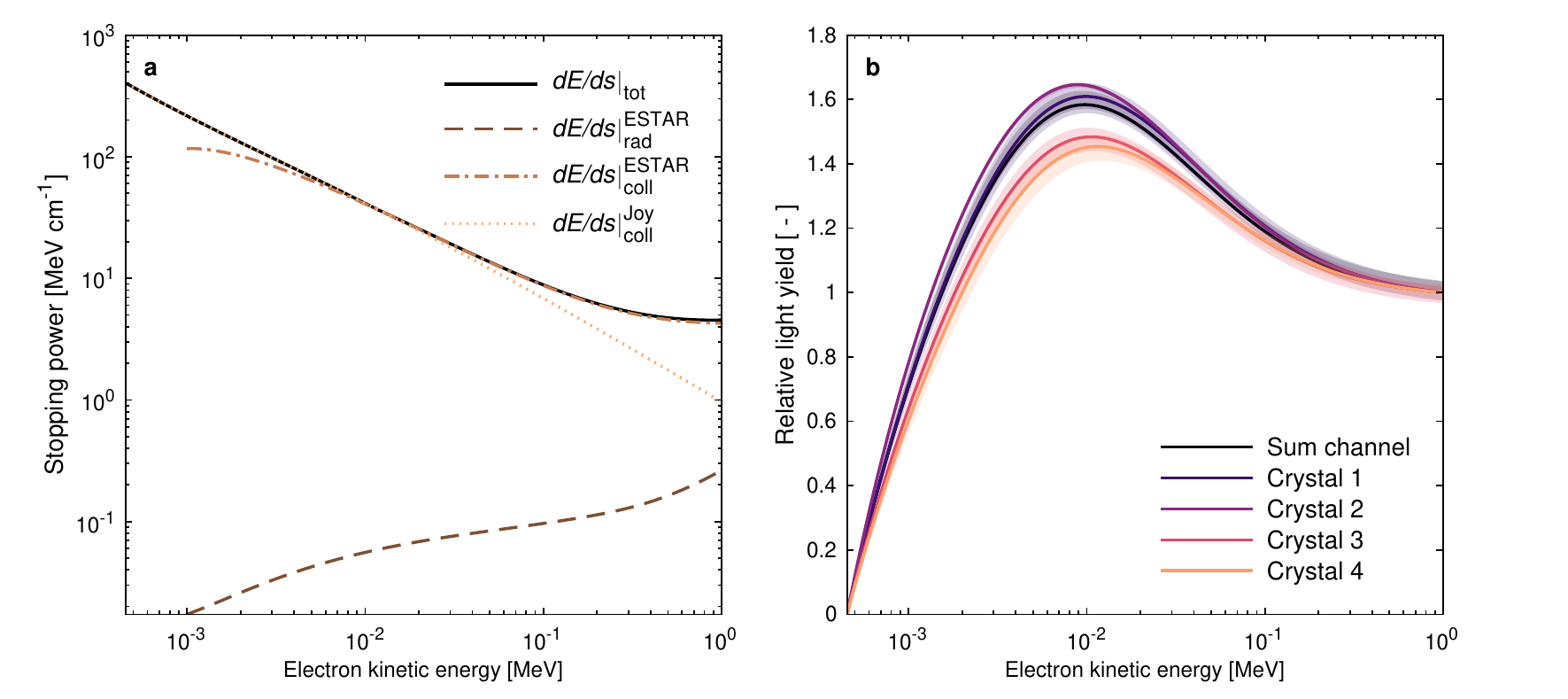}}
\caption[Light yield analysis]{\textbf{ Light yield analysis.} Here, we highlight the stopping power models for electrons in NaI(Tl) alongside the resulting relative light yield curves as a function of the electron kinetic energy $E_{\text{k}}$. \textbf{a}~Adopted stopping power model $dE/ds\mid_{\text{tot}}$ based on a modified Bethe-Bloch model $dE/ds\mid_{\text{coll}}^{\text{Joy}}$ for collisional losses at low energies derived by Joy and Luo \citep{Joy1989AnElectrons} as well as radiative and collisional losses at higher energies predicted by the \texttt{ESTAR} database \citep{Berger2017ESTAR2.0.1}. \textbf{b}~Relative light yield $\mathcal{L}(E_{\text{k}})/E_{\text{k}}$ as a function of $E_{\text{k}}$ for both, the sum channel and the individual scintillation crystals associated with the sum and single mode inversion pipelines, respectively. We applied the maximum a posteriori (MAP) probability point estimates for the individual model parameters, i.e. $\eta_{e/h}$, $dE/ds\mid_{\text{Ons}}$, $dE/ds\mid_{\text{Trap}}$ and $dE/ds\mid_{\text{Birks}}$, derived in the main study to compute the mean relative light yield function normalized at \qty{1}{MeV} according to \hyperref[eq:LYintegral]{Eq.~\ref{eq:LYintegral}}. In addition, we present \num{99}\% central credible intervals for each individual relative light yield function using the full set of posterior samples obtained by the sum and single mode inversion pipelines. A list of all material properties for NaI(Tl) used to compute the stopping power predictions as well as the resulting relative light yield functions can be found in \hyperref[tab:NaI]{Table~\ref{tab:NaI}}.}\label{fig:LightYield}
\end{figure}

\newpage

\begin{figure}[h!]%
\centerline{
\includegraphics[]{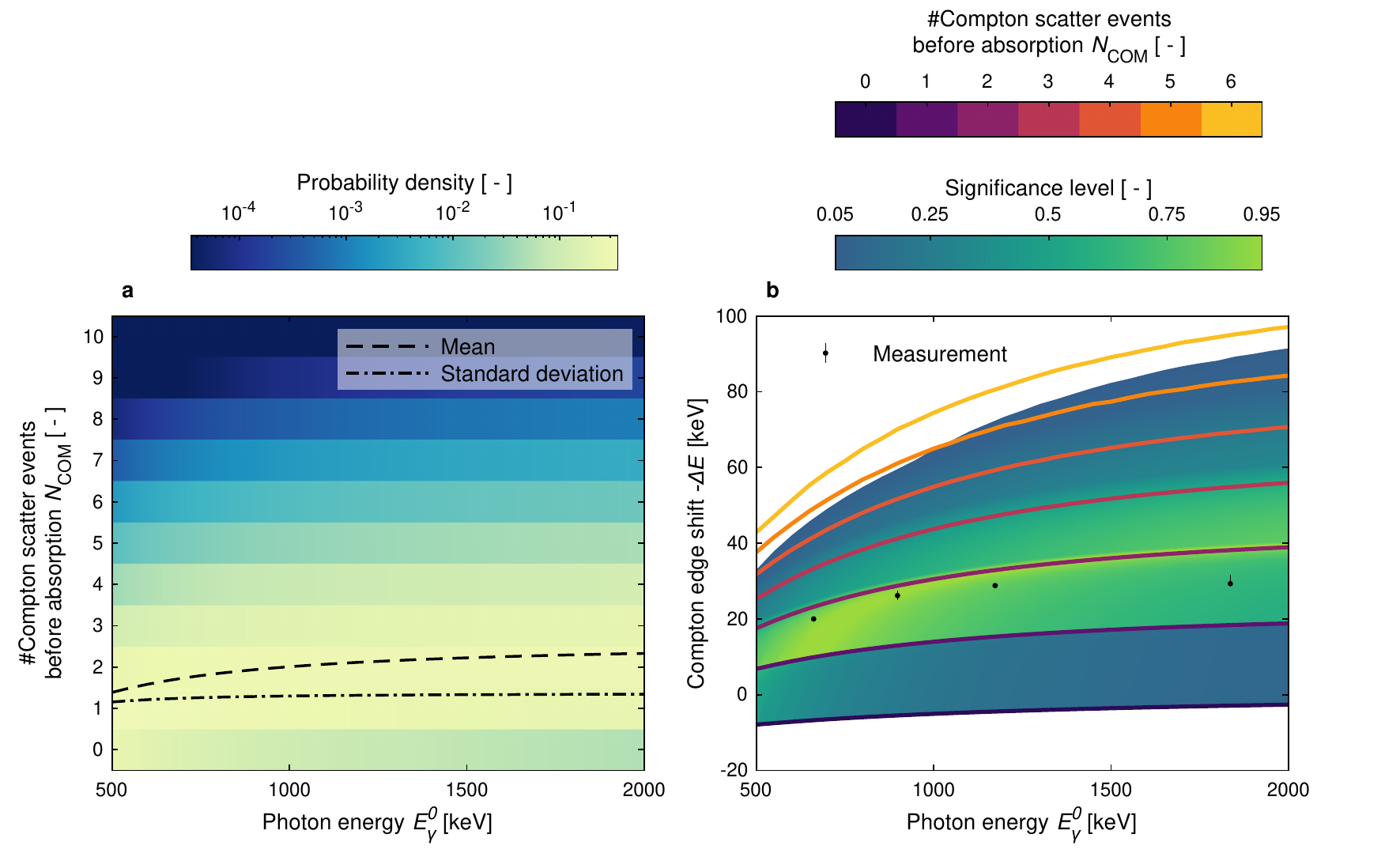}}
\caption[Semi-analytical model results]{\textbf{ Semi-analytical model results.} Here, we present a Monte Carlo based estimate of the number of Compton scatter (COM) events before absorption ($N_{\text{COM}}$) together with Compton edge shift predictions using a simplified semi-analytical model derived in \hyperref[subsec:CEshift]{Section~\ref{subsec:CEshift}} for a \qtyproduct[list-units = single]{10.2 x 10.2 x 40.6}{cm} prismatic NaI(Tl) scintillation crystal as function of the initial photon energy $E_{\gamma}^0$. \textbf{a}~Probability density for $N_{\text{COM}}$ as a function of $E_{\gamma}^0$ together with the mean and standard deviation values estimated by Monte Carlo simulations using the multi-purpose code \texttt{FLUKA} \citep{Ahdida2022NewCode}. \textbf{b}~Predicted median of the negative Compton edge shift $-\Delta E$ discriminated for individual number of COM events as well as the confidence interval based on the full distribution for different significance levels in the range \numrange{0.05}{0.95} adopting the relative light yield function for the sum channel. In addition, we show the measured mean Compton edge shifts for the sum channel together with the \num{99}\% confidence intervals for four different Compton edges, i.e. [477.334(3)\,,\,699.133(3)\,,\,963.419(3)\,,\,1611.77(1)]\,\unit{keV} associated with the photon emission lines of the radionuclides \{$^{137}\text{Cs}$\,,\,$^{88}\text{Y}$\,,\,$^{60}\text{Co}$\,,\,$^{88}\text{Y}$\} at [661.657(3)\,,\,898.042(3)\,,\,1173.228(3)\,,\,1836.063(3)]\,\unit{keV}, respectively.}\label{fig:CEshiftPred}
\end{figure}

\newpage

\begin{figure}[h!]%
\centerline{
\includegraphics[]{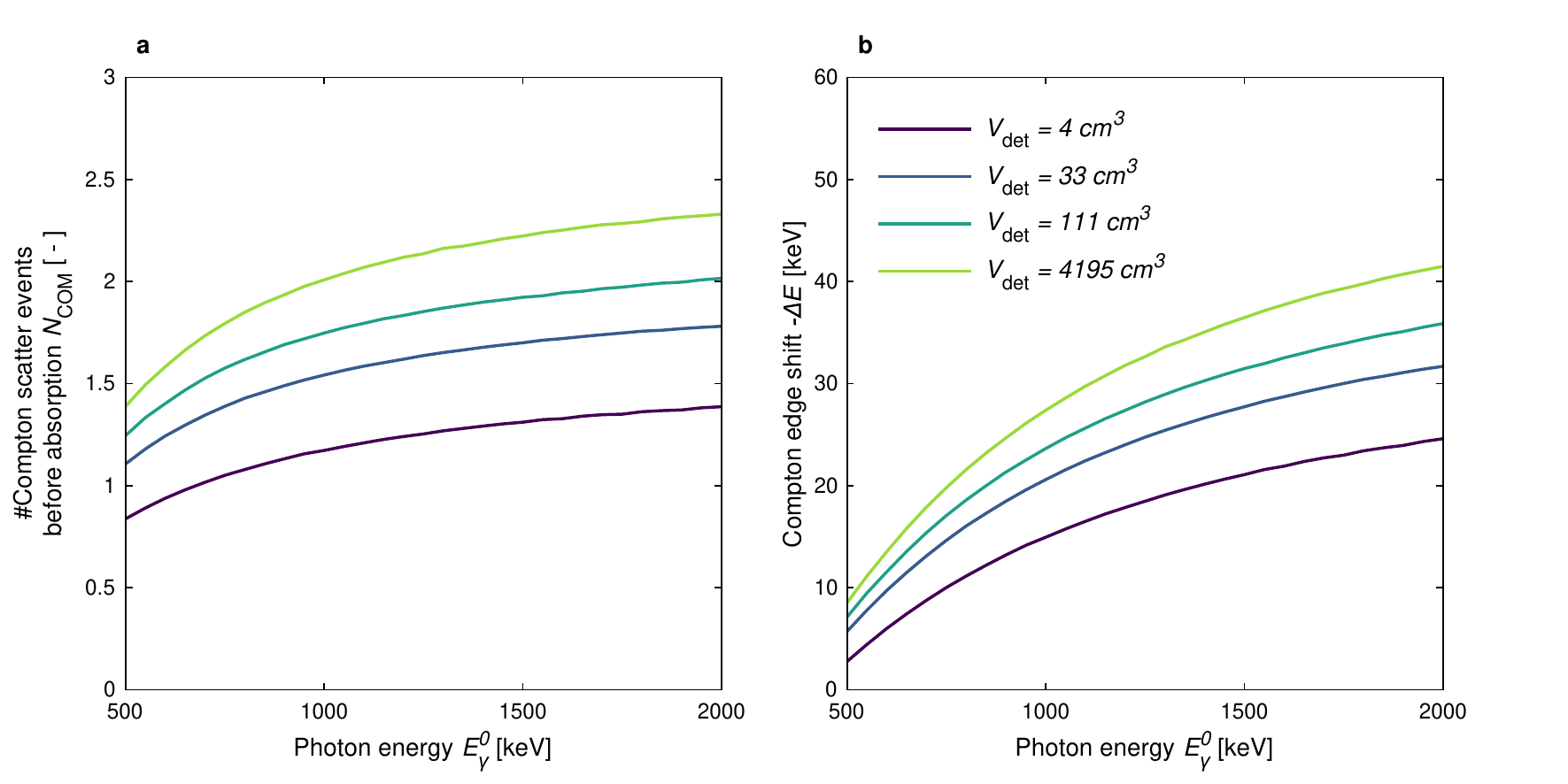}}
\caption[Predicted Compton edge shift for different scintillator sizes]{\textbf{ Predicted Compton edge shift for different scintillator sizes.} Here, we show Monte Carlo based estimates of the mean number of Compton scatter (COM) events before absorption ($N_{\text{COM}}$) together with mean Compton edge shift predictions as function of the initial photon energy $E_{\gamma}^0$ adopting a simplified semi-analytical model derived in \hyperref[subsec:CEshift]{Section~\ref{subsec:CEshift}} for four different NaI(Tl) scintillation crystals, i.e. three equilateral cylindrical crystals with characteristic lengths [2.54\,,\,5.08\,,\,7.62]\,\unit{cm} and associated volumes ($V_{\text{det}}$) [4\,,\,33\,,\,111]\,\unit{cm^{3}} as well as a \qtyproduct[list-units = single]{10.2 x 10.2 x 40.6}{cm} prismatic crystal with a volume of \qty{4195}{cm^{3}}. \textbf{a}~Mean number of COM events before absorption ($N_{\text{COM}}$) for four different NaI(Tl) scintillation crystals as a function of $E_{\gamma}^0$ obtained by Monte Carlo simulations using the multi-purpose code \texttt{FLUKA} \citep{Ahdida2022NewCode}. \textbf{b}~Mean negative Compton edge shift $-\Delta E$ as function of the initial photon energy $E_{\gamma}^0$ for four different NaI(Tl) scintillation crystals predicted by the semi-analytical model derived in \hyperref[subsec:CEshift]{Section~\ref{subsec:CEshift}}.}\label{fig:CEdetsize}
\end{figure}

\newpage


\section{Supplementary Tables}\label{sec:Tables}

\vspace{20mm}

\begin{table}[ht]
\begin{center}
\begin{minipage}{\textwidth}
\caption[Prior distribution summary]{\textbf{ Prior distribution summary.} This table summarizes the adopted prior distributions applied to the sum and single mode inversion pipelines for the individual model parameters, i.e. the Birks related stopping power parameter $dE/ds\mid_{\text{Birks}}$, the trapping related stopping power parameter $dE/ds\mid_{\text{Trap}}$, the free carrier fraction $\eta_{e/h}$ as well as the discrepancy model variance $\sigma^2_{\varepsilon}$. In addition, we list the consulted studies, which motivated the individual priors.}
\begin{tabular*}{\textwidth}{l l l p{0.1mm} p{1.58cm} p{0.1mm} l l l l}
\toprule
Pipeline & Variable & Prior\footnotemark[1] & \multicolumn{4}{l}{Prior parameters\footnotemark[2]} & Truncation & Unit & References\\
\midrule
 sum & $dE/ds\mid_{\text{Birks}}$ & Uniform  & $x_l$ & $=1.5\cdot10^{2}$ & $x_u$ & $=4.5\cdot10^{2}$ & \unit{-}                 & \unit{MeV~cm^{-1}}   & \citep{Payne2009,Payne2011,Payne2014} \\
& $dE/ds\mid_{\text{Trap}}$ & Uniform   & $x_l$ & $=1.0\cdot10^{1}$ & $x_u$ & $=1.5\cdot10^{1}$ & \unit{-}                      & \unit{MeV~cm^{-1}}   & \citep{Payne2014} \\
& $\eta_{e/h}$ & Uniform & $x_l$ & $=4.5\cdot10^{-1}$ & $x_u$ & $=6.5\cdot10^{-1}$ & \unit{-}                   & \unit{-}             & \citep{Payne2009,Payne2011,Payne2014}\\
\vspace{2mm}
& $\sigma_{\varepsilon}^{2}$ & Uniform & $x_l$ & $=0$ & $x_u$ & $=\sigma_{\varepsilon,\text{max}}^{2}$  & \unit{-}       & \unit{cps^{2}~Bq^{-2}}  & \unit{-}\\
 single & $dE/ds\mid_{\text{Birks}}$ & Gaussian & $\mu$ & $=3.23\cdot10^{2}$ & $\sigma$ & $=2.28\cdot10^{1}$ &  $[1.5,4.5]\cdot10^{2}\phantom{^{-}}$                  & \unit{MeV~cm^{-1}}   & \citep{Payne2009,Payne2011,Payne2014} \\
& $dE/ds\mid_{\text{Trap}}$ & Gaussian & $\mu$ & $=1.43\cdot10^{1}$ & $\sigma$ & $=7.51\cdot10^{-1}$ & $[1.0,1.8]\cdot10^{1}\phantom{^{-}}$                       & \unit{MeV~cm^{-1}}   & \citep{Payne2014} \\
& $\eta_{e/h}$ & Gaussian & $\mu$ & $=5.94\cdot10^{-1}$ & $\sigma$ & $=6.83\cdot10^{-3}$ & $[4.5,6.5]\cdot10^{-1}$                   & \unit{-}             & \citep{Payne2009,Payne2011,Payne2014}\\
& $\sigma_{\varepsilon}^{2}$ & Uniform  & $x_l$ & $=0$ & $x_u$ & $=\sigma_{\varepsilon,\text{max}}^{2}$ & \unit{-}       & \unit{cps^{2}~Bq^{-2}}  & \unit{-}\\
\botrule
\end{tabular*}
\footnotetext[1]{We use the continuous uniform distribution $\mathcal{U}\left(x_l,x_u\right)$ with the lower and upper boundary parameters $x_l$ and $x_u$ to denote the uniform prior. By the Gaussian prior, we refer to the truncated univariate normal distribution $\mathcal{N}\left(\mu,\sigma,x_l,x_u\right)$ with mean $\mu$, standard deviation $\sigma$ and truncation $[x_l,x_u]$.}
\footnotetext[2]{We define the upper limit for the discrepancy model variance $\sigma_{\varepsilon,\text{max}}^{2}$ as $\langle c_{\text{exp}}^{2}\rangle$ with $c_{\text{exp}}$ being the measured net count rate over the spectral Compton edge domain $\mathcal{D}_E$ for the corresponding detection channel (cf. Methods in the main study).}
\label{tab:prior}
\end{minipage}
\end{center}
\end{table}

\newpage

\begin{table}[ht]
\begin{center}
\begin{minipage}{\textwidth}
\caption[Posterior statistics summary]{\textbf{Posterior statistics summary.} This table includes posterior point and dispersion estimators for the Bayesian inverted non-proportional scintillation models obtained by the sum and single mode inversion pipelines. The listed estimators are the maximum a posteriori (MAP) probability estimate $\boldsymbol{x}_{\mathrm{MAP}}$, the posterior mean $\boldsymbol{x}_{\mathrm{Mean}}$ and the posterior median $\boldsymbol{x}_{\mathrm{Median}}$ together with the 95\% credible interval and the posterior standard deviation $\boldsymbol{\sigma}_{\boldsymbol{x}}$ for the parameters $\boldsymbol{x}\coloneqq\left( dE/ds\mid_{\text{Birks}},~dE/ds\mid_{\text{Trap}},~\eta_{e/h},~\sigma^2_{\varepsilon} \right)^{\intercal}$, i.e. the Birks related stopping power parameter $dE/ds\mid_{\text{Birks}}$, the trapping related stopping power parameter $dE/ds\mid_{\text{Trap}}$, the free carrier fraction $\eta_{e/h}$ as well as the discrepancy model variance $\sigma^2_{\varepsilon}$.}
\begin{tabular*}{\textwidth}{@{\extracolsep{\fill}}l l l l l l l l}
\toprule
Pipeline & Variable & $\boldsymbol{x}_{\mathrm{MAP}}$ & $\boldsymbol{x}_{\mathrm{Mean}}$ & $\boldsymbol{x}_{\mathrm{Median}}$ & 95\% credible interval\footnotemark[1] & $\boldsymbol{\sigma}_{\boldsymbol{x}}$ & Unit \\
\midrule
sum & $dE/ds\mid_{\text{Birks}}$     & $3.22\cdot10^{2}$      & $3.23\cdot10^{2}$     & $3.22\cdot10^{2}$     & $\left[2.78, 3.68\right]\cdot10^{2}$  & $2.28\cdot10^{1}$  & \unit{MeV~cm^{-1}}  \\
&$dE/ds\mid_{\text{Trap}}$      & $1.46\cdot10^{1}$      & $1.43\cdot10^{1}$     & $1.44\cdot10^{1}$     & $\left[1.15, 1.48\right]\cdot10^{1}$  & $7.51\cdot10^{-1}$ & \unit{MeV~cm^{-1}}   \\
&$\eta_{e/h}$                   & $5.96\cdot10^{-1}$     & $5.94\cdot10^{-1}$    & $5.95\cdot10^{-1}$    & $\left[5.79, 6.06\right]\cdot10^{-1}$ & $6.83\cdot10^{-3}$ & \unit{-}             \\ 
\vspace{2mm}
&$\sigma_{\varepsilon}^{2}$     & $1.24\cdot10^{-1}$     & $1.37\cdot10^{-1}$    & $1.34\cdot10^{-1}$    & $\left[0.98, 1.92\right]\cdot10^{-1}$ & $2.40\cdot10^{-2}$ & \unit{cps^{2}~Bq^{-2}} \\
single & $dE/ds\mid_{\text{Birks}}$     & $3.17\cdot10^{2}$      & $3.17\cdot10^{2}$     & $3.16\cdot10^{2}$     & $\left[2.87, 3.49\right]\cdot10^{2}$  & $1.57\cdot10^{1}$  & \unit{MeV~cm^{-1}}  \\
(crystal 1) & $dE/ds\mid_{\text{Trap}}$      & $1.33\cdot10^{1}$      & $1.32\cdot10^{1}$     & $1.32\cdot10^{1}$     & $\left[1.22, 1.43\right]\cdot10^{1}$  & $5.48\cdot10^{-1}$ & \unit{MeV~cm^{-1}}   \\
&$\eta_{e/h}$                   & $6.05\cdot10^{-1}$     & $6.05\cdot10^{-1}$    & $6.05\cdot10^{-1}$    & $\left[5.95, 6.14\right]\cdot10^{-1}$ & $4.96\cdot10^{-3}$ & \unit{-}             \\
\vspace{2mm}
&$\sigma_{\varepsilon}^{2}$     & $2.61\cdot10^{-2}$     & $2.83\cdot10^{-2}$    & $2.83\cdot10^{-2}$    & $\left[2.03, 3.94\right]\cdot10^{-2}$ & $4.92\cdot10^{-3}$ & \unit{cps^{2}~Bq^{-2}} \\

single & $dE/ds\mid_{\text{Birks}}$     & $4.15\cdot10^{2}$      & $4.15\cdot10^{2}$     & $4.15\cdot10^{2}$     & $\left[3.93, 4.36\right]\cdot10^{2}$  & $1.08\cdot10^{1}$  & \unit{MeV~cm^{-1}}  \\
(crystal 2) & $dE/ds\mid_{\text{Trap}}$      & $1.41\cdot10^{1}$      & $1.41\cdot10^{1}$     & $1.41\cdot10^{1}$     & $\left[1.38, 1.43\right]\cdot10^{1}$  & $1.08\cdot10^{-1}$ & \unit{MeV~cm^{-1}}   \\
&$\eta_{e/h}$                   & $5.94\cdot10^{-1}$     & $5.94\cdot10^{-1}$    & $5.94\cdot10^{-1}$    & $\left[5.88, 5.99\right]\cdot10^{-1}$ & $2.73\cdot10^{-3}$ & \unit{-}             \\
\vspace{2mm}
&$\sigma_{\varepsilon}^{2}$     & $6.92\cdot10^{-3}$     & $7,64\cdot10^{-3}$    & $7.50\cdot10^{-3}$    & $\left[5.44, 10.68\right]\cdot10^{-3}$ & $1.36\cdot10^{-3}$ & \unit{cps^{2}~Bq^{-2}} \\

single & $dE/ds\mid_{\text{Birks}}$     & $2.84\cdot10^{2}$      & $2.81\cdot10^{2}$     & $2.81\cdot10^{2}$     & $\left[2.47, 3.19\right]\cdot10^{2}$  & $1.85\cdot10^{1}$  & \unit{MeV~cm^{-1}}  \\
(crystal 3) & $dE/ds\mid_{\text{Trap}}$      & $1.50\cdot10^{1}$      & $1.53\cdot10^{1}$     & $1.52\cdot10^{1}$     & $\left[1.47, 1.61\right]\cdot10^{1}$  & $3.71\cdot10^{-1}$ & \unit{MeV~cm^{-1}}   \\
&$\eta_{e/h}$                   & $5.75\cdot10^{-1}$     & $5.75\cdot10^{-1}$    & $5.76\cdot10^{-1}$    & $\left[5.65, 5.86\right]\cdot10^{-1}$ & $5.47\cdot10^{-3}$ & \unit{-}             \\
\vspace{2mm}
&$\sigma_{\varepsilon}^{2}$     & $2.15\cdot10^{-2}$     & $2.31\cdot10^{-2}$    & $2.30\cdot10^{-2}$    & $\left[1.65, 3.23\right]\cdot10^{-2}$ & $4.08\cdot10^{-3}$ & \unit{cps^{2}~Bq^{-2}} \\

single & $dE/ds\mid_{\text{Birks}}$     & $2.54\cdot10^{2}$      & $2.56\cdot10^{2}$     & $2.56\cdot10^{2}$     & $\left[2.23, 2.88\right]\cdot10^{2}$  & $1.63\cdot10^{1}$  & \unit{MeV~cm^{-1}}  \\
(crystal 4) & $dE/ds\mid_{\text{Trap}}$      & $1.37\cdot10^{1}$      & $1.37\cdot10^{1}$     & $1.37\cdot10^{1}$     & $\left[1.33, 1.40\right]\cdot10^{1}$  & $1.96\cdot10^{-1}$ & \unit{MeV~cm^{-1}}   \\
&$\eta_{e/h}$                   & $5.75\cdot10^{-1}$     & $5.74\cdot10^{-1}$    & $5.74\cdot10^{-1}$    & $\left[5.66, 5.82\right]\cdot10^{-1}$ & $3.95\cdot10^{-3}$ & \unit{-}             \\
&$\sigma_{\varepsilon}^{2}$     & $1.05\cdot10^{-2}$     & $1.15\cdot10^{-2}$    & $1.14\cdot10^{-2}$    & $\left[0.82, 1.60\right]\cdot10^{-2}$ & $2.00\cdot10^{-3}$ & \unit{cps^{2}~Bq^{-2}} \\
\botrule
\end{tabular*}
\label{tab:post}
\footnotetext[1]{Central credible interval with a probability mass of 95\%.}
\end{minipage}
\end{center}
\end{table}

\newpage

\newpage

\begin{table}[ht]
\begin{center}
\begin{minipage}{\textwidth}
\caption[Compton edge domain sensitivity]{\textbf{Compton edge domain sensitivity.} To investigate the sensitivity of the selected Compton edge domain $\mathcal{D}_E\coloneqq \left\{E:E_{\text{CE}}-3\cdot\sigma_{\text{tot}}\left(E_{\text{CE}}\right)\leq E \leq E_{\text{FEP}}-2\cdot\sigma_{\text{tot}}\left(E_{\text{FEP}}\right)\right\}$ (cf. Methods in the main study) on the Bayesian inversion results, we have altered the domain size by 2.5\% symmetrically with respect to the domain boundaries and performed the emulator training and Bayesian inversion computation on this new domain using the sum mode inversion pipeline. This alteration corresponds to $\approx 20\%$ of the observed Compton edge shift (cf. Methods in the main study). These tables summarizes the posterior point and dispersion estimator results for these additional computations, i.e. the maximum a posteriori (MAP) probability estimate $\boldsymbol{x}_{\mathrm{MAP}}$, the posterior mean $\boldsymbol{x}_{\mathrm{Mean}}$ and the posterior median $\boldsymbol{x}_{\mathrm{Median}}$ together with the 95\% credible interval and the posterior standard deviation $\boldsymbol{\sigma}_{\boldsymbol{x}}$ for the parameters $\boldsymbol{x}\coloneqq\left( dE/ds\mid_{\text{Birks}},~dE/ds\mid_{\text{Trap}},~\eta_{e/h},~\sigma^2_{\varepsilon} \right)^{\intercal}$. These parameters are the Birks related stopping power parameter $dE/ds\mid_{\text{Birks}}$, the trapping related stopping power parameter $dE/ds\mid_{\text{Trap}}$, the free carrier fraction $\eta_{e/h}$ as well as the discrepancy model variance $\sigma^2_{\varepsilon}$.}
\begin{subtable}[h]{\textwidth}
    \centering
    \caption{2.5\%  decrease in $\mathcal{D}_E$}
    \begin{tabular*}{\textwidth}{@{\extracolsep{\fill}}c l l l l l l}
    \toprule
    Parameter & $\boldsymbol{x}_{\mathrm{MAP}}$ & $\boldsymbol{x}_{\mathrm{Mean}}$ & $\boldsymbol{x}_{\mathrm{Median}}$ & 95\% credible interval\footnotemark[1] & $\boldsymbol{\sigma}_{\boldsymbol{x}}$ & Unit \\
    \midrule
    $dE/ds\mid_{\text{Birks}}$     & $3.08\cdot10^{2}$      & $3.10\cdot10^{2}$     & $3.08\cdot10^{2}$     & $\left[2.79, 3.48\right]\cdot10^{2}$  & $2.14\cdot10^{1}$  & \unit{MeV~cm^{-1}}  \\
    $dE/ds\mid_{\text{Trap}}$      & $1.50\cdot10^{1}$      & $1.46\cdot10^{1}$     & $1.47\cdot10^{1}$     & $\left[1.33, 1.50\right]\cdot10^{1}$  & $6.24\cdot10^{-1}$ & \unit{MeV~cm^{-1}}   \\
    $\eta_{e/h}$                   & $5.93\cdot10^{-1}$     & $5.92\cdot10^{-1}$    & $5.92\cdot10^{-1}$    & $\left[5.82, 6.01\right]\cdot10^{-1}$ & $5.85\cdot10^{-3}$ & \unit{-}             \\
    $\sigma_{\varepsilon}^{2}$     & $1.05\cdot10^{-1}$     & $1.12\cdot10^{-1}$    & $1.18\cdot10^{-1}$    & $\left[0.89, 1.56\right]\cdot10^{-1}$ & $2.08\cdot10^{-2}$ & \unit{cps^{2}~Bq^{-2}} \\
    \botrule
    \end{tabular*}
    \label{tab:sensA}
\end{subtable}
\begin{subtable}[h]{\textwidth}
   \centering
   \caption{2.5\% increase in $\mathcal{D}_E$}
   \begin{tabular*}{\textwidth}{@{\extracolsep{\fill}}c l l l l l l}
   \toprule
   Parameter & $\boldsymbol{x}_{\mathrm{MAP}}$ & $\boldsymbol{x}_{\mathrm{Mean}}$ & $\boldsymbol{x}_{\mathrm{Median}}$ & 95\% credible interval\footnotemark[1] & $\boldsymbol{\sigma}_{\boldsymbol{x}}$ & Unit \\
   \midrule
   $dE/ds\mid_{\text{Birks}}$     & $3.34\cdot10^{2}$      & $3.30\cdot10^{2}$     & $3.31\cdot10^{2}$     & $\left[2.90, 3.70\right]\cdot10^{2}$  & $2.46\cdot10^{1}$  & \unit{MeV~cm^{-1}}  \\
   $dE/ds\mid_{\text{Trap}}$      & $1.46\cdot10^{1}$      & $1.42\cdot10^{1}$     & $1.43\cdot10^{1}$     & $\left[1.22, 1.48\right]\cdot10^{1}$  & $8.70\cdot10^{-1}$ & \unit{MeV~cm^{-1}}   \\
   $\eta_{e/h}$                   & $5.95\cdot10^{-1}$     & $5.94\cdot10^{-1}$    & $5.95\cdot10^{-1}$    & $\left[5.82, 6.04\right]\cdot10^{-1}$ & $7.10\cdot10^{-3}$ & \unit{-}             \\
   $\sigma_{\varepsilon}^{2}$     & $1.42\cdot10^{-1}$     & $1.58\cdot10^{-1}$    & $1.54\cdot10^{-1}$    & $\left[1.19, 2.01\right]\cdot10^{-1}$ & $2.75\cdot10^{-2}$ & \unit{cps^{2}~Bq^{-2}} \\
   \botrule
   \end{tabular*}
   \label{tab:sensB}
\end{subtable}
\label{tab:sens}
\footnotetext[1]{Central credible interval with a probability mass of 95\%.}
\end{minipage}
\end{center}
\end{table}

\newpage

\begin{table}[ht]
\begin{center}
\begin{minipage}{\textwidth}
\caption[Empirical model summary]{\textbf{Empirical model summary.} This table summarizes the adopted marginal distributions for the empirical model parameters discussed in \hyperref[subsec:Uncertainty]{Section~\ref{subsec:Uncertainty}} to quantify the systematic uncertainties for the Bayesian calibrated NPSM simulations. These parameters are the calibration factor $D_1$ as well as the empirical resolution parameters $B_1^{\ast}$ and $B_2$.}
\begin{tabular*}{\textwidth}{@{\extracolsep{\fill}}l l l l l l l l}
\toprule
Crystal & Variable & Distribution  & \multicolumn{2}{l}{Distribution parameters\footnotemark[1]} & Truncation & Unit \\
\midrule
1 & $D_1$   & Gaussian & $\mu=\phantom{-}3.31\cdot 10^{-1}$ & $\sigma=3.6\cdot 10^{-4}$ &        \unit{-}                   & \unit{keV^{-1}} \\
&$B_1^{\ast}$   & Gaussian & $\mu=-5.71\cdot 10^{-1}$ & $\sigma=5.5\cdot 10^{-2}$ &            \unit{-}            & \unit{-} \\
\vspace{2mm}
&$B_2$   & Gaussian & $\mu=\phantom{-}6.39\cdot 10^{-1}$ & $\sigma=1.1\cdot 10^{-2}$ & $\left[0,\infty\right)$                  & \unit{-} \\

2 & $D_1$   & Gaussian & $\mu=\phantom{-}3.32\cdot 10^{-1}$ & $\sigma=4.3\cdot 10^{-4}$ &        \unit{-}                   & \unit{keV^{-1}} \\
&$B_1^{\ast}$   & Gaussian & $\mu=-6.22\cdot 10^{-1}$ & $\sigma=6.1\cdot 10^{-2}$ &            \unit{-}            & \unit{-} \\
\vspace{2mm}
&$B_2$   & Gaussian & $\mu=\phantom{-}6.31\cdot 10^{-1}$ & $\sigma=1.2\cdot 10^{-2}$ & $\left[0,\infty\right)$                  & \unit{-} \\

3 & $D_1$   & Gaussian & $\mu=\phantom{-}3.35\cdot 10^{-1}$ & $\sigma=2.1\cdot 10^{-4}$ &         \unit{-}                 & \unit{keV^{-1}} \\
&$B_1^{\ast}$   & Gaussian & $\mu=-6.23\cdot 10^{-1}$ & $\sigma=6.0\cdot 10^{-2}$ &             \unit{-}           & \unit{-} \\
\vspace{2mm}
&$B_2$   & Gaussian & $\mu=\phantom{-}6.54\cdot 10^{-1}$ & $\sigma=1.1\cdot 10^{-2}$ & $\left[0,\infty\right)$                  & \unit{-} \\

4 & $D_1$   & Gaussian & $\mu=\phantom{-}3.34\cdot 10^{-1}$ & $\sigma=1.9\cdot 10^{-4}$ &        \unit{-}                   & \unit{keV^{-1}} \\
&$B_1^{\ast}$   & Gaussian & $\mu=-6.86\cdot 10^{-1}$ & $\sigma=5.7\cdot 10^{-2}$ &             \unit{-}           & \unit{-} \\
&$B_2$   & Gaussian & $\mu=\phantom{-}6.44\cdot 10^{-1}$ & $\sigma=1.0\cdot 10^{-2}$ & $\left[0,\infty\right)$                  & \unit{-} \\
\botrule
\end{tabular*}
\footnotetext[1]{By the Gaussian distribution, we refer to the (truncated) univariate normal distribution $\mathcal{N}\left(\mu,\sigma,x_l,x_u\right)$ with mean $\mu$, standard deviation $\sigma$ and optional truncation $[x_l,x_u]$.}
\label{tab:marginal}
\end{minipage}
\end{center}
\end{table}

\newpage

\begin{table}[ht]
\begin{center}
\begin{minipage}{14cm}
\caption[Material properties for NaI(Tl) scintillator]{\textbf{Material properties for NaI(Tl) scintillator.} In this table, we summarize all material properties associated with the inorganic scintillator NaI(Tl), which were used for the Compton edge shift analysis in \hyperref[subsec:CEshift]{Section~\ref{subsec:CEshift}}. Moreover, we list the references, which were consulted to retrieve the individual numerical values.}
\begin{tabular*}{14cm}{@{\extracolsep{\fill}}l l l l l}
\toprule
Quantity & Symbol & Numerical value & Unit & Reference\\
\midrule
Atomic number & $Z$ & 64 & \unit{-} & \citep{Berger2017ESTAR2.0.1} \\
Mass density & $\rho$ & 3.667 & \unit{g.cm^{-3}} & \citep{Berger2017ESTAR2.0.1} \\
Mean excitation energy & $I$ & 452 & \unit{eV} & \citep{Berger2017ESTAR2.0.1} \\
Molecular weight &$A$ & 149.89424 & \unit{-} & \citep{Berger2017ESTAR2.0.1} \\
Stopping power correction factor & $c$ & 2.8 & \unit{-} & \citep{Payne2009,Payne2011} \\
\botrule
\end{tabular*}
\label{tab:NaI}
\end{minipage}
\end{center}
\end{table}

\newpage


\section{Supplementary Algorithms}\label{sec:Codes}

\vspace{20mm}

\begin{algorithm}
\caption{\texttt{COMSCW}($dE/ds\mid_{\text{Birks}}$\,,\,$dE/ds\mid_{\text{Ons}}$\,,\,$dE/ds\mid_{\text{Trap}}$\,,\,$\eta_{e/h}$)\newline \texttt{COMSCW} is a custom user-routine for the multi-purpose Monte Carlo code \texttt{FLUKA} \citep{Ahdida2022NewCode} called at each energy deposition event in the scintillation crystal. We adapt this routine by weighting each electron or positron energy deposition event by the adopted non-proportional scintillation model (NPSM) \citep{Payne2014}. The algorithm accounts for both continuous as well as local energy deposition events. As described in the Methods section in the main study, we set a kinetic energy threshold of 1~\unit{keV} below which the 
electrons and positrons as well as particles generated by atomic deexcitation are no longer transported and their energy is deposited 
on the spot. We refer to these events as ”local” energy deposition events. On the other hand, above this threshold, ionization losses are evenly distributed along the particle step \citep{Ferrari1992AnTransport,Battistoni2015}. Hence, we call 
these events ”continuous”. The pseudo-code added below is a simplified version of the one implemented in our forward model. For more details, we kindly refer to the actual routine deposited on the ETH Research Collection repository: \url{https://doi.org/10.3929/ethz-b-000595727} \citep{Breitenmoser2023}.}
\label{algo1} 
\begin{algorithmic}[1]
\State \textbf{load} EventID \Comment{EventID = \{"continuous"\,,\,"local"\}}
\State \textbf{load} ParticleID \Comment{Particle type}
\State \textbf{load} $dE$ \Comment{Deposited energy}
\If{EventID = "continuous" AND ( ParticleID = "electron" OR \newline 
 \indent ParticleID = "positron" )}
        \State \textbf{load} $ds$ \Comment{Curved particle path}
        \State $S = dE / ds$ \Comment{Estimate stopping power $S$}
\ElsIf{EventID = "local"}
\State \textbf{load} $E_{\text{k}}$ \Comment{Kinetic particle energy}
 \If{ParticleID = "electron"}
        \State $S$ = \texttt{EDEDXT}($E_{\text{k}}$)  
        \Comment{\parbox[t]{.4\linewidth}{\raggedleft Call built-in stopping power function \texttt{EDEDXT} for electrons}}\raggedright
\ElsIf{ParticleID = "positron"}
       \State $S$ = \texttt{PDEDXT}($E_{\text{k}}$)  
        \Comment{\parbox[t]{.4\linewidth}{\raggedleft Call built-in stopping power function \texttt{PDEDXT} for positrons}}\raggedright
\EndIf
\EndIf

\State $dL = dE \times \frac{1-\eta_{e/h}\exp{ \left[ -\frac{S}{dE/ds\mid_{\text{Ons}}} \exp{\left( - \frac{dE/ds\mid_{\text{Trap}}}{S} \right)}\right] }}{1+\frac{S}{dE/ds\mid_{\text{Birks}}}}  $
\Comment{Weight deposited energy by NPSM}







\end{algorithmic}
\end{algorithm}

\newpage

